\documentclass[lettersize,journal]{IEEEtran}

\usepackage{amsmath,amsfonts,amssymb}
\usepackage[nointegrals]{wasysym}   % avoid clash with amsmath integrals
\usepackage{pifont}                 % \ding{} check/cross
\usepackage{textcomp}
\usepackage[T1]{fontenc}
\usepackage{fontawesome5}

\usepackage{algorithm}
\usepackage{algorithmic}

\usepackage{graphicx}
\usepackage{adjustbox}
\usepackage{rotating}
\usepackage[dvipsnames]{xcolor}
\usepackage[caption=false,font=normalsize,labelfont=sf,textfont=sf]{subfig}
\usepackage{caption}

\usepackage[dvipsnames,svgnames]{xcolor}
\usepackage{tikz}
\usepackage{pgfplots}
\usepackage{pgf-pie}
\pgfplotsset{compat=1.18}
\usetikzlibrary{%
  arrows.meta, calc, backgrounds, fit, positioning, shapes,
  shapes.geometric, shadows, shadows.blur, trees,
  decorations.pathreplacing, decorations.markings, matrix
}

\usepackage{array}
\usepackage{booktabs}
\usepackage{multirow}
\usepackage{threeparttable}
\usepackage{tabularx}
\usepackage{xltabular}
\usepackage{ltablex}                % adds longtable features to tabularx
\keepXColumns                       % <-- restore normal tabularx (fixes the error)
\usepackage{enumitem}

\usepackage{stfloats}
\usepackage{xurl}                   % better line-breaking than url
\usepackage{cite}
\usepackage{verbatim}
\usepackage{comment}
\usepackage{pdflscape}
\usepackage{siunitx}
\usepackage{soul}
\usepackage[normalem]{ulem}
\usepackage{balance}
\usepackage{fontawesome5}

\usepackage[acronym]{glossaries}
\usepackage{glossaries-prefix}
\usepackage{glossary-mcols}
\setglossarystyle{mcolindexgroup}
\makeglossaries
\newacronym[firstplural=Independent and identically distributed (i.i.d.)]{i.i.d.}{i.i.d.}{independent and identically distributed}
\newacronym[plural=ARIMAs, firstplural=AutoRegressive integrated moving averages (ARIMAs)]{ARIMA}{ARIMA}{autoregressive integrated moving average}
\newacronym[plural=PIEs, firstplural=Proportional integral enhanced (PIE)]{PIE}{PIE}{proportional integral enhanced}
\newacronym[plural=CoDels]{CoDel}{CoDel}{controlled delay}
\newacronym[plural=DRLs]{DRL}{DRL}{deep reinforcement learning}
\newacronym[plural=AQMs, firstplural=Active queue management schemes (AQMs)]{AQM}{AQM}{active queue management}
\newacronym[plural=RED, firstplural=Random early detection (RED)]{RED}{RED}{random early detection}
\newacronym[plural=URLLCs, firstplural=Ultra-reliable low-latency communications (URLLCs)]{URLLC}{URLLC}{ultra-reliable low-latency communication}
\newacronym[plural=mMTCs, firstplural=Massive machine type communications (mMTCs)]{mMTC}{mMTC}{massive machine type communication}
\newacronym[plural=UE, firstplural=User equipment (UE)]{UE}{UE}{user equipment}
\newacronym[plural=gNBs, firstplural=Next generation NodeBs (gNBs)]{gNB}{gNB}{next generation NodeB}
\newacronym[plural=MCSs, firstplural=Modulation and coding schemes (MCSs)]{MCS}{MCS}{modulation and coding scheme}
\newacronym[plural=FDD, firstplural=Frequency fivision fuplex (FDD)]{FDD}{FDD}{frequency division duplex}
\newacronym[plural=TDD, firstplural=Time division duplex (TDD)]{TDD}{TDD}{time division duplex}
\newacronym[plural=OFDM, firstplural=Orthogonal frequency division multiplexing (OFDM)]{OFDM}{OFDM}{orthogonal frequency division multiplexing}
\newacronym[plural=HARQs, firstplural=Hybrid automatic repeat requests (HARQs)]{HARQ}{HARQ}{hybrid automatic repeat request}
\newacronym[plural=PDFs, firstplural=Probability density functions (PDFs)]{PDF}{PDF}{probability density function}
\newacronym[plural=MDNs, firstplural=Mixture density detworks (MDNs)]{MDN}{MDN}{mixture density network}
\newacronym[plural=GMMs, firstplural=Gaussian mixture models (GMMs)]{GMM}{GMM}{Gaussian mixture model}
\newacronym[plural=KDEs, firstplural=Kernel density estimations (KDEs)]{KDE}{KDE}{kernel density estimation}
\newacronym[plural=GEVs, firstplural=Generalized extreme value distributions (GEVs)]{GEV}{GEV}{generalized extreme value distribution}
\newacronym[plural=GPDs, firstplural=Generalized Pareto distributions (GPDs)]{GPD}{GPD}{generalized pareto distribution}
\newacronym[plural=RMSEs, firstplural=Root mean square errors (RMSEs)]{RMSE}{RMSE}{root mean square error}
\newacronym[plural=NLLs, firstplural=Negative log-likelihoods (NLLs)]{NLL}{NLL}{negative log-likelihood}
\newacronym[plural=LSTMs, firstplural=Long short-term memory networks (LSTMs)]{LSTM}{LSTM}{long short-term memory}
\newacronym[plural=GRUs, firstplural=Gated recurrent units (GRUs)]{GRU}{GRU}{gated recurrent unit}
\newacronym{Bi-LSTM}{Bi-LSTM}{bidirectional LSTM}
\newacronym[plural=CNNs, firstplural=Convolutional neural networks (CNNs)]{CNN}{CNN}{convolutional neural network}
\newacronym[plural=SVMs, firstplural=Support vector machines (SVMs)]{SVM}{SVM}{support vector machine}
\newacronym[plural=SVRs, firstplural=Support vector regressor (SVRs)]{SVR}{SVR}{support vector regressor}
\newacronym[plural=KNNs, firstplural=k-nearest neighbors (KNNs)]{KNN}{KNN}{k-Nearest Neighbor}
\newacronym[plural=GBDTs, firstplural=Gradient boosted decision trees (GBDTs)]{GBDT}{GBDT}{gradient boosted decision tree}
\newacronym[plural=RFs, firstplural=Random forests (RFs)]{RF}{RF}{random forest}
\newacronym{XGBoost}{XGBoost}{extreme gradient boosting}
\newacronym{LightGBM}{LightGBM}{light gradient boosting machine}
\newacronym[plural=RNNs, firstplural=Recurrent neural networks (RNNs)]{RNN}{RNN}{recurrent neural network}
\newacronym[plural=GNNs, firstplural=Graph neural networks (GNNs)]{GNN}{GNN}{graph neural network}
\newacronym[plural=ACKs, firstplural=Acknowledgements (ACKs)]{ACK}{ACK}{acknowledgement}
\newacronym[plural=NACKs, firstplural=Negative acknowledgements (NACKs)]{NACK}{NACK}{negative acknowledgement}
\newacronym[plural=EVTs, firstplural=Extreme value theories (EVTs)]{EVT}{EVT}{extreme value theory}
\newacronym[plural=i.i.d., firstplural=Independent and identically distributed (i.i.d.)]{iid}{i.i.d.}{independent and identically distributed}
\newacronym[plural=CDEs,firstplural=Conditional density estimation (CDE)]{CDE}{CDE}{conditional density estimation}
\newacronym[plural=CDFs,firstplural=Cumulative density functions (CDF)]{CDF}{CDF}{cumulative density function}
\newacronym[plural=MLEs,firstplural=Maximum likelihood estimations (MLE)]{MLE}{MLE}{maximum likelihood estimation}
\newacronym[plural=KL-divergences,firstplural=Kullback-Leibler divergences (KL-divergence)]{KL-divergence}{KL-divergence}{Kullback-Leibler divergence}
\newacronym[]{PoT}{PoT}{peaks over thresholds}
\newacronym[]{FIFO}{FIFO}{first-in, first-out}
\newacronym[plural=EVMs,firstplural=Extreme value mixture models (EVM)]{EVM}{EVM}{extreme value mixture model}
\newacronym[plural=HITLs,firstplural=Human-in-the-loop (HITL)]{HITL}{HITL}{human-in-the-loop}
\newacronym[plural=CPSs,firstplural=Cyber-physical systems (CPS)]{CPS}{CPS}{cyber-physical system}
\newacronym[plural=SLAs,firstplural=Service level aggrements (SLA)]{SLA}{SLA}{service level aggrement}
\newacronym[plural=DVPs,firstplural=Delay violation probabilities (DVP)]{DVP}{DVP}{delay violation probability}
\newacronym{MA}{MA}{moving average}

\newacronym[]{TSN}{TSN}{time sensitive networking}
\newacronym{ARQ}{ARQ}{automatic repeat request}
\newacronym{V2X}{V2X}{vehicle-to-everything}
\newacronym{C-V2X}{C-V2X}{cellular V2X}
\newacronym{FEC}{FEC}{forward error correction}
\newacronym{IoT}{IoT}{internet of things}
\newacronym{MLP}{MLP}{multi-layer perceptron}
\newacronym[plural=SDRs, firstplural=Software-defined radios]{SDR}{SDR}{software-defined radio}
\newacronym{OAI}{OAI}{OpenAirInterface}
\newacronym{RTT}{RTT}{round-trip time}
\newacronym{PTP}{PTP}{precision time protocol}
\newacronym[plural=PRBs,firstplural=Physical resource blocks (PRB)]{PRB}{PRB}{physical resource block}
\newacronym[plural=RANs,firstplural=Radio access networks (RAN)]{RAN}{RAN}{radio access network}
\newacronym{COTS}{COTS}{commercial off-the-shelf}
\newacronym[]{CCDF}{CCDF}{complementary cumulative censity function}
\newacronym[]{mmWave}{mmWave}{millimeter-wave}
\newacronym{OTA}{OTA}{over-the-air}
\newacronym[plural=VMs, firstplural=Virtual machines]{VM}{VM}{virtual machine}
\newacronym{IP}{IP}{internet protocol}
\newacronym{K8S}{K8S}{Kubernetes}
\newacronym{SRS}{SRS}{software radio systems}
\newacronym{FPS}{FPS}{frames per second}
\newacronym[user1={Kungliga Tekniska Högskolan}, prefixfirst={the\ }]{KTH}{KTH}{Royal Institute of Technology of Sweden}
\newacronym{CHI}{CHI}{chameleon infrastructure}
\newacronym[plural=UAVs, firstplural=Unmanned aerial vehicles]{UAV}{UAV}{unmanned aerial vehicle}
\newacronym{RAM}{RAM}{random access memory}
\newacronym{HI}{HI}{hierarchical inference}
\newacronym{FL}{FL}{federated learning}
\newacronym{CL}{CL}{centralized learning}
\newacronym[plural=NWDAFs, firstplural=Network data analytics functions (NWDAFs)]{NWDAF}{NWDAF}{network data analytics function}
\newacronym[plural=AMFs, firstplural=Access and mobility management Functions (AMFs)]{AMF}{AMF}{access and mobility management function}
\newacronym[plural=SMFs, firstplural=Session management functions (SMFs)]{SMF}{SMF}{session management function}
\newacronym[plural=NFs, firstplural=Network functions (NFs)]{NF}{NF}{network function}
\newacronym[plural=OWDs, firstplural=One-way delays (OWDs)]{OWD}{OWD}{one-way delay}
\newacronym[plural=UPFs, firstplural=User plane functions (UPFs)]{UPF}{UPF}{user plane function}
\newacronym[plural=BLERs, firstplural=Block error rates (BLERs)]{BLER}{BLER}{block error rate}
\newacronym[plural=BERs, firstplural=Bit error rates (BERs)]{BER}{BER}{bit error rate}
\newacronym[plural=PERs, firstplural=Packet error rates (PERs)]{PER}{PER}{packet error rate}
\newacronym[plural=TBSs, firstplural=Transport block sizes (TBSs)]{TBS}{TBS}{transport block size}
\newacronym[plural=PDUs, firstplural=Protocol data units (PDUs)]{PDU}{PDU}{protocol data unit}
\newacronym[plural=NICs, firstplural=Network interface cards (NICs)]{NIC}{NIC}{network interface card}
\newacronym[plural=PDCPs, firstplural=Packet data convergence protocols (PDCPs)]{PDCP}{PDCP}{packet data convergence protocol}
\newacronym[plural=SNs, firstplural=Sequence numbers (SNs)]{SN}{SN}{sequence number}
\newacronym[plural=E2Es, firstplural=End-to-end connections (E2Es)]{E2E}{E2E}{end-to-end}
\newacronym[plural=NLMTs, firstplural=Network latency measurement tools (NLMTs)]{NLMT}{NLMT}{network latency measurement tool}
\newacronym[plural=3GPPs, firstplural=3rd generation partnership projects (3GPPs)]{3GPP}{3GPP}{3rd generation partnership project}
\newacronym[plural=RLCs, firstplural=Radio link control protocols (RLCs)]{RLC}{RLC}{radio link control}
\newacronym[plural=MACs, firstplural=Medium access control protocols (MACs)]{MAC}{MAC}{medium access control}
\newacronym[plural=IIoTs, firstplural=Industrial Internet of things systems (IIoTs)]{IIoT}{IIoT}{industrial Internet of things}
\newacronym[plural=TVs, firstplural=Total variations (TVs)]{TV}{TV}{total variation}
\newacronym[plural=PEs, firstplural=Permutation entropies (PEs)]{PE}{PE}{permutation entropy}
\newacronym[plural=OOMs, firstplural=Observable markov models (OMMs)]{OMM}{OMM}{observable markov model}
\newacronym[plural=5GAAs, firstplural=5G automotive associations (5GAAs)]{5GAA}{5GAA}{5G automotive association}
\newacronym[longplural={Wearable cognitive assistants}]{WCA}{WCA}{wearable cognitive assistance}
\newacronym[longplural={Augmented realities}]{AR}{AR}{augmented reality}
\newacronym[longplural={virtual realities}]{VR}{VR}{virtual reality}
\newacronym[longplural={extended realities}]{XR}{XR}{extended reality}
\newacronym{GPU}{GPU}{graphical processing unit}
\newacronym{DNN}{DNN}{deep neural network}
\newacronym[prefixfirst={a\ }, prefix={an\ }]{NCS}{NCS}{networked control system}
\newacronym{UDP}{UDP}{user datagram protocol}
\newacronym{CI}{CI}{confidence interval}
\newacronym{AP}{AP}{access point}
\newacronym{API}{API}{application programming interface}
\newacronym{LTE}{LTE}{long-term evolution}
\newacronym{O-RAN}{O-RAN}{open {RAN}}
\newacronym{MEC}{MEC}{mobile edge computing}
\newacronym{USRP}{USRP}{universal software radio peripheral}
\newacronym{NR}{NR}{new radio}
\newacronym{FPGA}{FPGA}{field-programmable gate array}
\newacronym{ARA}{ARA}{agricultural and rural communities}
\newacronym{UHF}{UHF}{ultra-high frequency}
\newacronym{MIMO}{MIMO}{multi-input multi-output}
\newacronym{GPS}{GPS}{global positioning system}
\newacronym{FSM}{FSM}{finite state machine}
\newacronym{QoE}{QoE}{quality of experience}
\newacronym{QoS}{QoS}{quality of service}
\newacronym{PCA}{PCA}{principal component analysis}
\newacronym{IEEE}{IEEE}{institute of electrical and electronics engineers}
\newacronym{CPU}{CPU}{central processing unit}
\newacronym{ECDF}{ECDF}{empirical {CDF}}
\newacronym{ML}{ML}{machine learning}
\newacronym{ExPECA}{ExPECA}{experimental platform for edge computing applications}
\newacronym{EDAF}{EDAF}{end-to-end delay analytics framework}
\newacronym{KPI}{KPI}{key performance indicator}
\newacronym{EMM}{EMM}{extreme mixture model}
\newacronym{GMEVM}{GMEVM}{Gaussian mixture extreme value model}
\newacronym{SGD}{SGD}{stochastic gradient descent}
\newacronym{EM}{EM}{expectation maximization}
\newacronym{MAE}{MAE}{mean absolute error}
\newacronym[plural=LLMs, firstplural=Large language models(LLMs)]{LLM}{LLM}{large language model}
\newacronym{CQI}{CQI}{channel quality indicator}
\newacronym{RSRP}{RSRP}{reference signal received power}
\newacronym{RSRQ}{RSRQ}{reference signal received quality}
\newacronym{RSSI}{RSSI}{received signal strength indicator}
\newacronym{SNR}{SNR}{signal-to-noise ratio}
\newacronym{SINR}{SINR}{signal-to-interference-plus-noise ratio}
\newacronym[plural=REMs, firstplural=Radio environment maps (REMs)]{REM}{REM}{radio environment map}
\newacronym{SHAP}{SHAP}{SHapley Additive exPlanations}
\newacronym{LASSO}{LASSO}{least absolute shrinkage and selection operator}
\newacronym{CAM}{CAM}{connected and automated mobility}
\newacronym{PPO}{PPO}{proximal policy optimization}
\newacronym{LOS}{LOS}{line-of-sight}
\newacronym{NLOS}{NLOS}{non-LOS}
\newacronym{LiDAR}{LiDAR}{light detection and ranging}
\newacronym{VHF}{VHF}{very high frequency}
\newacronym{NB-IoT}{NB-IoT}{narrowband IoT}
\newacronym{MDT}{MDT}{minimization of drive tests}
\newacronym{TCP}{TCP}{transmission control protocol}
\newacronym{PM}{PM}{performance management}
\newacronym{OSS}{OSS}{operations support systems}
\newacronym{MDAF}{MDAF}{management data analytics function}
\newacronym{MTLF}{MTLF}{model training function}
\newacronym{AnLF}{AnLF}{analytics function}
\newacronym{MnS}{MnS}{service-based management service}
\newacronym{RIC}{RIC}{RAN intelligent controller}
\newacronym{SMO}{SMO}{service management and orchestration}
\newacronym{Non-RT}{Non-RT}{non-real-time}
\newacronym{Near-RT}{Near-RT}{near-real-time}
\newacronym{TSCTSF}{TSCTSF}{time sensitive communication and time synchronization function}
\newacronym{MSE}{MSE}{mean squared error}
\newacronym{ROC}{ROC}{receiver operating characteristic}
\newacronym{AUC}{AUC}{area under the curve}
\newacronym{PHY}{PHY}{physical layer}
\newacronym{ROI}{ROI}{return on investment}
\newacronym{AI}{AI}{artificial intelligence}
\newacronym[plural=SONs, firstplural=Self-organizing networks (SONs)]{SON}{SON}{self-organizing network}
\newacronym{ISAC}{ISAC}{Integrated Sensing and Communication}
\newacronym{RSS}{RSS}{received signal strength}
\newacronym{B5G}{B5G}{beyond 5G}
\newacronym{MAPE-K}{MAPE-K}{monitor, analyze, plan, execute, and knowledge}
\newacronym{SLO}{SLO}{service level objective}
\newacronym{ETSI}{ETSI}{European telecommunications standards institute}
\newacronym{PL}{PL}{path-loss}
\newacronym{CSI}{CSI}{channel state information}
\newacronym{AdaBoost}{AdaBoost}{adaptive boosting}
\newacronym{CatBoost}{CatBoost}{categorical boosting}
\newacronym{ToD}{ToD}{tele-operated driving}
\newacronym{GCformer}{GCformer}{graph-convolutional transformer}
\newacronym{NGBoost}{NGBoost}{natural gradient boosting}
\newacronym[plural=ANNs, firstplural=Artificial neural networks (ANNs)]{ANN}{ANN}{artificial neural network}
\newacronym{LS}{LS}{least squares}
\newacronym{LMMSE}{LMMSE}{Linear Minimum Mean Squared Error}
\newacronym{RL}{RL}{reinforcement learning}
\newacronym{SDN}{SDN}{software-defined networking}
\newacronym{LEO}{LEO}{low earth orbit}
\newacronym{PCF}{PCF}{point coordination function}

\newacronym[plural=VGGNets, firstplural=Visual geometry group networks(VGGNets)]{VGGNet}{VGGNet}{visual geometry group network}
\newacronym[plural=FCNs, firstplural=Fully convolutional networks (FCNs)]{FCN}{FCN}{fully convolutional network}

\newacronym[plural=rApps, firstplural=RAN intelligent controller applications(rApps)]{rApp}{rApp}{RAN intelligent controller application}

\newacronym[plural=xApps, firstplural=eXtended applications(xApps)]{xApp}{xApp}{extended application}

\newacronym[plural=BNNs, firstplural=Bayesian neural networks (BNNs)]{BNN}{BNN}{Bayesian neural network}

\newcommand{\cmark}{\ding{51}}  % check
\newcommand{\xmark}{\ding{55}}  % cross
\renewcommand{\arraystretch}{1.15}

\definecolor{accessblue}{cmyk}{1, 0.3, 0, 0.2}
\definecolor{greycolor}{cmyk}{0, 0, 0, 0.8}
\definecolor{bggray}{HTML}{F8F9FA}
\definecolor{textgray}{HTML}{333333}

\definecolor{centercol}{RGB}{52, 73, 94}      % center node - dark blue-grey
\definecolor{covcol}{RGB}{150, 168, 96}       % Coverage
\definecolor{capcol}{RGB}{76, 96, 55}         % Capacity
\definecolor{latcol}{RGB}{189, 142, 64}       % Latency
\definecolor{relcol}{RGB}{180, 93, 42}        % Reliability

\definecolor{corecol}{HTML}{0072B2}
\definecolor{rancol}{HTML}{009E73}
\definecolor{uecol}{HTML}{D55E00}
\definecolor{mlcol}{HTML}{CC79A7}

\definecolor{meascol}{RGB}{167, 136, 96}
\definecolor{simcol}{RGB}{179, 182, 160}
\definecolor{syncol}{RGB}{241, 213, 176}
\definecolor{hybcol}{RGB}{217, 84, 64}

\definecolor{model1col}{RGB}{211, 210, 199}
\definecolor{model2col}{RGB}{91, 137, 174}
\definecolor{model3col}{RGB}{173, 196, 215}
\definecolor{model4col}{RGB}{203, 142, 102}
\definecolor{model5col}{RGB}{118, 151, 125}
\definecolor{model6col}{RGB}{220, 114, 96}
\definecolor{model7col}{RGB}{118, 148, 171}

\definecolor{netdark}{HTML}{1A2530}   % Dark Slate
\definecolor{covcol}{HTML}{2980B9}    % Premium Blue
\definecolor{capcol}{HTML}{D35400}    % Deep Orange
\definecolor{latcol}{HTML}{8E44AD}    % Rich Purple
\definecolor{relcol}{HTML}{16A085}    % Teal/Green

\definecolor{modelUnsup}{RGB}{211, 210, 199}
\definecolor{modelClass}{RGB}{91, 137, 174}
\definecolor{modelEns}{RGB}{173, 196, 215}
\definecolor{modelCNN}{RGB}{203, 142, 102}
\definecolor{modelSeq}{RGB}{118, 151, 125}
\definecolor{modelFL}{RGB}{118, 148, 171}
\definecolor{modelGNN}{RGB}{220, 114, 96}

\begin{document}

\title{AI-Based KPI Prediction Methods in Future 6G Networks: A Survey}

\author{Niloofar~Mehrnia,~\IEEEmembership{Member, IEEE},
        Gourav~Prateek~Sharma,~\IEEEmembership{Member, IEEE},
        Samie~Mostafavi,~\IEEEmembership{Member, IEEE},
        Andreas~Johnsson,~\IEEEmembership{Senior Member, IEEE},
        Sinem~Coleri,~\IEEEmembership{Fellow, IEEE}, 
        Carlo~Fischione,~\IEEEmembership{Fellow, IEEE}, and % <-this % stops a space
        James~Gross,~\IEEEmembership{Senior Member, IEEE}
\thanks{Niloofar~Mehrnia, Carlo~Fischione, and James~Gross are with the Department of Electrical Engineering and Computer Sciences, KTH Royal Institute of Technology, Stockholm, Sweden (e-mail:\{nilome,carlofi,jamesgr\}@kth.se). (Corresponding author: Niloofar Mehrnia)}
\thanks{Carlo~Fischione and James~Gross are also with Digital Futures, KTH Royal Institute of Technology, Stockholm, Sweden.}
\thanks{Gourav~Prateek~Sharma is with the Electronics and Communications Engineering Department at National Institute of Technology Kurukshetra, Haryana, India (e-mail:gourav.sharma@nitkkr.ac.in).}
\thanks{Samie~Mostafavi and Andreas~Johnsson are with Ericsson Research,
%, Research Area AI, 
Stockholm, Sweden (e-mail:\{samie.mostafavi,andreas.a.johnsson\}@ericsson.com).}
\thanks{Andreas~Johnsson is also with the Department of Information Technology, Uppsala University, Uppsala, Sweden (e-mail:andreas.johnsson@it.uu.se).}
\thanks{Sinem~Coleri is with the Department of Electrical and Electronics Engineering,
Koc University, Istanbul, Turkey (e-mail:scoleri@ku.edu.tr).}
        % <-this % stops a space
%\thanks{This paper was produced by the IEEE Publication Technology Group. They are in Piscataway, NJ.}% <-this % stops a space
%\thanks{Manuscript received April 19, 2021; revised August 16, 2021.}
}

\markboth{Journal of \LaTeX\ Class Files,~Vol.~14, No.~8, August~2021}%
{Shell \MakeLowercase{\textit{et al.}}: A Sample Article Using IEEEtran.cls for IEEE Journals}


\maketitle

\begin{abstract}
The evolution from 5G to 5G-Advanced and the vision of 6G demand unprecedented levels of network performance, in which meeting stringent network Key Performance Indicators (KPIs), including capacity, latency, coverage, and reliability, is critical to supporting emerging applications such as autonomous driving, industrial automation, and immersive communications. Traditional reactive network management is insufficient in this context, driving the need for predictive, data-driven approaches. Machine Learning (ML) has emerged as a key enabler, enabling the forecasting of KPI trends from diverse data sources and thereby enabling proactive, AI-native automation in mobile networks. This survey provides the first comprehensive and systematic review of data-driven KPI prediction methods for future 6G networks. We introduce a multi-dimensional taxonomy that classifies prediction approaches by KPI type, data source, the network protocol stack at which the KPI is predicted, prediction horizon, model family, and prediction objective. Using this taxonomy, we analyze the state of the art across various KPIs, highlighting representative methods ranging from classical statistical models to deep learning and reinforcement learning. We further discuss enabling system aspects, including data collection and learning architectures, and examine deployment challenges, including data availability, scalability, privacy, and sustainability. Finally, we outline open research directions spanning new KPI definitions, probabilistic and explainable predictions. This survey aims to provide researchers and practitioners with a structured understanding of the KPI prediction landscape and a roadmap toward predictive network automation in future 6G systems. 
\end{abstract}


\section{Introduction}
The advent of 6G networks represents a significant shift in ambition from previous generations, moving beyond traditional connectivity platforms toward \gls{AI}-native architectures in which intelligence is intrinsically embedded across all layers, from the \gls{RAN} to the core and orchestration domains. This native \gls{AI} integration enables networks to perceive, learn, reason, adapt, and act autonomously in real time, shifting toward fully autonomous networks capable of zero-touch or intent-driven operations with minimal human intervention. By design, 6G incorporates \gls{AI} as a foundational element, which is expected to enable key use cases such as autonomous zero-touch network management, proactive KPI assurance \cite{shi2023machine}, and \gls{AI}-enhanced \gls{ISAC} for digital twins and environmental awareness, all to manage the network's extreme complexity, dynamic conditions, and massive scale requirements \cite{ericsson2024cocreation, ericsson2023defining, 6gia2024vision}.

The realization of future 6G networks is contingent upon the integration of \gls{AI}-native design and self-governing architectures \cite{yang2020artificial}. This evolution is necessitated by the inherent limitations of legacy manual and reactive frameworks, which lack the scalability required to manage the complexity of next-generation networks. By shifting from active troubleshooting to an autonomous predictive-control network paradigm, the network can effectively leverage ML to optimize control and management within the complex design spaces of 6G \cite{jiangMachineLearningParadigms2017}. One of the primary objectives of this autonomy is to enable accurate performance gauging, which serves as the foundation for both operational efficiency and the fulfillment of application requirements. These performance targets are quantified through network KPIs, including latency, capacity, and coverage, which serve as key metrics for assessing not only traditional network performance but also the network's efficiency and reliability. 

\subsection{Motivation}
Beyond basic data analytics frameworks, current 6G research increasingly embraces the concept of \emph{\gls{AI}-native networks}, in which learning-based intelligence is embedded as a fundamental design principle rather than a post hoc optimization tool. This vision is reflected in emerging notions such as \emph{\gls{AI}-\gls{RAN}}, where radio access functions, including scheduling, mobility management, and resource adaptation, are continuously driven by data and learning-based inference \cite{niknam2022intelligent}. In such architectures, the network operates as a closed-loop cognitive system that senses its state, anticipates future conditions, and proactively adapts its behavior \cite{giannopoulos2022supporting}. The transition from conventional control and management to \emph{\gls{AI}-native} network operation is fundamentally enabled by KPI prediction, which acts as the network's predictive sensing layer and bridges raw data with autonomous, learning-driven control decisions. This predictive layer is what allows the network to project ahead, processing raw data into the actionable foresight needed to provide service guarantees with efficiency.

Forecasting KPI metrics enables the system to preemptively allocate resources, reroute traffic, or trigger handovers, thereby ensuring dependable connectivity \cite{sharma2023toward}. \textcolor{black}{The role of KPI prediction/forecasting becomes particularly tangible when considered within concrete operational contexts. In \gls{SON} frameworks, predicted cell-level throughput degradation can trigger proactive parameter tuning, such as adjusting antenna tilt or handover thresholds, before congestion happens \cite{aliu2012survey}. In \gls{O-RAN} deployments, a \gls{Near-RT} \gls{RIC} \gls{xApp} can use per-\gls{UE} latency forecasts to preemptively reallocate \glspl{PRB} ahead of a predicted \gls{URLLC} violation, operating within the millisecond control loop without waiting for a measurement \cite{polese2023understanding}. At the core network level, \gls{NWDAF} can ingest predicted slice-level reliability trends to drive proactive admission control and traffic steering before an \gls{SLA} breach occurs \cite{3gpp-ts-28105}.} It is important to note that while \gls{AI}-driven closed-loop control can, in principle, operate without an explicit prediction step, maintaining a distinct \gls{KPI} prediction module offers several advantages. First, it provides interpretability and modularity, allowing forecasts to be validated and reused across multiple control and management loops operating at timescales ranging from millisecond radio resource scheduling to hour-long energy-saving cycles. Second, the prediction layer aligns with standardized architectures such as \gls{3GPP}'s \gls{MnS} and \gls{O-RAN}'s \gls{RIC}, which separate analytics from control to ensure scalability and stability \cite{3gpp-ts-28105,oran-architecture}. Finally, explicit \gls{KPI} prediction enables anticipatory decisions with quantifiable confidence levels, reducing the risk of instability or unintended behavior in fully automated systems. 
The predictive control cycle in mobile networks operates as a continuous feedback loop that transforms raw network data into proactive management actions. It begins with data collection from various sources across the mobile network, including \gls{RAN}, core, and edge elements. This data is stored in a repository, where it is processed and used to train and infer data-driven \gls{KPI} prediction models. The predicted \gls{KPI} values are then fed into the network control and management functions, enabling anticipatory decisions such as resource reallocation, load balancing, or admission control. Finally, the resulting network behavior is evaluated against \gls{SLA} assurance and validation mechanisms, closing the loop and generating new data for continuous refinement of prediction models and control policies. It is important to note that, in addition to enabling dependable connectivity through proactive control, predicting/forecasting \glspl{KPI} such as latency, throughput, reliability, and availability also allows for long-term resource orchestration, strategic traffic engineering, and proactive mobility management \cite{du2020machine}.    

\subsection{Related Surveys}

\begin{table*}[t]
\centering
\caption{Prior Reviews/Surveys on Data-Driven KPI Prediction in Mobile Networks and Their Scope.}
\label{table:list_surveys}
\footnotesize
\begin{tabular}{p{1.0cm}p{4.5cm}p{2.5cm}p{7cm}p{1cm}}
\hline
\textbf{Year} & \textbf{Title} & \textbf{Scope / KPI Types Covered} & \textbf{Main Limitation} & \textbf{Reference} \\
\hline
2018
& Machine Learning for Performance Prediction in Mobile Cellular Networks
& Coverage and throughput
& Classic temporal and spatial models considered for selected KPIs.
& \cite{riihijarviMachineLearningPerformance2018}
\\ \hline
2019
& Machine Learning for Wireless Communication Channel Modeling: An Overview
& Channel modeling (PHY layer)
& Restricted to propagation/channel KPIs; excludes higher-layer metrics (latency, capacity, energy).
& \cite{aldossariMachineLearningWireless2019}
\\ \hline
2021
& A Survey on Client Throughput Prediction Algorithms in Wired and Wireless Networks
& Throughput (TCP, LTE/5G, vehicular)
& Restricted to throughput only; omits latency, coverage, and reliability KPIs; no taxonomy of prediction horizon or output formulation.
& \cite{schmid2021survey}
\\ \hline

2021
& An Overview of Machine Learning Techniques for Radiowave Propagation Modeling
& Path loss, RSS/RSRP (PHY layer)
& Restricted to propagation/coverage KPIs at PHY layer; excludes higher-layer metrics such as throughput, latency, and reliability; no prediction horizon or output taxonomy.
& \cite{seretis2021overview}
\\ \hline

2022
& An Architecture and Performance Evaluation Framework for AI Solutions in Beyond-5G RANs
& RAN AI architecture; generic KPIs
& Proposes an evaluation framework for ML deployment in 5G; does not classify works by KPI type or network stack.
& \cite{koudouridisArchitecturePerformanceEvaluation2022a}
\\ \hline

2022
& Cellular Traffic Prediction with Machine Learning: A Survey
& Traffic/throughput forecasting
& Narrow KPI scope (traffic only) and prediction approach; omits model comparison along taxonomy dimensions such as data source and prediction attributes.
& \cite{jiangCellularTrafficPrediction2022}
\\ \hline

2022
& Deep Learning for B5G Open Radio Access Network: Evolution, Survey, Case Studies, and Challenges
& Traffic load, throughput, latency, energy efficiency (O-RAN context)
& Architectural emphasis; does not classify works by KPI type or provide a prediction-focused taxonomy; KPI prediction treated as one component among many RAN functions.
& \cite{brik2022deep}
\\ \hline
2023
& Machine Learning for QoS Prediction in Vehicular Communication: Challenges and Solution Approaches
& Throughput (DL/UL), RSRP, RSRQ, SINR, CQI, latency
& Restricted to vehicular (V2X/C-V2X) scenarios; no cross-KPI taxonomy; evaluation methodology focus limits generalization to broader cellular contexts.
& \cite{palaios2023machine}
\\ \hline

2023
& Mobility Prediction in Cellular Networks: A Survey
& Mobility KPI
& Single-KPI focus; no cross-KPI taxonomy or evaluation of model generalization to other metrics.
& \cite{rajuleMobilityPredictionCellular2023}
\\ \hline

2024
& A Survey on Deep Learning for Cellular Traffic Prediction
& Cellular traffic volume, throughput/capacity
& Narrow KPI scope (traffic/capacity only); focused exclusively on deep learning, omitting classical and ensemble models; no coverage of latency, reliability, or cross-KPI prediction.
& \cite{wang2024survey}
\\ \hline

2024
& A comparison of neural networks for wireless channel prediction
& CSI prediction
& Narrow KPI scope; Restricted to coverage/propagation KPIs; omits capacity, latency, and reliability.
& \cite{stenhammar2024comparison}
\\ \hline

2024
& Fault Prediction for Heterogeneous Telecommunication Networks Using Machine Learning: A Survey
& Network fault KPI (availability/reliability)
& Focuses on anomaly/fault detection.
& \cite{murphyFaultPredictionHeterogeneous2024}
\\ \hline

2024
& Machine Learning for Radio Propagation Modeling: A Comprehensive Survey
& Path loss, RSS/RSRP/RSSI, radio coverage maps
& Restricted to coverage/propagation KPIs; omits capacity, latency, and reliability; no discussion of prediction horizon, data source realism, or deployment in live networks.
& \cite{vasudevan2024machine}
\\ \hline

\end{tabular}
\end{table*}

The research landscape for \gls{KPI} prediction has evolved substantially over the past decade, reflecting a shift in both \textit{what} is predicted and \textit{how} it is predicted. A review of existing works, summarized in Table~\ref{table:list_surveys}, reveals that the literature can be organized into two broad groups, each leaving a distinct gap that motivates the present survey.

\textcolor{black}{\emph{Narrow-scope surveys} provide depth on one prediction target, but cannot capture cross-\gls{KPI} dependencies or methodological trade-offs that span families. On the throughput side, Schmid~\emph{et al.}~\cite{schmid2021survey} systematically review client throughput prediction across wired and cellular (\gls{LTE}/5G) scenarios, covering over multiple algorithms from simple smoothing to \gls{LSTM}-based predictors, yet omit latency, coverage, and reliability entirely, and provide no taxonomy of prediction horizon or output formulation. Jiang~\cite{jiangCellularTrafficPrediction2022} and Wang~\emph{et al.}~\cite{wang2024survey} survey cellular traffic prediction with classical \gls{ML} and deep learning, respectively, but both restrict their scope to traffic volume and throughput, leaving the latency and reliability \gls{KPI} families unaddressed. On the coverage side, Aldossari and Chen~\cite{aldossariMachineLearningWireless2019}, Seretis and Sarris~\cite{seretis2021overview}, Stenhammar~\emph{et al.}~\cite{stenhammar2024comparison}, and Vasudevan and Yuksel~\cite{vasudevan2024machine} each survey \gls{ML} for radio propagation and coverage prediction (including path loss, \gls{RSRP}, \gls{RSS}, and \gls{CSI}), but restrict their scope to the \gls{PHY} layer and do not discuss higher-layer \glspl{KPI} or deployment considerations. Additionally, none of these survey papers, except \cite{stenhammar2024comparison}, considered different prediction horizons. Riihijarvi and Mahonen~\cite{riihijarviMachineLearningPerformance2018} consider both coverage and throughput prediction using classical \gls{ML} but limit their treatment to a narrow set of temporal and spatial models, without a structured taxonomy. For mobility, Rajule~\emph{et al.}~\cite{rajuleMobilityPredictionCellular2023} survey handover and location prediction, but do not extend to performance \glspl{KPI} or evaluate generalization across metric families. Murphy~\emph{et al.}~\cite{murphyFaultPredictionHeterogeneous2024} address network fault and anomaly prediction (availability, reliability), but their focus is binary fault detection rather than continuous \gls{KPI} prediction/forecasting, and no regression-based prediction taxonomy is provided. Palaios~\emph{et al.}~\cite{palaios2023machine} offer a tutorial-style treatment of \gls{ML} for \gls{QoS} prediction in vehicular networks, covering throughput, \gls{RSRP}, \gls{SINR}, \gls{CQI}, and latency together and providing practical guidance on train/test splitting and feature engineering. However, their findings are explicitly scoped to \gls{V2X} scenarios and do not generalize to infrastructure-centric or multi-service cellular deployments, nor do they provide a cross-KPI taxonomy.}

\textcolor{black}{\emph{Architecture-oriented surveys} embed \gls{KPI} prediction within a broader system discussion but do not treat prediction as the primary object of study. Koudouridis~\emph{et al.}~\cite{koudouridisArchitecturePerformanceEvaluation2022a} propose a framework for evaluating \gls{AI} solutions in beyond-5G \glspl{RAN}, but do not classify prediction works by \gls{KPI} type, data source, or network stack. Brik~\emph{et al.}~\cite{brik2022deep} survey deep learning for \gls{B5G} \gls{O-RAN}, touching on throughput, latency, traffic load, and energy efficiency prediction as sub-topics within a broader architectural treatment, without a prediction-focused taxonomy.}

\subsection{Contributions and Survey Structure}
\textcolor{black}{This survey provides a comprehensive and systematic study of data-driven \gls{KPI} prediction for future 6G networks. The primary contributions are as follows:}

\begin{itemize}

    \item \textcolor{black}{\emph{Multi-dimensional taxonomy:} We introduce
    a unified taxonomy that characterizes and compares prediction approaches along four orthogonal dimensions: \gls{KPI} type (capacity, latency, coverage, reliability), data source, prediction attributes (protocol stack layer, horizon, output formulation), and \gls{ML}/\gls{AI} model family. Unlike prior surveys, this taxonomy enables consistent cross-study comparison and reveals methodological biases and underexplored research areas.}

    \item \textcolor{black}{\emph{Quantitative landscape analysis:} Using the taxonomy as a coding framework, we annotate state-of-the-art data-driven KPI-prediction studies and report empirical distributions across \gls{KPI} families, model families, data source types, prediction horizons, and artifact availability, providing the first quantitative map of the field.}

    \item \textcolor{black}{\emph{\gls{KPI}-centric comparative analysis:} We provide a systematic, \gls{KPI}-by-\gls{KPI} review of state-of-the-art prediction approaches for capacity, latency, coverage, and reliability,       dedicated taxonomy tables, and network-domain figures that map methods, data sources, and use cases across the RAN, core, and application domains.}

    \item \textcolor{black}{\emph{System-level and deployment analysis:} We analyze the practical integration of KPI predictors within \gls{3GPP} and \gls{O-RAN} lifecycle management frameworks, covering data collection, model training, deployment constraints, and the challenges of telemetry granularity, edge resource limits, and model drift.}

    \item \textcolor{black}{\emph{Open challenges and research roadmap:} We identify and discuss various cross-cutting limitations spanning reproducibility, evaluation metric inconsistency, offline-to-live integration gaps, inference-latency trade-offs, data lifecycle management, explainability, sustainability, and privacy, and outline concrete future research directions toward operationally deployable, uncertainty-aware KPI predictors for \gls{AI}-native 6G networks.}

\end{itemize}

\begin{figure*}[htbp]
    \centering
    
    % Define the sophisticated, minimal nude palette
    \definecolor{rootcol}{HTML}{9BA9B4} % Deep Warm Taupe / Mocha
    \definecolor{nudecol}{HTML}{D9CBBF} % Soft Nude / Sand
    \definecolor{outercol}{HTML}{F7F5F2} % Very light warm background

    \begin{tikzpicture}[
      >=Latex,
      % Header blocks (Deep Taupe)
      header/.style={
        fill=rootcol, text=white, font=\footnotesize\bfseries, align=center,
        rounded corners=2pt, minimum height=5mm, minimum width=8.2cm,
        inner sep=2pt, text width=8.0cm
      },
      % Base style for internal sub-boxes
      subbox/.style={
        fill=nudecol!35, draw=rootcol!40, thick, text=black!85,
        font=\scriptsize\bfseries, align=center, rounded corners=2pt,
        minimum height=6.5mm, inner sep=1.5mm
      },
      % Dynamic grid sizing for 2-column TikZ Layout (Column Width = 8.2cm)
      box1/.style={subbox, minimum width=8.2cm, text width=7.8cm},       % Full column span
      box2/.style={subbox, minimum width=4.0cm, text width=3.7cm},       % Half column span
      % Outer section wrapper with soft modern shadow
      outer/.style={
        fill=outercol, draw=rootcol!15, thick, rounded corners=3pt, inner sep=4.5pt,
        blur shadow={shadow blur steps=4, shadow xshift=0pt, shadow yshift=-1pt}
      }
    ]

    % Spacing configurations for compact layout
    \def\Hsep{2mm}     % Horizontal gap between boxes
    \def\Vsep{0.5mm}   % Vertical gap between rows
    \def\BlockSep{1mm} % Vertical gap between major sections

    % =========================================================
    % LEFT COLUMN (X = 0cm)
    % =========================================================
    
    % --- Section I: Introduction ---
    \node[header, anchor=north west] (S1H) at (0, 0) {Section I. Introduction};
    
    \node[box2, below=\Vsep of S1H.south west, anchor=north west] (S1A) {Motivation};
    \node[box2, right=\Hsep of S1A] (S1B) {Related Surveys};
    \node[box1, below=\Vsep of S1A.south west, anchor=north west] (S1C) {Contributions and Survey Structure};
    \begin{scope}[on background layer] \node[outer, fit=(S1H)(S1A)(S1B)(S1C)] (S1O) {}; \end{scope}

    % --- Section II: Methodology ---
    \node[header, below=\BlockSep of S1O.south west, xshift=0.2cm, yshift=-0.3cm, anchor=north west] (S2H) {Section II. Scope, Problem Formulation, \& Taxonomy};
    
    \node[box2, below=\Vsep of S2H.south west, anchor=north west] (S2A) {Survey Scope};
    \node[box2, right=\Hsep of S2A] (S2B) {KPI Families \& Definitions};
    \node[box2, below=\Vsep of S2A.south west, anchor=north west] (S2C) {Prediction, Estimation, Forecasting};
    \node[box2, right=\Hsep of S2C] (S2D) {Taxonomy};
    \begin{scope}[on background layer] \node[outer, fit=(S2H)(S2A)(S2D)] (S2O) {}; \end{scope}

    % --- Section III: Overview of methods ---
    \node[header, below=\BlockSep of S2O.south west, xshift=0.2cm, yshift=-0.25cm, anchor=north west] (S3H) {Section III. Overview of Data-driven Prediction Methods};
    
    \node[box2, below=\Vsep of S3H.south west, anchor=north west] (S3A) {Unsupervised Learning};
    \node[box2, right=\Hsep of S3A] (S3B) {Supervised Learning};
    \node[box2, below=\Vsep of S3A.south west, anchor=north west] (S3C) {Ensemble Learning};
    \node[box2, right=\Hsep of S3C] (S3D) {Deep Learning};
    \node[box2, below=\Vsep of S3C.south west, anchor=north west] (S3E) {Distributed \& Structure-Aware};
    \node[box2, right=\Hsep of S3E] (S3F) {Prediction Uncertainty};
    %\node[box1, below=\Vsep of S3E.south west, anchor=north west] (S3G) {Prevalence and Evolution of Methods};
    \begin{scope}[on background layer] \node[outer, fit=(S3H)(S3A)(S3B)(S3F)] (S3O) {}; \end{scope}

    % --- Section IV: Comparative analysis ---
    \node[header, below=\BlockSep of S3O.south west, xshift=0.2cm, yshift=-0.2cm, anchor=north west] (S4H) {Section IV. Comparative Analysis of Approaches};
    
    \node[box2, below=\Vsep of S4H.south west, anchor=north west] (S4A) {Capacity};
    \node[box2, right=\Hsep of S4A] (S4B) {Latency};
    \node[box2, below=\Vsep of S4A.south west, anchor=north west] (S4C) {Coverage};
    \node[box2, right=\Hsep of S4C] (S4D) {Reliability};
    \begin{scope}[on background layer] \node[outer, fit=(S4H)(S4A)(S4D)] (S4O) {}; \end{scope}

    % =========================================================
    % RIGHT COLUMN (X = 8.6cm)
    % =========================================================

    % --- Section V: System Aspects ---
    \node[header, anchor=north west] (S5H) at (8.6, 0) {Section V. System Aspects of KPI Predictions};
    
    \node[box2, below=\Vsep of S5H.south west, anchor=north west] (S5A) {Data Collection \& Preprocessing};
    \node[box2, right=\Hsep of S5A] (S5B) {Model Training \& Deployment};
    \node[box2, below=\Vsep of S5A.south west, anchor=north west] (S5C) {Monitoring and Model Update};
    \node[box2, right=\Hsep of S5C] (S5D) {Practical Guidelines};
    \node[box1, below=\Vsep of S5C.south west, anchor=north west] (S5E) {Challenges};
    \begin{scope}[on background layer] \node[outer, fit=(S5H)(S5A)(S5B)(S5E)] (S5O) {}; \end{scope}

    % --- Section VI: Limitations & Challenges ---
    \node[header, below=\BlockSep of S5O.south west, xshift=0.2cm, yshift=-0.05cm, anchor=north west] (S6He) {Section VI. Limitations, Challenges, and Future Works};
    
    \node[box2, below=\Vsep of S6He.south west, anchor=north west] (S6A) {Maturity across KPIs};
    \node[box2, right=\Hsep of S6A] (S6B) {Reproducibility \& Comparability};
    \node[box2, below=\Vsep of S6A.south west, anchor=north west] (S6C) {Evaluation Metrics};
    \node[box2, right=\Hsep of S6C] (S6D) {Generalizability \& Robustness};
    
    % Grouped logically to allow the long "Trade-offs" box its own full width
    \node[box2, below=\Vsep of S6C.south west, anchor=north west] (S6E) {Offline Prediction};
    \node[box2, right=\Hsep of S6E] (S6G) {Data Lifecycle};  
    \node[box1, below=\Vsep of S6E.south west, anchor=north west] (S6F) {Trade-offs among Inference Latency and Accuracy in Real-time RAN};
    
    \node[box2, below=\Vsep of S6F.south west, anchor=north west] (S6H_node) {Explainability};
    \node[box2, right=\Hsep of S6H_node] (S6I) {Sustainability};
    \node[box2, below=\Vsep of S6H_node.south west, anchor=north west] (S6J) {Privacy \& Security};
    \node[box2, right=\Hsep of S6J] (S6K) {Future Works};
    \begin{scope}[on background layer] \node[outer, fit=(S6He)(S6A)(S6B)(S6K)] (S6O) {}; \end{scope}

    % --- Section VII: Conclusions ---
    \node[header, below=\BlockSep of S6O.south west, xshift=0.2cm, yshift=-0.03cm, anchor=north west] (S7H) {Section VII. Conclusions};
    \begin{scope}[on background layer] \node[outer, fit=(S7H)] (S7O) {}; \end{scope}

    % --- Section VIII: Appendix ---
    \node[header, below=\BlockSep of S7O.south west, xshift=0.2cm, anchor=north west] (S8H) {Appendix};
    
    \node[box1, below=\Vsep of S8H.south west, anchor=north west] (S8A) {Literature Collection \& Coding};
    \begin{scope}[on background layer] \node[outer, fit=(S8H)(S8A)] (S8O) {}; \end{scope}

    \end{tikzpicture}

    \caption{Summary of the structure of this survey.}
    \label{fig:orgenization}
\end{figure*}
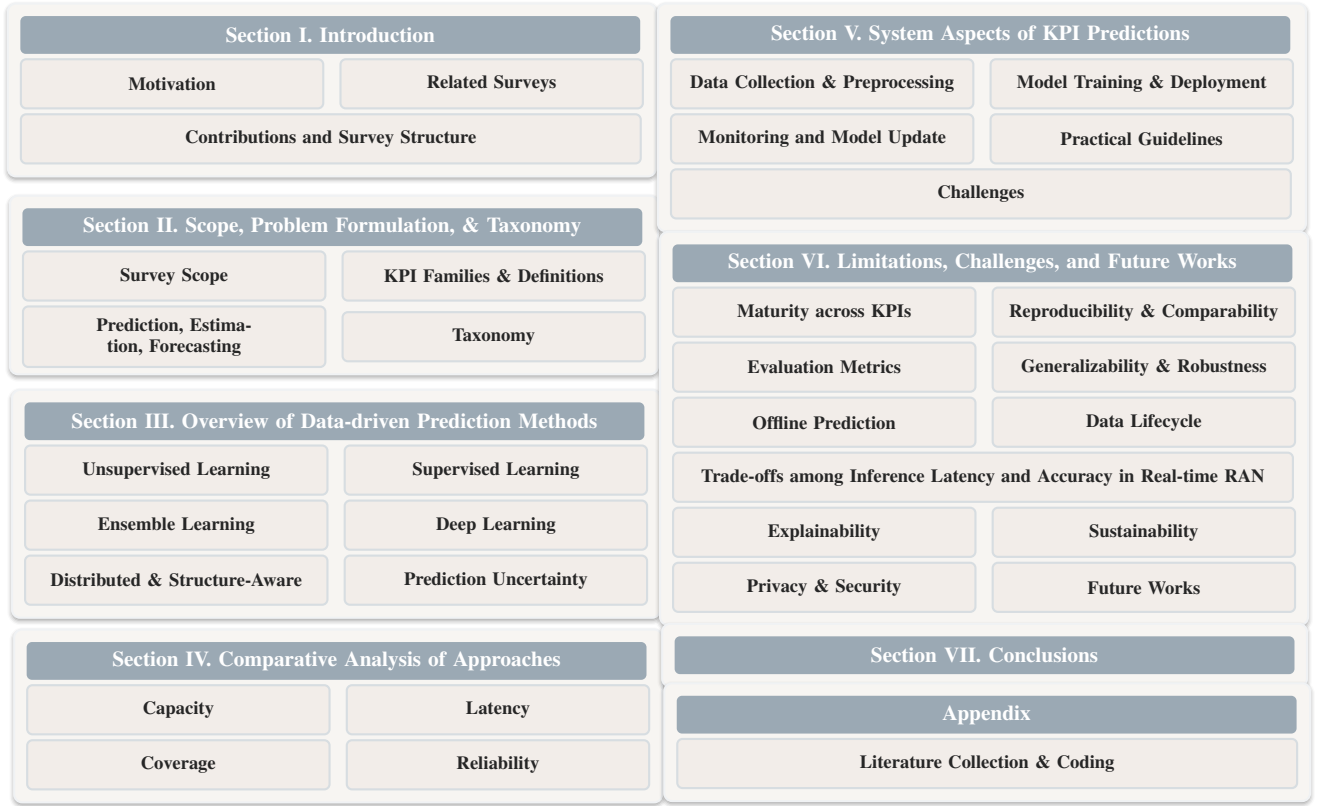

The rest of this survey is organized as follows (Fig.\ref{fig:orgenization}). Section~\ref{sec:scope_taxonomy} presents the survey scope, problem formulation, and the taxonomy construction used for analysis. Section~\ref{sec:methods} provides an overview of data-driven prediction methods, summarizing major \gls{ML}/\gls{AI} model families and their roles in \gls{KPI} prediction. Section~\ref{sec:comparisons} offers a comparative, \gls{KPI}-centric analysis of prediction approaches across capacity, latency, coverage, and reliability metrics. Section~\ref{sec:enablers} discusses system-level enablers for \gls{KPI} prediction, such as data collection, training pipelines, deployment considerations, and operational challenges. Limitations, challenges, and open research directions for advancing predictive automation in future mobile networks are presented in Section~\ref{sec:challenges}. Finally, Section~\ref{sec:conclusions} concludes the paper.

\section{Scope, problem formulation, and taxonomy}
\label{sec:scope_taxonomy}
\textcolor{black}{This section establishes the technical framing that underpins the remainder of the survey. We (i) delineate the scope of ``data-driven \gls{KPI} prediction'' in cellular networks, (ii) formalize the prediction/estimation/forecasting problem and its outputs, and (iii) introduce the taxonomy used to structure both the qualitative review and the quantitative landscape analysis. The methodology, literature search, and screening protocol are documented separately in Appendix~\ref{app:screening} to keep the main narrative focused on concepts and comparability.}

\subsection{Survey scope}
\label{subsec:scope}
\textcolor{black}{We focus on \emph{data-driven KPI prediction} in mobile cellular networks studies published from January 2000 up to early January 2026 (inclusive). The studies include learning-based approaches that map network-related data to (i) infer a \gls{KPI} at the current decision time $t$ (\emph{prediction/estimation}), or (ii) forecast a \gls{KPI} (or a \gls{KPI}-relevant indicator) at a future time, interval, or operating state (\emph{prediction/forecasting}). Such approaches leverage heterogeneous inputs, including measurement-based telemetry data and \glspl{KPI}, as well as simulation-based, synthetic, or hybrid data sources. Since many cellular network prediction tasks are inherently sequential, we cast the problem as a time-series problem.}

\textcolor{black}{The survey centers on \gls{3GPP}-based mobile networks (4G/5G/5G-Advanced) and their evolution toward 6G, spanning \gls{PHY}-, \gls{RAN}-, network-, and application-level viewpoints, focusing on operationally relevant metrics for mobile network performance and management. To compare heterogeneous studies, we group prediction targets into four \gls{KPI} families used throughout the paper: \emph{Capacity}, \emph{Latency}, \emph{Coverage/Link Quality}, and \emph{Reliability}. Notably, under \gls{3GPP}'s standardized KPI grouping (availability, accessibility, retainability, mobility, and integrity), all \glspl{KPI} considered in this survey fall within the integrity category \cite{3gpp-TS32-450}.}

\subsection{Prediction, estimation, and forecasting}
\label{subsec:formulation}
\textcolor{black}{Let $y_t\in\mathbb{R}$ denote a \gls{KPI} of interest at time $t$, and let $x_t\in\mathcal{X}^d$ denote the $d$-dimensional feature vector available at time $t$ (e.g., counters, measurements, and contextual variables), where $\mathcal{X}$ may include heterogeneous components such as real-valued, categorical, or other non-numeric features. At decision time $t$, the predictor has access to an information set
\[
\mathcal{I}_t^k \triangleq \{x_t,x_{t-1},\ldots,x_{t-k}\}.
\]
A learned model $f_\theta$ produces either a \emph{point prediction}
$\hat{y}_{t+\tau}=f_\theta(\mathcal{I}_t^k)$,
or a \emph{probabilistic prediction}, e.g., a predictive distribution
$p_\theta\!\left(y_{t+\tau}\mid \mathcal{I}_t^k\right)$.}

We use \emph{prediction} as the umbrella term for inferring an unknown \gls{KPI} value $y_{t+\tau}$ from available information set $\mathcal{I}_t^k$. \emph{Estimation} is the special case of prediction at the current decision time ($\tau = 0$), i.e., inferring the present \gls{KPI}/state from partial or noisy measurements. \emph{Forecasting} is the special case of prediction for a future time index ($\tau > 0$). This survey focuses on data-driven approaches for both (i) estimation ($\tau = 0$) and (ii) forecasting ($\tau > 0$), with forecasting further characterized by the prediction horizon.

The primary motivation is to enable \gls{AI}-native, proactive network operations. Specifically, accurate \gls{KPI} estimates and forecasts can be integrated into closed-loop optimization and control processes (such as \gls{MAPE-K} or observe-decide-act loops). These forecasts inform decisions related to traffic steering, admission and rate control, mobility and radio-parameter adjustments, energy-saving initiatives, and the scaling of slices and cloud resources, all with the explicit objective of minimizing anticipated \gls{SLA} or \gls{SLO} violations within operational constraints \cite{ericsson_review_toward_5g}. In this context, probabilistic predictions are especially valuable, as the associated uncertainty (expressed through predictive distributions or intervals) can be incorporated into risk-aware objectives and safety margins. This approach enables more conservative actions when confidence is low, and more aggressive optimization when confidence is high. Such a perspective is consistent with ongoing industrial and standardization efforts to advance data-driven network automation and analytics in cellular networks, as well as broader closed-loop and intent-driven management frameworks being developed by standards organizations and industry consortia~\cite{ericsson_blog_ai_native}.

\subsection{KPI families and operational definitions}
\label{subsec:kpi_definitions}

\textcolor{black}{Each study is assigned to a single primary \gls{KPI}, $y$, family, including capacity, latency, coverage, or reliability, based on its main prediction target.}

    \begin{figure*}[htbp]
\centering

% Scales to exactly the width of the page, ensuring minimal vertical height
\resizebox{\textwidth}{!}{

\begin{tikzpicture}[
  font=\small,
  >=Stealth, 
  kpi/.style={
    rounded corners=4pt, thick, align=left,
    inner sep=6pt, text width=48mm, 
    drop shadow={shadow xshift=1.5pt, shadow yshift=-1.5pt, opacity=0.15}
  },
  leading/.style={kpi, top color=blue!5, bottom color=blue!15, draw=blue!20!black},
  lagging/.style={kpi, top color=orange!5, bottom color=orange!15, draw=orange!70!black},
  driver/.style={
    rounded corners=3pt, thin, top color=gray!2, bottom color=gray!12, draw=gray!50, align=left,
    inner sep=5pt, text width=50mm, 
    drop shadow={shadow xshift=1.5pt, shadow yshift=-1.5pt, opacity=0.15}
  },
  strong/.style={->, very thick, draw=black!80},
  impact/.style={->, thick, dashed, draw=gray!70!black}, 
  groupbox/.style={rounded corners=6pt, thick, dashed},
  labelbox/.style={fill opacity=0.9, text opacity=1, font=\scriptsize, rounded corners=2pt, inner sep=2pt}
]

% =========================================================
%  Extracted Gray Boxes (Effective Inputs) - Moved down to y=3
% =========================================================

\node[driver] (env) at (-5.0, 3) {%
  \textbf{Propagation \& interference}\\[1pt]
  \scriptsize mobility, geometry, fading, interferers};

\node[driver] (load) at (1.5, 3) {%
  \textbf{Traffic load \& demand}\\[1pt]
  \scriptsize arrivals, hot-spots, backhaul/core load};

\node[driver] (ctrl) at (8.0, 3) {%
  \textbf{Control \& configuration}\\[1pt]
  \scriptsize scheduler, MCS/HARQ, routing, queueing};

% =========================================================
%  Main KPI nodes - Vertically compressed
% =========================================================

% Left column: Upstream Link Quality
\node[leading] (cov) at (-6.5, 0.0) {%
  \textbf{Coverage / Link quality}\\[1pt]
  \scriptsize PL, RSSI, RSRP, RSRQ, SNR/SINR, CQI};

% Middle column: Reliability as the central bridge
\node[leading] (rel) at (1.5, 0) {%
  \textbf{Reliability}\\[1pt]
  \scriptsize BER, BLER, PER, PLR};

% Right column: Outcomes / Lagging KPIs (Moved closer to center at ±1.1)
\node[lagging] (lat) at (9.5, 1.1) {%
  \textbf{Latency}\\[1pt]
  \scriptsize E2E delay, RTT, jitter\\[1pt]
  \scriptsize RAN delay: queueing, scheduling, HARQ};

\node[lagging] (cap) at (9.5, -1.1) {%
  \textbf{Capacity}\\[1pt]
  \scriptsize UE \& cell throughput, slice rate\\[1pt]
  \scriptsize (goodput / spectral efficiency)};

% =========================================================
%  Main coupling structure (KPI to KPI - Solid)
% =========================================================

% Coverage -> Reliability
\draw[strong] (cov.east) --
  node[midway, above, sloped, labelbox, text=black!80] {decoding quality} (rel.west);

% Reliability -> Latency & Capacity
\draw[strong] (rel.east) to[out=30, in=180]
  node[midway, above, sloped, labelbox, text=black!80] {reTx \& queueing} (lat.west);

\draw[strong] (rel.east) to[out=-30, in=180]
  node[midway, below, sloped, labelbox, text=black!80] {goodput reduction} (cap.west);

% Coverage -> Latency (KPI to KPI, solid)
% Arced gently to bridge exactly between the load node and the reliability node
\draw[strong] (cov.north east) to[out=10, in=175] (lat.north west);

% =========================================================
%  Requested Explicit Connections (Parameters to KPIs - Dashed)
% =========================================================

% 1. propagation & interference -> coverage/link quality
\draw[impact] (env.south) to[out=-90, in=90] (cov.north);

% 2. propagation & interference -> reliability
\draw[impact] (env.south east) to[out=-45, in=135] (rel.north west);

% 3. Traffic load & demand -> reliability
\draw[impact] (load.south) -- (rel.north);

% 4. Traffic load & demand -> capacity
\draw[impact] (load.south east) to[out=-45, in=135] (cap.north west);

% 5. Control & Configuration -> reliability
\draw[impact] (ctrl.south west) to[out=-135, in=45] (rel.north east);

% =========================================================
%  Group boxes and labels (Drawn behind the nodes)
% =========================================================

\begin{scope}[on background layer]
  % Extracted Inputs box
  \node[groupbox, draw=gray!40, fill=gray!5, fit=(env) (load) (ctrl)] (gInputs) {};

  % Upstream box containing leading indicators + reliability bridge + lagging KPIs
  \node[groupbox, draw=blue!30, fill=blue!3, fit=(cov) (rel) (lat) (cap)] (gLead) {};
\end{scope}

\node[font=\small\bfseries, fill=gray!5, text=gray!80!black, inner sep=2pt,
      anchor=south west] at ($(gInputs.north west)+(0.2,0)$)
  {Effective parameters};

\node[font=\small\bfseries, fill=blue!3, text=gray!80!black, inner sep=2pt,
      anchor=south west] at ($(gLead.north west)+(0.2,0)$)
  {KPIs};

\end{tikzpicture}
}
\caption{\textcolor{black}{Interconnected architecture of network KPIs.}}
\label{fig:system_model}
\end{figure*}
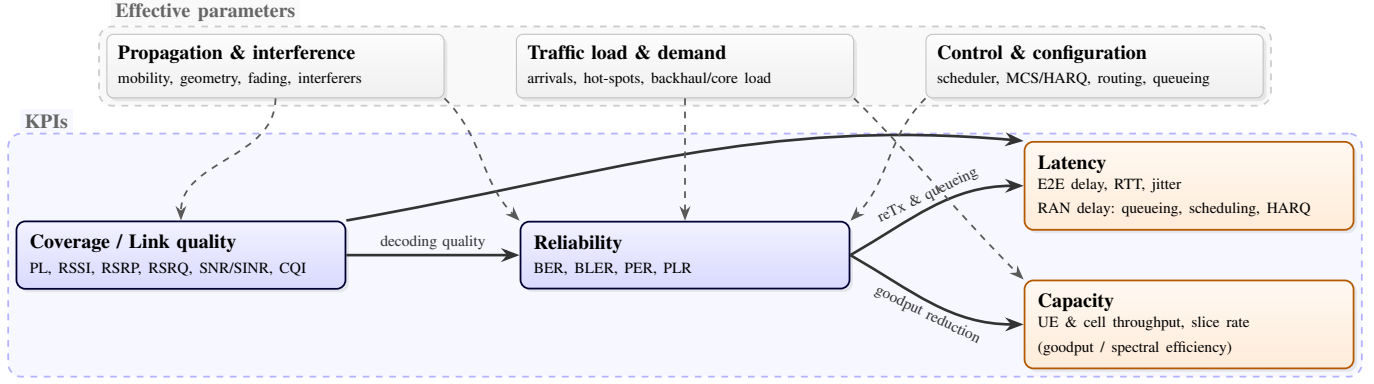

\textcolor{black}{Capacity reflects the network's capability to consistently deliver user data over time, representing the sustained data-delivery rate at a specified reference point and protocol layer. In practical terms, capacity is most often measured as throughput or goodput at the user, cell, or slice level over a defined interval (e.g., bits/s), and also include related efficiency metrics such as spectral efficiency. Since the reported value of capacity depends on factors such as the measurement point (\gls{MAC}/\gls{RLC}/\gls{PDCP}/\gls{IP}), the aggregation level (\gls{UE}, cell, slice), and the averaging window, we classify as capacity those \glspl{KPI} whose primary interpretation is the achieved rate under the prevailing resource allocation, interference, and load conditions~\cite{3GPP_TS_28_554_v16_07_00,3GPP_TS_28_552_v16_9_0}.}

\textcolor{black}{Latency characterizes the time required for a data packet (or protocol data unit) to travel between two specified reference points within a network. This is typically reported as either one-way delay or round-trip time and is commonly summarized using statistical measures such as the mean or percentiles; delay variation, or jitter, serves as a related descriptor~\cite{ITU_Y1540}. The total end-to-end latency can be decomposed into several components, including processing, queueing or buffering, transmission and scheduling, retransmissions, propagation, and routing. The relative significance of each component depends on whether the dominant segment is the \gls{RAN}, transport, or core network. We classify as latency targets those \glspl{KPI} whose primary interpretation is packet transfer delay, whether end-to-end or component-wise, as formalized in packet-network performance definitions and cellular management specifications~\cite{ETSI_TS_128_554_v16_07_00,3GPP_TS_28_552_v16_9_0}.}

\textcolor{black}{Coverage represents the spatial availability and quality of radio service, indicating whether the received signal at a given location and time satisfies the minimum strength and quality thresholds necessary for reliable connectivity and desired performance. In practice, coverage is assessed using a range of radio measurements and derived indicators, such as \gls{PL}, \gls{RSSI}, \gls{RSRP}, \gls{RSRQ}, \gls{SNR}/\gls{SINR}, and \gls{CQI}. These metrics collectively capture aspects of received power, interference-plus-noise conditions, and link quality relevant for adaptive transmission. 
We classify as coverage or link-quality targets those \glspl{KPI} whose primary interpretation is radio signal strength or quality, including spatial field reconstructions such as radio environment maps (\gls{REM} grids) or coverage probability maps. This approach aligns with standardized definitions for radio measurement and reporting~\cite{3GPP_TS_38_215,3GPP_TS_36_214}.}

\textcolor{black}{Reliability measures the likelihood of delivering accurate, timely information across the communication stack. Depending on the protocol layer and service context, reliability is typically assessed using error rates (e.g., \gls{BER}, \gls{BLER}, and \gls{PER}), retransmission statistics, and end-to-end packet loss. For mission-critical applications, reliability is often defined as the probability of successful delivery within a specified latency constraint, reflecting the stringent requirements for deadline-bound communication. We classify as reliability targets those KPIs whose primary interpretation centers on delivery correctness and/or delivery completion, in alignment with cellular service requirements and established network performance criteria for loss and successful transfer~\cite{3GPP_TS_22_261,ITU_R_M2410,ITU_Y1540}.}

\textcolor{black}{Fig.~\ref{fig:system_model} summarizes an interconnected KPI architecture in which three effective parameter sets, propagation $\&$ interference, traffic load $\&$ demand, and control $\&$ configuration, jointly determine service performance. Propagation $\&$ interference mainly affects coverage and link quality, which in turn influence decoding quality and overall reliability. When reliability drops, retransmissions and extra buffering become necessary. This leads to higher latency and reduced capacity, as more resources are spent maintaining data integrity rather than delivering new data. At the same time, increased traffic load increases contention and queue occupancy, which further increases latency and can also impact reliability and capacity, especially during congestion. Control $\&$ configuration settings, such as scheduling and \gls{HARQ} parameters, help balance these trade-offs by translating radio conditions and network load into actual throughput, delay, and error rates. This hierarchical structure highlights the sequential nature of performance degradation, where issues in upstream indicators propagate through the system and manifest as service-level challenges further downstream.}

\subsection{Taxonomy}
\label{subsec:taxonomy}
\begin{figure*}[htbp]
    \centering

    % Define a sophisticated, minimal nude palette
    \definecolor{rootcol}{HTML}{9BA9B4} % Deep Warm Taupe / Mocha
    \definecolor{nudecol}{HTML}{D9CBBF} % Soft Nude / Sand

    \begin{tikzpicture}[
      >=Latex,
      % Set the default style for the tree structure and edges
      edge from parent fork down,
      level 1/.style={sibling distance=4.2cm, level distance=1.3cm}, % Reduced from 2.0cm
      level 2/.style={sibling distance=4.2cm, level distance=0.8cm}, % Reduced from 1.5cm
      edge from parent/.style={draw=rootcol!80, thick, rounded corners=3pt}, 
      every node/.style={
          rectangle, rounded corners=2.5mm,
          blur shadow={shadow blur steps=4, shadow xshift=0pt, shadow yshift=-1.5pt},
          align=left, font=\footnotesize
      },
      % Root Node Style (Deep Taupe)
      root/.style={
          fill=rootcol!80, text=white, font=\small\bfseries, align=center, 
          minimum height=0.7cm, text width=9.5cm, draw=rootcol!80!black, line width=1.2pt % Reduced height
      },
      % Branch Style (Solid Nude)
      branch/.style={
          fill=nudecol!90, draw=rootcol!70, text=black!90, font=\small\bfseries, 
          align=center, minimum height=0.6cm, text width=3.6cm, line width=1pt % Reduced height
      },
      % Leaf Style (Soft nude background, anchored at the north to prevent overlap)
      leaf/.style={
          fill=nudecol!20, draw=rootcol!40, text=black!85, font=\footnotesize, 
          text width=3.2cm, inner sep=5pt, line width=0.8pt, % Reduced inner sep from 8pt to 5pt
          anchor=north % <--- This aligns the tops of all leaves and prevents edge overlap!
      }
    ]

    % Root node
    \node[root] {Data-driven KPI Prediction in Mobile Networks}
      % First branch - KPI type
      child {
        node[branch] {KPI Type}
        child {node[leaf] {
            \textbullet~Capacity\\[2pt]
            \textbullet~Latency\\[2pt]
            \textbullet~Coverage\\[2pt]
            \textbullet~Reliability
        }}
      }
      % Second branch - Data source
      child {
        node[branch] {Data Source}
        child {node[leaf] {
            \textbullet~Measurements\\[2pt]
            \textbullet~Simulated\\[2pt]
            \textbullet~Synthetic\\[2pt]
            \textbullet~Hybrid
        }}
      }
      % Third branch - Prediction Attributes
      child {
        node[branch] {Prediction Attributes}
        child {node[leaf] {
            \textbullet~Protocol stack\\[2pt]
            \textbullet~Horizon\\[2pt]
            \textbullet~Output / metric
        }}
      }
      % Fourth branch - Model Families
      child {
        node[branch] {Model Families}
        child {node[leaf] {
            \textbullet~Unsupervised ML\\[2pt]
            \textbullet~Classic Supervised\\[2pt]
            \textbullet~Ensemble models\\[2pt]
            \textbullet~Deep Spatial (CNN)\\[2pt]
            \textbullet~Deep Sequence\\[2pt]
            \textbullet~Other
        }}
      };

    \end{tikzpicture}
    \caption{Multi-dimensional taxonomy of data-driven KPI prediction approaches in mobile networks.}
    \label{fig:taxonomy}
\end{figure*}
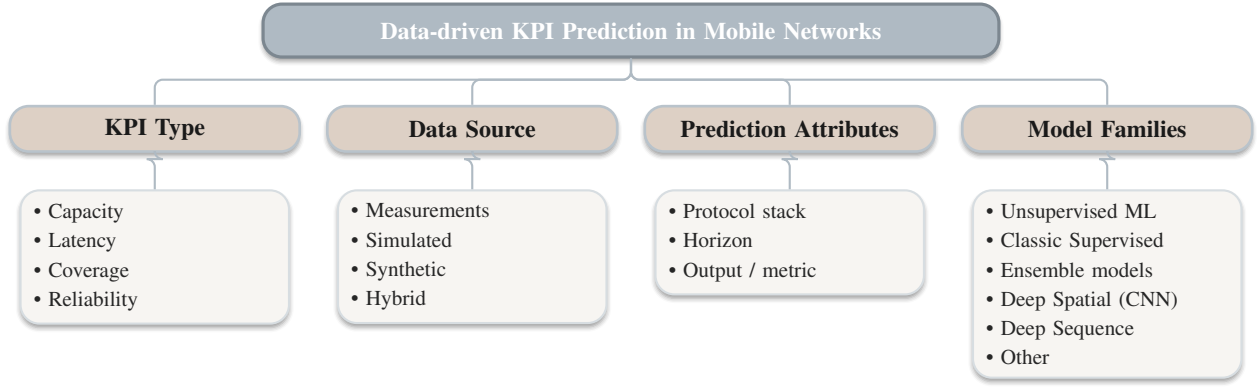

\textcolor{black}{The survey taxonomy organizes each study along four dimensions: (a) \gls{KPI} family, (b) data source type, (c) prediction attributes (network protocol stack, horizon, and output), and (d) model family.} 
Fig.~\ref{fig:taxonomy} serves as the conceptual anchor, enabling two complementary uses of the taxonomy: a) Taxonomy as an organizing framework, which provides a consistent structure to compare studies along \gls{KPI} family, data source realism, prediction formulation, and modeling choices; and b) Taxonomy as a quantitative lens, based on which, the same taxonomy axes generate the empirical landscape figures used throughout the paper, enabling us to identify dominant trends (e.g., \gls{KPI} and model families with the largest presence) and systematic gaps. After screening, the final corpus comprises \textbf{151} data-driven \gls{KPI}-prediction studies.

Building on this methodology, we first provide background on widely used \gls{ML}/\gls{AI} model families for \gls{KPI} prediction, as we believe this foundation is helpful to establish before discussing the taxonomy. We then proceed to review the model families in detail and present \gls{KPI}-wise comparative analyses and open research directions, grounded in the coded corpus and the quantitative landscape.
\section{Overview of data-driven prediction methods}
\label{sec:methods}
Building on the prediction and estimation formulation introduced in Section~\ref{sec:scope_taxonomy}, data-driven \gls{KPI} prediction learns from an available information set \(\mathcal{I}_t^k\) to estimate a target \(y_{t+\tau}\) or infer a latent state. The structure of the current observation \(x_t\) varies significantly, shaping the choice of \gls{ML} method. In wireless networks, \(x_t\) is typically derived from \gls{UE} reports, \gls{RAN} counters, and management telemetry (e.g., \gls{RSRP}/\gls{SINR}, traffic load, mobility statistics, and \gls{QoE} metrics).
Depending on the task, \(x_t\) takes different forms as follows:
\begin{itemize}
    \item {Tabular}: Feature vectors representing instantaneous or aggregated metrics, common in classical supervised and ensemble models.
    \item {Sequential}: Historical observation windows used for temporal \gls{KPI} forecasting.
    \item {Spatial}: Grid or tensor representations (\(x_t \in \mathcal{X}^{H\times W\times C}\)) capturing maps, imagery, or transmitter layouts, heavily utilized by \glspl{CNN} for coverage prediction.
    \item {Graph-structured}: Nodes and edges representing network topologies, routing paths, or interference relations, making them suited for \glspl{GNN}.
    \item {Unlabeled}: Raw observations used to learn latent representations, clusters, or anomaly scores.
\end{itemize}

In the following, we provide an overview of the main \gls{ML} families used in the reviewed literature.

\subsection{Unsupervised Learning}
\label{sec:unsupervised}
Unsupervised learning uncovers intrinsic data structures without external labels, focusing on clustering, dimensionality reduction, and anomaly detection. Common methods include \(k\)-means and \glspl{GMM} for clustering~\cite{macqueen1967multivariate, dempster1977em}, and \gls{PCA} for dimensionality reduction~\cite{pearson1901liii}. Anomaly detection identifies out-of-distribution events by learning the normal data distribution~\cite{breunig2000lof}, a capability crucial for root cause analysis and the detection of abnormal network behavior.

\begin{figure}[t]
    \centering
    \includegraphics[width=0.9\linewidth]{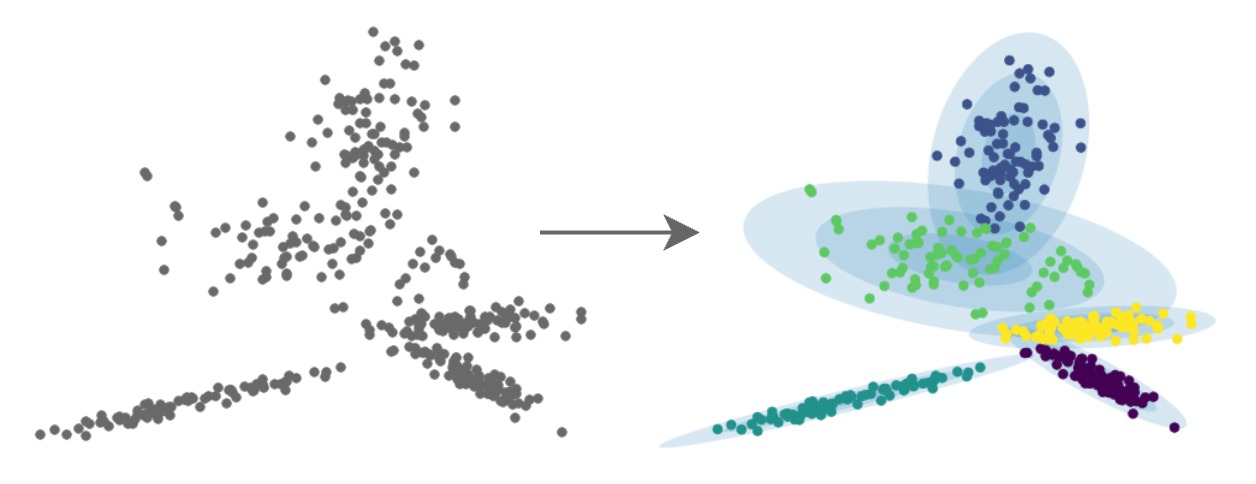}
    \caption{Illustration of unsupervised GMM clustering. Left: unlabeled observations. Right: data fitted to a GMM, accurately capturing the underlying Gaussian components.}
    \label{fig:gmm_clustering_example}
\end{figure}

\subsection{Supervised Learning}
\label{sec:supervised}
Supervised learning maps labeled inputs to \gls{KPI} targets. Traditional parametric models serve as strong, interpretable baselines when data is scarce. Linear regression (often regularized via Ridge or Lasso) and Generalized Linear Models are widely used \cite{tibshirani1996regression}. Logistic regression provides a convex, calibrated baseline for classification~\cite{cox1972regression}. Other fundamental approaches include \glspl{SVM} and \gls{SVR} for margin-based non-linear prediction~\cite{cortes1995support}, as well as \gls{KNN} and Naïve Bayes~\cite{cover1967nearest}. The limited representational capacity of these models for complex network dynamics motivates the adoption of ensemble and deep learning architectures.

\subsection{Ensemble Learning}
\label{sec:ensemble}
Ensemble methods combine multiple base learners to improve accuracy and stability. Bagging methods, such as \glspl{RF}~\cite{Breiman2001}, average decorrelated decision trees to reduce variance. Boosting methods sequentially correct errors in predecessors to minimize bias. \gls{AdaBoost}~\cite{freund1997decision} and Gradient boosting frameworks~\cite{friedman2001greedy}, particularly \gls{XGBoost}~\cite{chen2016xgboost}, \gls{LightGBM}, and \gls{CatBoost}~\cite{ke2017lightgbm}, are highly popular for tabular \gls{KPI} prediction due to their scalability, robust categorical feature handling, and strong performance.

\subsection{Deep Learning}
\label{subsec:deep}
Deep neural networks extract hierarchical features from raw data, relying on optimizers like Adam~\cite{kingma2014adam}, batch normalization~\cite{ioffe2015batch}, and dropout`\cite{srivastava2014dropout} to stabilize training and improve generalization. Different network architectures are tailored to specific input structures:

\subsubsection{Spatial Modeling with Convolutional Networks}
\label{subsubsec:cnn}
\glspl{CNN} are ideal for spatial data like radio environment maps and sensor grids. Using local filters, weight sharing, and encoder-decoder designs (e.g., U-Net), \glspl{CNN}`\cite{lecun1998, krizhevsky2012imagenet} efficiently capture spatial correlations, enabling dense, per-pixel predictions for \glspl{KPI} such as \gls{PL} or \gls{SINR}.

\subsubsection{Sequence Modeling}
\label{subsubsec:sequence}
\begin{figure}[ht]
    \centering
    \includegraphics[width=0.9\linewidth]{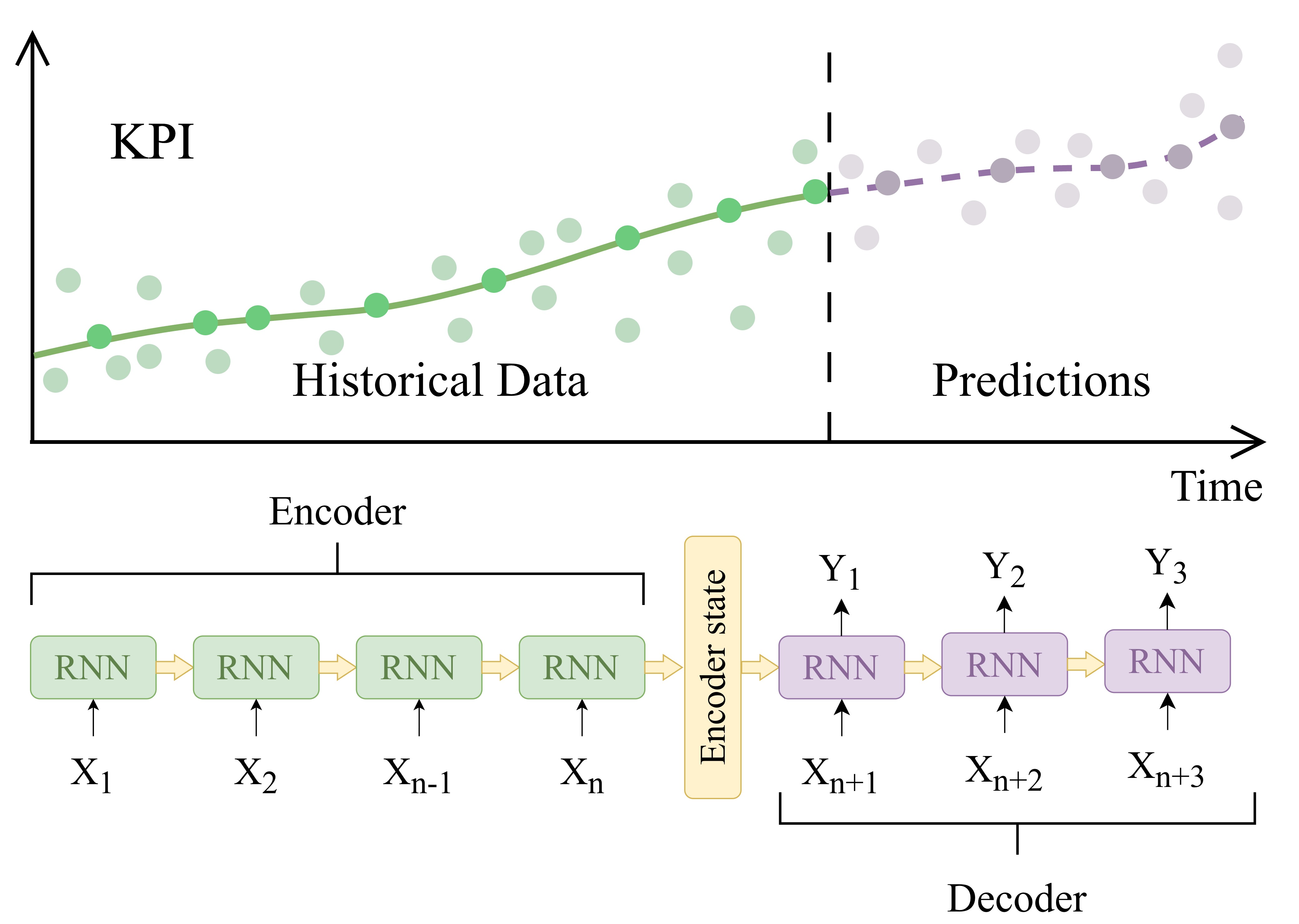}
    \caption{Sequence-to-sequence RNN architecture for time-series KPI prediction.}
    \label{fig:rnn_seq2seq_kpi}
\end{figure}

For temporal \gls{KPI} forecasting, \glspl{RNN}~\cite{elman1990finding} and \glspl{LSTM}~\cite{hochreiter1997long} process inputs autoregressively, maintaining hidden states to capture temporal dependencies (Fig.~\ref{fig:rnn_seq2seq_kpi}). However, wireless \glspl{KPI} often exhibit quasi-stationarity and strong periodic structures (e.g., diurnal traffic cycles). Models must account for regime shifts and seasonality, often requiring careful split design (e.g., rolling-origin evaluation) or seasonal normalization.

\begin{figure}[t]
    \centering
    \includegraphics[width=\linewidth]{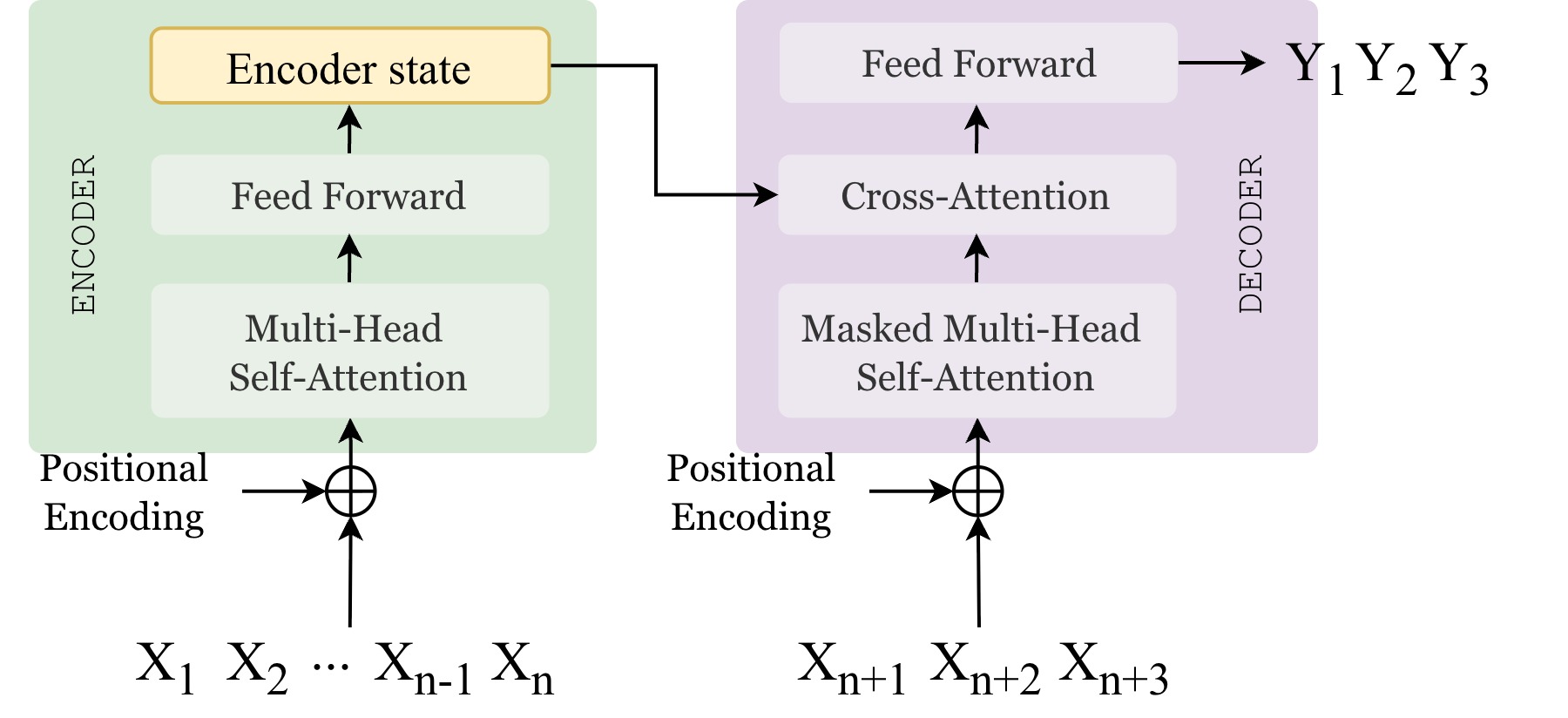}
    \caption{Transformer sequence-to-sequence model for KPI forecasting.}
    \label{fig:transformer_seq2seq_kpi}
\end{figure}

Recently, Transformer architectures~\cite{vaswani2017attention} (Fig.~\ref{fig:transformer_seq2seq_kpi}) have largely superseded standard \glspl{RNN} for complex tasks. By tokenizing raw inputs~ \cite{devlin2018bert, radford2019language} and relying on self-attention instead of recurrence, Transformers process data in parallel to capture long-range dependencies and multi-resolution temporal patterns, making them a robust standard for multi-step sequence prediction.

\subsubsection{Distributed and Structure-Aware Learning}
\label{subsubsec:fl_gnn}
Federated learning~\cite{mcmahan2017communication} enables collaborative model training across network entities without exchanging raw data, preserving privacy and reducing backhaul load. Concurrently, \glspl{GNN}~\cite{battaglia2018relational} process network topologies by passing messages across nodes and edges, effectively capturing multi-hop interference and routing dynamics that standard grid or node-level architectures miss.

\subsection{Prediction uncertainty and probabilistic outputs}
\label{subsec:forecasting_uncertainty}
While many predictors output deterministic point forecasts, \gls{AI}-native networks inherently require uncertainty-aware outputs for risk-aware control. To streamline the prediction of uncertainty, methods fall into four key paradigms:
\begin{itemize}
    \item \emph{Probabilistic forecasting}: Outputs full predictive distributions or parameters, useful for capturing latency tails in \gls{URLLC}~\cite{sharma2023toward, mostafavi2025probabilistic, lat_09, lat_12, lat_22, lat_23}.
    \item \emph{Confidence and prediction intervals}: Quantifies uncertainty ranges around mean estimates to bound future unobserved values~\cite{albert2025characterizing}.
    \item \emph{Bayesian learning}: Decomposes aleatoric and epistemic uncertainty using \glspl{BNN} for robust, out-of-distribution planning~\cite{ericsson2023defining, rsrp_dl_01, lat_07, carlobayesian}.
    \item \emph{Conformal prediction}: Constructs distribution-free prediction sets that guarantee finite-sample coverage under distribution shifts, ensuring strict \gls{SLA} compliance~\cite{carloconformal,conformal_01,conformal_02,conformal_03}.
\end{itemize}

In this section, we highlighted the major algorithms used for \gls{KPI} prediction. While classical models offer interpretability and low computational cost, sequence and deep learning architectures provide strong representational power for complex temporal and spatial dynamics. Building on this methodological overview, the next section shifts to a \gls{KPI}‑centric analysis, synthesizing the implementation of these methods across capacity, latency, coverage, and reliability tasks.
\section{Comparative Analysis of \gls{KPI} Prediction Approaches}
\label{sec:comparisons}
In this section, we provide an overview of the \gls{ML}/\gls{AI} models used for predicting \glspl{KPI} in mobile networks, focusing on four major KPI families: coverage/link quality, latency, capacity, and reliability. 
Accordingly, each \gls{KPI}-specific subsection follows a consistent structure. We begin with a brief “warm-up” discussion that motivates the importance of the \gls{KPI}, its operational relevance in the \gls{RAN}, and its role in network planning, optimization, and control. This is followed by a concise overview of traditional prediction and estimation techniques, which are typically based on analytical models or classical statistical and time-series methods. The main body of each subsection then reviews the evolution of data-driven approaches, progressing from early, relatively simple \gls{ML} models to more recent, expressive architectures, including ensemble methods, deep learning, and domain-aware designs. Within this discussion, particular attention is given to novel modeling ideas introduced in recent works, as well as to ablation and comparative studies that systematically analyze the contributions of individual model components, features, or data modalities and provide insights into what drives performance, robustness, and generalization in practice.
To facilitate comparison across studies, each \gls{KPI}-specific subsection includes a table of matrix taxonomy of KPI prediction approaches. Additionally, in Fig.~\ref{fig:unified_sys_model}, we provide an overview of how the proposed machine learning models and target use cases are mapped directly to the network domains where they are deployed, per \gls{KPI}.

\input{figures/KPI_MLdeploy_fig}

\subsection{Capacity}
Capacity-related network \glspl{KPI} are fundamental to understanding how efficiently a mobile network delivers data across different operational levels. These include the throughput experienced by an individual user, the aggregate throughput of a base-station \cite{hussien2025machine}, or the overall rate achieved by a network slice \cite{raca2020leveraging}. The relationships among these \glspl{KPI} are inherently hierarchical and interdependent. For instance, variations in radio link quality directly impact per-user throughput, which in turn influences aggregate cell capacity, while end-to-end slice throughput is often constrained by bottlenecks in either the \gls{RAN} or the transport domain. 

\begin{table*}[ht!]
\centering
\caption{Matrix Taxonomy of Capacity-Related KPI Prediction Approaches.}
\label{tab:cap_matrix}
\scriptsize
\renewcommand{\arraystretch}{1.1}
\setlength{\tabcolsep}{4pt}
\begin{tabular}{@{} l | c c | c c | c c c | c c | c c c @{}}
\toprule
\multirow{2}{*}{\textbf{Reference}} & \multicolumn{2}{c|}{\textbf{Data Source}} & \multicolumn{2}{c|}{\textbf{Network Type}} & \multicolumn{3}{c|}{\textbf{Target KPI}} & \multicolumn{2}{c|}{\textbf{Horizon}} & \multicolumn{3}{c}{\textbf{Model Family}} \\
\cmidrule(lr){2-3} \cmidrule(lr){4-5} \cmidrule(lr){6-8} \cmidrule(lr){9-10} \cmidrule(lr){11-13}
 & \textbf{Real Trace} & \textbf{Sim./Emu.} & \textbf{4G/LTE} & \textbf{5G/B5G} & \textbf{DL Tput} & \textbf{UL Tput} & \textbf{Other} & \textbf{Short-term} & \textbf{Long/Static} & \textbf{Stat/Math} & \textbf{Class. ML} & \textbf{Deep L.} \\
\midrule
\cite{lv2025accurate} & $\checkmark$ & & & $\checkmark$ & $\checkmark$ & $\checkmark$ & $\checkmark$ & $\checkmark$ & & $\checkmark$ & $\checkmark$ & \\
\cite{mostafa2022downlink} & $\checkmark$ & & $\checkmark$ & & $\checkmark$ & & & & $\checkmark$ & $\checkmark$ & & $\checkmark$ \\
\cite{mismar2021unsupervised} & $\checkmark$ & & $\checkmark$ & $\checkmark$ & & & $\checkmark$ & & $\checkmark$ & & $\checkmark$ & \\
\cite{biernacki2024throughput} & $\checkmark$ & & & $\checkmark$ & $\checkmark$ & & & $\checkmark$ & & & & $\checkmark$ \\
\cite{elsherbiny20204g} & $\checkmark$ & & $\checkmark$ & & $\checkmark$ & & & $\checkmark$ & & $\checkmark$ & $\checkmark$ & $\checkmark$ \\
\cite{malekzadeh2023performance} & & $\checkmark$ & & $\checkmark$ & $\checkmark$ & $\checkmark$ & $\checkmark$ & & $\checkmark$ & & $\checkmark$ & \\
\cite{tran2023ml} & & $\checkmark$ & & $\checkmark$ & $\checkmark$ & & $\checkmark$ & $\checkmark$ & & $\checkmark$ & & $\checkmark$ \\
\cite{yue2017linkforecast} & $\checkmark$ & & $\checkmark$ & & $\checkmark$ & & & $\checkmark$ & & & $\checkmark$ & \\
\cite{raca2020leveraging} & $\checkmark$ & & $\checkmark$ & $\checkmark$ & $\checkmark$ & & & $\checkmark$ & & & $\checkmark$ & $\checkmark$ \\
\cite{alshawki2024congestion} & & $\checkmark$ & & $\checkmark$ & & & $\checkmark$ & & $\checkmark$ & & $\checkmark$ & \\
\cite{azmin2022bandwidth} & $\checkmark$ & & & $\checkmark$ & $\checkmark$ & & & $\checkmark$ & & & $\checkmark$ & $\checkmark$ \\
\cite{lee2020perceive} & $\checkmark$ & & $\checkmark$ & & & $\checkmark$ & & $\checkmark$ & & & & $\checkmark$ \\
\cite{riihijarviMachineLearningPerformance2018}& $\checkmark$ & & $\checkmark$ & & $\checkmark$ & & $\checkmark$ & $\checkmark$ & $\checkmark$ & $\checkmark$ & $\checkmark$ & \\
\cite{kim5G23} & $\checkmark$ & & & $\checkmark$ & $\checkmark$ & & & & $\checkmark$ & & $\checkmark$ & $\checkmark$ \\
\cite{barmpounakis2021lstm} & & $\checkmark$ & & $\checkmark$ & & $\checkmark$ & & $\checkmark$ & & & & $\checkmark$ \\
\cite{cap_02} & $\checkmark$ & & $\checkmark$ & $\checkmark$ & $\checkmark$ & & & $\checkmark$ & & & & $\checkmark$ \\
\cite{cap_03} & $\checkmark$ & & $\checkmark$ & $\checkmark$ & $\checkmark$ & & $\checkmark$ & $\checkmark$ & & & $\checkmark$ & $\checkmark$ \\
\cite{cap_04} & $\checkmark$ & & $\checkmark$ & & $\checkmark$ & & & $\checkmark$ & & $\checkmark$ & & $\checkmark$ \\
\cite{cap_07} & $\checkmark$ & & $\checkmark$ & $\checkmark$ & $\checkmark$ & & $\checkmark$ & $\checkmark$ & & & $\checkmark$ & $\checkmark$ \\
\cite{cap_01} & $\checkmark$ & & & & $\checkmark$ & & $\checkmark$ & $\checkmark$ & $\checkmark$ & & & $\checkmark$ \\
\cite{cap_08} & $\checkmark$ & & & $\checkmark$ & & $\checkmark$ & & $\checkmark$ & & & & $\checkmark$ \\
\cite{cap_22} & $\checkmark$ & & $\checkmark$ & & $\checkmark$ & $\checkmark$ & & $\checkmark$ & & & $\checkmark$ & \\
\cite{cap_23} & $\checkmark$ & & $\checkmark$ & $\checkmark$ & & $\checkmark$ & & & $\checkmark$ & & $\checkmark$ & \\
\cite{cap_24} & $\checkmark$ & & $\checkmark$ & $\checkmark$ & & & $\checkmark$ & & $\checkmark$ & & $\checkmark$ & $\checkmark$ \\
\cite{cap_25} & & $\checkmark$ & $\checkmark$ & $\checkmark$ & & & $\checkmark$ & $\checkmark$ & & & & $\checkmark$ \\
\cite{cap_26} & $\checkmark$ & & $\checkmark$ & $\checkmark$ & $\checkmark$ & $\checkmark$ & & & $\checkmark$ & & $\checkmark$ & \\
\cite{cap_27} & $\checkmark$ & & & $\checkmark$ & $\checkmark$ & & & $\checkmark$ & & & & $\checkmark$ \\
\cite{cap_28} & $\checkmark$ & & $\checkmark$ & & $\checkmark$ & $\checkmark$ & & $\checkmark$ & & & $\checkmark$ & \\
\cite{cap_29} & $\checkmark$ & & $\checkmark$ & $\checkmark$ & $\checkmark$ & & & & $\checkmark$ & & $\checkmark$ & \\
\cite{cap_30} & $\checkmark$ & & $\checkmark$ & $\checkmark$ & $\checkmark$ & & $\checkmark$ & & $\checkmark$ & & $\checkmark$ & \\
\cite{cap_31} & $\checkmark$ & & $\checkmark$ & $\checkmark$ & $\checkmark$ & & $\checkmark$ & & $\checkmark$ & & & $\checkmark$ \\
\cite{cap_05} & $\checkmark$ & & $\checkmark$ & & $\checkmark$ & & & $\checkmark$ & $\checkmark$ & $\checkmark$ & & \\
\cite{cap_09} & $\checkmark$ & & $\checkmark$ & $\checkmark$ & & $\checkmark$ & & $\checkmark$ & & & & $\checkmark$ \\
\cite{cap_10} & $\checkmark$ & & & $\checkmark$ & & $\checkmark$ & & & $\checkmark$ & & $\checkmark$ & $\checkmark$ \\
\cite{cap_11} & $\checkmark$ & & $\checkmark$ & $\checkmark$ & $\checkmark$ & $\checkmark$ & & $\checkmark$ & & & $\checkmark$ & $\checkmark$ \\
\cite{cap_12} & $\checkmark$ & & & $\checkmark$ & & $\checkmark$ & & $\checkmark$ & & & & $\checkmark$ \\
\cite{cap_14} & $\checkmark$ & & $\checkmark$ & $\checkmark$ & $\checkmark$ & $\checkmark$ & & $\checkmark$ & & & $\checkmark$ & \\
\cite{cap_13,cap_20,cap_21,cap_15,cap_18} & & $\checkmark$ & & & & & $\checkmark$ & $\checkmark$ & & & $\checkmark$ & $\checkmark$ \\
\bottomrule
\multicolumn{13}{p{17.5cm}}{\textit{Note:} Characteristics marked `Other' include ACK/NACK, CQI, RAT handoffs, or transport congestion states. `Stat/Math' includes statistical, linear programming, or algorithmic rules. `Short-term' denotes sub-minute horizons, whereas `Long/Static' applies to horizons ranging from hours to weeks, or offline architectural mappings.}
\end{tabular}
\end{table*}

Traditional throughput prediction approaches have heavily relied on simple linear models (e.g., \gls{MA}) and statistical time-series forecasting methods, such as \gls{ARIMA} and its seasonal or exponential variants \cite{elsherbiny20204g,mostafa2022downlink,lv2025accurate,cap_05}. While these methods provide decent baseline estimates and are highly interpretable, their simplified assumptions about linearity and sensitivity to parameter tuning render them inadequate for the highly dynamic, non-linear environments typical of 5G networks \cite{santos2025statistics}. 

To overcome these limitations, a broad spectrum of classical \gls{ML} algorithms has been widely adopted. Methods such as linear regression, \gls{SVM}, and decision trees provide a balance between computational efficiency and the ability to capture complex mappings between input features (e.g., channel quality, load, mobility indicators) and throughput \cite{raca2020leveraging,yue2017linkforecast,elsherbiny20204g,malekzadeh2023performance,mostafa2022downlink,cap_23,cap_29}. As computational capacities grew, ensemble models like \glspl{RF}, \gls{XGBoost}, and \gls{LightGBM} became the industry standard for operational throughput forecasting. These boosting and bagging frameworks are highly valued for their robustness to feature noise and their scalability for deployment in \gls{Near-RT} prediction pipelines, with applications ranging from standard link throughput prediction to proactive \gls{QoS} warnings in \gls{C-V2X} scenarios \cite{riihijarviMachineLearningPerformance2018,kim5G23,lv2025accurate,cap_22,cap_32}. 

The increasing heterogeneity of 5G deployments and cross-domain dependencies has driven a shift toward deep learning approaches, which excel at modeling complex spatio-temporal dynamics. Recurrent models, particularly \gls{LSTM} and \gls{GRU} architectures, are extensively used to learn temporal dependencies in volatile throughput traces, capturing bursty fluctuations and repetitive traffic patterns \cite{tran2023ml,raca2020leveraging,elsherbiny20204g,lee2020perceive,lv2025accurate,azmin2022bandwidth,cap_25}. To deploy these models on resource-constrained devices, researchers have developed compact hybrid architectures (e.g., stacked \gls{Bi-LSTM} or Conv\gls{LSTM} networks) tailored for short-term, on-device forecasting \cite{cap_02, cap_12}. For more comprehensive network modeling, spatio-temporal frameworks combining \glspl{CNN} or \glspl{GNN} with \glspl{LSTM} explicitly integrate base-station topology, spatial dependencies across cells, and time-series similarity \cite{zhang2019deep, narayanan2020lumos5g, barmpounakis2021lstm, biernacki2024throughput}. Most recently, attention-based mechanisms and transformer variants (e.g., Informer, \gls{GCformer}) have demonstrated superior performance in multi-horizon and multi-cell traffic prediction by effectively learning long-range temporal dependencies and spatial relationships \cite{azmin2022bandwidth, shang2024long}.

Despite achieving state-of-the-art accuracy, modern deep learning models face operational challenges related to generalization, data imbalance, and limited predictability. Models often suffer performance degradation when transferred across different spectrum bands, hardware configurations, or mobility profiles \cite{santos2025statistics, al2023downlink}. Furthermore, fine-scale sub-second forecasting is fundamentally hindered by inherent network noise and ``conflicting samples'', instances in which identical \gls{RAN} feature sets yield divergent throughput outcomes \cite{cap_11}. Consequently, recent complementary research has shifted focus toward reliability and uncertainty estimation. Using uncertainty-aware predictors (e.g., via \gls{NGBoost}) and asymmetric loss functions, these capacity-aware frameworks penalize overprediction, ensuring forecasts remain strictly within achievable bounds and ultimately mitigating the risk of bufferbloat and latency spikes in safety-critical applications~\cite{albert2025characterizing, cap_08}.

To further optimize model efficiency and interpretability, extensive ablation and feature-importance analyses have been conducted. Research shows that traditional sequence encoders are often insufficient without explicit temporal modules that capture daily traffic periodicity \cite{shang2024long}. Additionally, incorporating an exhaustive set of external covariates does not guarantee improved performance and can introduce noise \cite{santos2025statistics}. To address this, algorithmic filtering strategies, including \gls{LASSO} regression, two-stage Pearson correlation, and \gls{SHAP} analysis, are increasingly applied to reduce dimensionality, trace dominant predictors across network layers, and make complex predictors operationally actionable \cite{mostafa2022downlink,azmin2022bandwidth,kim5G23,albert2025characterizing,lv2025accurate}. 

\subsection{Latency}
Latency-related \glspl{KPI} in cellular networks form a tightly coupled set across the \gls{RAN}, transport, and core, capturing the responsiveness experienced by applications and users \cite{ITU_Y1540, ETSI_TS_128_554_v16_07_00, 3GPP_TS_28_552_v16_9_0}. End-to-end latency aggregates multiple components along the \gls{UE}-to-endpoint path. The dominant contributors vary by segment. In the \gls{RAN}, latency is primarily shaped by buffer build-up and queueing, scheduling/grant timing, retransmissions, and radio conditions. In the transport/core, latency is driven by processing, routing, and congestion within the packet network and core infrastructure. These components vary with load, mobility, and user position, and infrastructure configuration, yielding both slow trends and abrupt regime changes (e.g., bursts and handovers). In practice, end-to-end latency is the operational quality metric at the service edge, while scheduling and queue management variables are internal levers. Accordingly, predicting latency at multiple granularities (\gls{UE}/cell/service) supports proactive resource allocation and service assurance \cite{ITU_Y1540, ETSI_TS_128_554_v16_07_00, 3GPP_TS_28_552_v16_9_0}.

\begin{table*}[h!]
\centering
\caption{Matrix Taxonomy of Latency-Related KPI Prediction Approaches.}
\label{tab:lat_matrix}
\scriptsize
\renewcommand{\arraystretch}{1.1}
\setlength{\tabcolsep}{2pt}
\begin{tabular}{@{} l | c c | c c | c c c | c c | c c c @{}}
\toprule
\multirow{2}{*}{\textbf{Reference}} & \multicolumn{2}{c|}{\textbf{Data Source}} & \multicolumn{2}{c|}{\textbf{Network Type}} & \multicolumn{3}{c|}{\textbf{Target KPI}} & \multicolumn{2}{c|}{\textbf{Horizon}} & \multicolumn{3}{c}{\textbf{Model Family}} \\
\cmidrule(lr){2-3} \cmidrule(lr){4-5} \cmidrule(lr){6-8} \cmidrule(lr){9-10} \cmidrule(lr){11-13}
 & \textbf{Real Trace} & \textbf{Sim/Hybrid} & \textbf{Cellular} & \textbf{IP/Other} & \textbf{E2E/Path} & \textbf{RAN/RTT} & \textbf{Dist/Tail} & \textbf{Point/Static} & \textbf{Multi/Seq.} & \textbf{Math/ML} & \textbf{Deep NN} & \textbf{Graph/Top.} \\
\midrule
\cite{rao2020generalizable, larsson2023domain, lat_03, lat_16, lat_26} & $\checkmark$ & & $\checkmark$ & & & $\checkmark$ & & $\checkmark$ & & & $\checkmark$ & \\
\cite{rao2022prediction, khatouni2019machine, lat_17} & $\checkmark$ & & $\checkmark$ & & & $\checkmark$ & & $\checkmark$ & & $\checkmark$ & & \\
\cite{ahmed2021predicting} & $\checkmark$ & & $\checkmark$ & & $\checkmark$ & & & $\checkmark$ & & $\checkmark$ & & \\
\cite{lat_13} & $\checkmark$ & & $\checkmark$ & & & $\checkmark$ & $\checkmark$ & $\checkmark$ & & $\checkmark$ & & \\
\cite{mostafavi2023data} & $\checkmark$ & & $\checkmark$ & & & $\checkmark$ & $\checkmark$ & $\checkmark$ & & & $\checkmark$ & \\
\cite{fadhil2023estimation} & $\checkmark$ & & $\checkmark$ & & $\checkmark$ & & $\checkmark$ & $\checkmark$ & & & $\checkmark$ & \\
\cite{lat_09} & $\checkmark$ & & $\checkmark$ & & $\checkmark$ & & & & $\checkmark$ & & $\checkmark$ & \\
\cite{lan2025multi, cai2024adaptive} & $\checkmark$ & & $\checkmark$ & & & $\checkmark$ & & & $\checkmark$ & & $\checkmark$ & \\
\cite{lat_10} & $\checkmark$ & & $\checkmark$ & & $\checkmark$ & $\checkmark$ & & & $\checkmark$ & & $\checkmark$ & \\
\cite{mostafavi2025probabilistic, lat_02} & $\checkmark$ & & $\checkmark$ & & & $\checkmark$ & $\checkmark$ & & $\checkmark$ & & $\checkmark$ & \\
\cite{lat_07} & $\checkmark$ & & $\checkmark$ & & & $\checkmark$ & $\checkmark$ & & $\checkmark$ & & & $\checkmark$ \\
\cite{zhou2024predictable, taghia2024congruent} & & $\checkmark$ & $\checkmark$ & & & $\checkmark$ & & $\checkmark$ & & & $\checkmark$ & \\
\cite{moreira2020qos} & & $\checkmark$ & $\checkmark$ & & $\checkmark$ & & & & $\checkmark$ & $\checkmark$ & $\checkmark$ & \\
\cite{she2020deep} & & $\checkmark$ & $\checkmark$ & & & $\checkmark$ & & & $\checkmark$ & & $\checkmark$ & \\
%\midrule
\cite{lat_01, lat_05, lat_08, lat_14, lat_24} & $\checkmark$ & & & $\checkmark$ & $\checkmark$ & & & $\checkmark$ & & $\checkmark$ & & \\
\cite{flinta2020predicting} & $\checkmark$ & & & $\checkmark$ & & $\checkmark$ & $\checkmark$ & $\checkmark$ & & $\checkmark$ & & \\
\cite{lat_06} & $\checkmark$ & & & $\checkmark$ & $\checkmark$ & & $\checkmark$ & $\checkmark$ & & & $\checkmark$ & \\
\cite{lat_12} & $\checkmark$ & & & $\checkmark$ & $\checkmark$ & & & $\checkmark$ & & & $\checkmark$ & \\
\cite{lat_11} & $\checkmark$ & & & $\checkmark$ & & $\checkmark$ & & & $\checkmark$ & & $\checkmark$ & \\
\cite{lat_04} & & $\checkmark$ & & $\checkmark$ & & $\checkmark$ & & $\checkmark$ & & $\checkmark$ & & \\
\cite{lat_21} & & $\checkmark$ & & $\checkmark$ & $\checkmark$ & & & $\checkmark$ & & & $\checkmark$ & \\
\cite{lat_18, lat_19, lat_20, lat_22, lat_23} & & $\checkmark$ & & $\checkmark$ & $\checkmark$ & & & $\checkmark$ & & & & $\checkmark$ \\
\cite{guemes2025bridging} & & $\checkmark$ & & $\checkmark$ & $\checkmark$ & & & $\checkmark$ & & & & $\checkmark$ \\
\cite{lat_27} & & $\checkmark$ & & $\checkmark$ & $\checkmark$ & & & & $\checkmark$ & & & $\checkmark$ \\
\bottomrule
\multicolumn{13}{p{17.5cm}}{\textit{Note:} 'Math/ML' encompasses statistical, heuristic, and classical machine learning (e.g., Random Forests). 'Deep NN' includes RNNs, CNNs, Transformers, and Reinforcement Learning architectures. 'IP/Other' captures MANET, WSN, SDN, and traditional IP topologies.}
\end{tabular}
\end{table*}

Analytical latency models remain useful for interpretability and, in some settings, conservative reasoning about latency under uncertainty. Queueing- and calculus-inspired formulations can represent queue build-up and derive probabilistic end-to-end latency bounds under modeled service processes and channel variability \cite{alzubaidy_2016_ton, champati_transient_2018, e2e_v2x}. However, their tractability often depends on assumptions such as stationarity, simplified channel abstractions, and limited dependence across components, assumptions that are difficult to guarantee in dynamic and heterogeneous cellular networks with bursty traffic and cross-layer feedback. This has accelerated the shift toward data-driven latency predictors that learn latency behavior from \gls{KPI}/telemetry streams and context, capturing nonlinear dependencies and temporal effects without relying on restrictive modeling assumptions. 

A major practical driver for these data-driven models is observability. Some quantities are difficult to measure directly at scale, most notably one-way latency, which can require tight clock synchronization, dedicated instrumentation, or intrusive probing. Learning-based approaches can approximate such hard-to-observe quantities from correlates available in the network data plane and control/management telemetry, including generalizable one-way delay prediction via domain adaptation \cite{rao2020generalizable}. As a pragmatic baseline for deployment, studies have focused on predictors that use readily available network indicators to forecast near-future latency or detect impending degradation events. Hybrid approaches for latency prediction in \gls{V2X} settings combine time-series trend information with non-linear dependence on network features \cite{moreira2020qos, lat_01, lat_04, lat_14}. Tree-based ensembles have been used to predict whether a user's latency will become ``high'' in the next interval, framing the task as regime classification that supports alarms and policy triggers \cite{ahmed2021predicting, lat_03, lat_17}. Beyond point forecasts, distribution-oriented formulations estimate the \gls{RTT} distribution by classifying features into histogram bins and mapping a feature vector describing network conditions to a probability mass over latency ranges \cite{flinta2020predicting, lat_05, lat_08, lat_24}. These approaches are typically efficient and robust on heterogeneous data, but they can under-represent sequential dependence and tail behavior unless temporal and probabilistic structure is modeled explicitly \cite{khatouni2019machine, lat_21}.

As latency requirements tighten, especially for real-time and \gls{URLLC}-like services, two modeling needs become central: temporal dependence and tail latency. Deep learning models address both by learning from sequences rather than \gls{iid} snapshots and also by producing distributional outputs rather than only means. Distributional neural predictors have been trained to output mixture-based representations of latency, enabling conditional latency distributions that vary with network state \cite{fadhil2023estimation, lat_06}. These were extended with heavy-tail-aware components to better capture rare but operationally critical extremes in the latency distribution \cite{mostafavi2023data, lat_13}.

For time dependence, sequence models such as \gls{LSTM}/\gls{GRU} are widely used to forecast short-horizon latency by exploiting autocorrelation, periodicity, and mobility-driven regime changes. Bidirectional \glspl{LSTM} have been applied to user-level latency prediction using sequences of past latency and load/context features \cite{dang2023time, lat_02, lat_11, lat_16}. \gls{LSTM}-based predictors have been also used in 5G \gls{CAM} scenarios to anticipate short-term latency variation for vehicular users \cite{barmpounakis2021lstm, lat_10, lat_26}. Transformer-style and attention-augmented models further improve sequence modeling by focusing on salient time points such as recent spikes or handover-adjacent samples, supporting probabilistic latency forecasting in 5G settings \cite{mostafavi2025probabilistic, lat_09, lat_12, lat_22, lat_23}. \gls{CNN}-based models have also been proposed for \gls{URLLC} latency prediction, including contrastive learning objectives that encourage separable representations for different latency regimes \cite{cai2024adaptive}. At the measurement interface, baseband-feature-driven classification of \gls{RTT} regimes has been explored and extended toward one-way latency prediction with domain adaptation~\cite{rao2022prediction, rao2020generalizable}.

Latency is not only a per-link phenomenon. It is often shaped by shared infrastructure and routing, where multiple services interact through common links, queues, and processing stages. Topology-aware learning, therefore, extends latency prediction into network-scale contexts. \glspl{GNN} and graph transformers have been explored for network-level latency modeling and digital-twin inference, capturing spatial coupling among nodes/links/paths that flat feature vectors can miss~\cite{lat_07, lat_18, lat_19, lat_20, lat_27}. While reinforcement learning is primarily used for control rather than pure prediction, latency-aware agents often learn an internal model of latency dynamics as part of decision-making. For example, a Lyapunov-guided \gls{PPO} agent has been applied to schedule resources for hybrid time-sensitive traffic, effectively embedding latency-dynamics learning to maintain latency constraints~\cite{zhou2024predictable}.

A recurring challenge for latency predictors is generalizability across cells, vendors, traffic mixes, and evolving configurations. To address distribution shift, domain adaptation methods have been proposed for out-of-distribution latency prediction~\cite{rao2020generalizable, larsson2023domain}. Transfer learning is a practical approach for reusing representations across scenarios~\cite{she2020deep, ghous2024deep}, including simulation-to-real transfer with fine-tuning on limited operational data~\cite{guemes2025bridging}. Continual learning has been studied to cope with evolving 5G environments without catastrophic forgetting~\cite{lan2025multi}. Additionally, federated learning has been explored to handle multi-agent settings and non-IID data expected in 6G-style deployments~\cite{taghia2024congruent}. Despite these efforts, robust cross-domain performance and dependable tail-risk estimation remain understudied relative to their operational importance.

Ablation studies and controlled comparisons show that across sequence-forecasting works, removing temporal context (e.g., replacing \gls{LSTM}/\gls{GRU} models with a static regressor or using only the most recent sample) typically degrades short-horizon accuracy and, more importantly, weakens the ability to anticipate bursty spikes and regime shifts~\cite{dang2023time, lat_02, lat_11, lat_16, barmpounakis2021lstm, lat_10, lat_26}. Similarly, attention mechanisms are commonly validated by comparing against plain recurrent baselines; emphasizing salient time points (e.g., recent spikes or handover-adjacent samples) is generally reported to improve probabilistic latency forecasting relative to non-attentive variants~\cite{mostafavi2025probabilistic, lat_09, lat_12, lat_22, lat_23}. For tail-aware prediction reliability, distributional heads are often assessed by ablating mixture/distribution outputs to point estimates, with tail-focused extensions explicitly targeting improved modeling of heavy-tailed latency behavior~\cite{fadhil2023estimation, lat_06, mostafavi2023data, lat_13}. At network scale, topology-aware models are typically compared against ``flattened'' tabular representations; encoding nodes/links/paths in \gls{GNN}/graph-transformer predictors is motivated and supported by gains when spatial coupling is strong (shared links, routing effects), which flat models may miss~\cite{lat_07, lat_18, lat_19, lat_20, lat_27}. Finally, deployment-focused ablations around robustness, training on one domain and testing on shifted domains, with/without adaptation, highlight the practical value of domain adaptation, transfer, and simulation-to-real fine-tuning for sustaining accuracy under configuration and traffic drift~\cite{rao2020generalizable, larsson2023domain, she2020deep, ghous2024deep, guemes2025bridging, lan2025multi, taghia2024congruent}.

\subsection{Coverage}
Coverage in cellular networks is fundamentally characterized by a constellation of interrelated \glspl{KPI}, each reflecting different aspects of signal strength and quality. \gls{PL} quantifies the reduction in signal power as it propagates, forming the basis for received signal metrics. \gls{RSRP} measures the average power of reference signals, while \gls{RSSI} captures overall received wideband power, including interference and noise. \gls{RSRQ} expresses the ratio of \gls{RSRP} to \gls{RSSI}, thereby linking absolute signal strength to interference conditions. Quality metrics like \gls{SNR} and \gls{SINR} compare received signal power to noise and interference, respectively, and \gls{CQI} distills \gls{SINR}/\gls{SNR} into discrete indices that inform link adaptation and scheduling. \textcolor{black}{More recent coverage-focused studies also treat ``coverage'' as a spatial field to be reconstructed, predicting not only \gls{RSRP}/\gls{PL} but also derived quantities such as coverage probability (e.g., \gls{SINR} above a threshold) and \glspl{REM} grids along roads or over city tiles.} Accurate and consistent prediction of these \glspl{KPI} is essential for effective coverage planning, optimization, and mobility management in modern wireless networks.

The evolution of machine learning for coverage-related \gls{KPI} prediction stemmed from the need to overcome the inherent trade-offs of statistical and deterministic \gls{PL} models. Statistical models are computationally efficient but environment-specific, whereas deterministic physical models (such as ray tracing) offer strong generalization but are computationally demanding and require high-resolution environmental data. \glspl{ANN} were recognized early on as a powerful compromise, leveraging adaptability and self-learning to model complex propagation environments without full 3D deterministic modeling~\cite{pl_mlp_03, pl_nonmlai, pl_survey}. These basic \gls{ANN}-based approaches quickly advanced from simple feedforward architectures to specialized models that integrated physical priors (e.g., diffraction calculations) or principal component analysis to enhance accuracy in challenging \gls{NLOS} and non-cellular (e.g., \gls{VHF}/\gls{UHF}) environments~\cite{pl_mlp_03_28, pl_mlp_03_32, pl_mlp_03_37, pl_mlp_03_37_33}.

As real-world deployments grew in density and complexity, attention shifted to interpretable, scalable, and efficient \gls{ML} methods. Ensemble approaches, particularly gradient-boosted trees (\gls{LightGBM}, \gls{XGBoost}) and \glspl{RF}, became prevalent due to their ability to ingest diverse radio, environmental, and context features while delivering scalable accuracy across large-scale, multi-band scenarios~\cite{rsrp_ann_01, rss_ml_01, rsrp_rsrq_ml_01, rss_ml_02, rsrp_ml_01}. Hybrid frameworks further improved prediction by coupling broad physical features with dimensionality reduction~\cite{rsrp_rsrq_ml_01} or with deep feature extractors such as \gls{VGGNet}~\cite{rsrp_rsrq_ml_01_16}. The availability of \emph{crowdsourced} data and open multi-operator measurement campaigns has since fortified these models, establishing strong baselines for reproducible \gls{REM} construction \cite{rsrp_dnn_01_13}.

\begin{table*}[h!]
\centering
\caption{Matrix Taxonomy of Coverage-Related KPI Prediction Approaches.}
\label{tab:cov_matrix}
\scriptsize
\renewcommand{\arraystretch}{1.1}
\setlength{\tabcolsep}{2pt}
\begin{tabular}{@{} l | c c | c c | c c c | c c | c c c @{}}
\toprule
\multirow{2}{*}{\textbf{Reference}} & \multicolumn{2}{c|}{\textbf{Data Source}} & \multicolumn{2}{c|}{\textbf{Network Type}} & \multicolumn{3}{c|}{\textbf{Target KPI}} & \multicolumn{2}{c|}{\textbf{Horizon}} & \multicolumn{3}{c}{\textbf{Model Family}} \\
\cmidrule(lr){2-3} \cmidrule(lr){4-5} \cmidrule(lr){6-8} \cmidrule(lr){9-10} \cmidrule(lr){11-13}
 & \textbf{Real Trace} & \textbf{Sim/Hybrid} & \textbf{Sub-6/LTE/5G} & \textbf{mmW/Other} & \textbf{Pwr/PL} & \textbf{Qual/SNR} & \textbf{REM/Map} & \textbf{Spatial/Stat.} & \textbf{Temp/Seq.} & \textbf{Class. ML} & \textbf{Deep Spat.} & \textbf{Deep Seq.} \\
\midrule
\cite{rsrp_ann_01, rsrp_ml_01, rsrp_dnn_01_13, rss_ml_01} & $\checkmark$ & & $\checkmark$ & & $\checkmark$ & & & $\checkmark$ & & $\checkmark$ & & \\
\cite{rsrp_dnn_01_22, rsrp_dnn_01_21, rsrp_dnn_01} & & $\checkmark$ & $\checkmark$ & & $\checkmark$ & & $\checkmark$ & $\checkmark$ & & & $\checkmark$ & \\
\cite{snr_dl_radiounet_2021, cqi_gnn_01, rsrp_dnn_01_18} & & $\checkmark$ & $\checkmark$ & & & & $\checkmark$ & $\checkmark$ & & & $\checkmark$ & \\
\cite{pl_mlp_03, pl_mlp_03_28, pl_mlp_03_37, pl_mlp_03_37_33} & $\checkmark$ & & & $\checkmark$ & $\checkmark$ & & & $\checkmark$ & & $\checkmark$ & & \\
\cite{cqi_2022, cqi_2022_27, cqi_2022_26, cqi_2022_22, cqi_2022_21} & & $\checkmark$ & $\checkmark$ & & & $\checkmark$ & & & $\checkmark$ & & & $\checkmark$ \\
\cite{rssi_urllc_2025_03, rssi_urllc_2025_02, cqi_2022_23} & & $\checkmark$ & $\checkmark$ & & & $\checkmark$ & & & $\checkmark$ & & & $\checkmark$ \\
\cite{snr_dl_01_ref11, snr_cnn_01, snr_dl_01} & & $\checkmark$ & $\checkmark$ & & & $\checkmark$ & & $\checkmark$ & & & $\checkmark$ & \\
\cite{snr_dl_02, sinr_drnn_01, rssi_urllc_2025} & $\checkmark$ & & $\checkmark$ & & $\checkmark$ & $\checkmark$ & & & $\checkmark$ & & & $\checkmark$ \\
\cite{snr_ml_01, snr_ml_urllc_01} & $\checkmark$ & $\checkmark$ & $\checkmark$ & & & $\checkmark$ & & & $\checkmark$ & $\checkmark$ & & \\
\cite{pl_ml_urbancanyon_28ghz_2022} & $\checkmark$ & & & $\checkmark$ & $\checkmark$ & & $\checkmark$ & $\checkmark$ & & $\checkmark$ & $\checkmark$ & \\
\cite{rsrp_dl_01, rsrp_rsrq_ml_01_16} & $\checkmark$ & & $\checkmark$ & & $\checkmark$ & & $\checkmark$ & $\checkmark$ & & & $\checkmark$ & \\
\cite{snr_dl_v2x_rem_lstm_2024} & & $\checkmark$ & $\checkmark$ & & & & $\checkmark$ & $\checkmark$ & $\checkmark$ & & & $\checkmark$ \\
\cite{cqi_urllc_01} & & $\checkmark$ & $\checkmark$ & & & $\checkmark$ & & $\checkmark$ & & & & $\checkmark$ \\
\cite{pl_rf_v2v_2019} & $\checkmark$ & & & $\checkmark$ & $\checkmark$ & & & $\checkmark$ & & $\checkmark$ & & \\
\cite{rss_ml_02} & & $\checkmark$ & & $\checkmark$ & $\checkmark$ & & & $\checkmark$ & & $\checkmark$ & & \\
\cite{ho_01} & & $\checkmark$ & $\checkmark$ & & & $\checkmark$ & & & $\checkmark$ & $\checkmark$ & & \\
\bottomrule
\multicolumn{13}{p{17.5cm}}{\textit{Note:} 'Pwr/PL' covers RSRP, RSSI, and Path Loss. 'Qual/SNR' covers SNR, SINR, and CQI. 'Deep Spat.' includes CNNs, semantic segmentation, and GNNs. 'Deep Seq.' includes LSTMs, GRUs, and Transformers. 'mmW/Other' captures mmWave, VHF/UHF, V2V, and non-standard bands.}
\end{tabular}
\end{table*}

A major leap in coverage modeling came with multi-modal deep learning, which bypassed hand-crafted features by operating directly on high-dimensional data such as satellite imagery, 3D building maps, and \gls{LiDAR} scans. \glspl{CNN} and semantic segmentation networks (e.g., UNet) extract rich structural representations, embedding environmental semantics directly into radio map prediction. Frameworks like \emph{DeepChannel} and \emph{SS-DeepChannel} fuse imagery context with numeric base-station metadata, achieving massive speedups over traditional ray-tracing \cite{rsrp_dnn_01, rsrp_dnn_01_21, rsrp_dnn_01_22}. Complementary spatial models operate as ``map-to-radio'' surrogates, generating dense \glspl{REM} or coverage probabilities directly from city layouts \cite{snr_dl_radiounet_2021, cqi_gnn_01, pl_ml_urbancanyon_28ghz_2022}. However, true generalization to unmeasured areas remains a hurdle. Therefore, recent Bayesian frameworks explicitly decompose predictive uncertainty into aleatoric and epistemic components. This risk-aware shift enables planners to handle out-of-distribution deployments more reliably by identifying regions that require further drive testing or crowdsourcing \cite{rsrp_dl_01, all_01}.

While static spatial modeling dominates long-term planning, dynamic operational tasks such as link adaptation and mobility management require temporal sequence modeling. Recurrent networks (\gls{LSTM}, \gls{GRU}) forecast short-term trajectories of channel states, \gls{CQI} and \gls{SNR}, thereby mitigating feedback aging and lowering block error rates in latency-sensitive applications like \gls{URLLC} and \gls{UAV} communications \cite{cqi_2022, snr_dl_02, sinr_drnn_01, cqi_2022_26, rssi_urllc_2025_03}. Moreover, these sequence models can perform spatial sequence completion, reconstructing 1D coverage trajectories along roads from sparse samples \cite{snr_dl_v2x_rem_lstm_2024}. Reliability in these highly dynamic regimes has been further reinforced using tail-focused learning objectives, such as \gls{EVT} losses and quantile regression, which penalize rare, critical power dips more heavily than standard mean-squared error \cite{cqi_urllc_01, rssi_urllc_2025}. 

Ablation studies validate that moving from coordinates alone to imagery-derived semantics drastically improves out-of-area performance \cite{rsrp_dnn_01}. Similarly, replacing standard \gls{MSE} with tail-sensitive quantile losses prevents catastrophic failure in \gls{URLLC} link adaptation \cite{rssi_urllc_2025, cqi_urllc_01}, while sequence-based deep encoders vastly outperform classical estimators in handling variable modulations and Doppler profiles \cite{snr_dl_01_ref11, snr_cnn_01}. These evaluations underscore a paradigm shift. Coverage prediction has transitioned from point-wise measurement regression to multi-modal, uncertainty-aware field estimation, bridging the gap between physical radio environments and proactive network orchestration.

\subsection{Reliability}
Reliability-related \glspl{KPI} in cellular networks are closely linked across all protocol layers. Metrics such as \gls{BER}, \gls{BLER}, \gls{PER}, and end-to-end packet loss each capture different aspects of reliability, from symbol errors to aggregated packet drops, and are influenced by factors like \gls{SINR}, \gls{MCS} selection, interference, and buffer dynamics. In practice, \gls{BLER} guides adaptive modulation and retransmissions, while packet loss signals service-level issues; accurate prediction of both is vital for link adaptation, congestion control, and \gls{QoE}, especially in latency-sensitive applications~\cite{rel_01, rel_02, rel_04, rel_12}.

Traditional reliability modeling relies on analytical \gls{BER} formulas, link abstraction, and queueing theory, but these approaches falter in the face of real-world complexities such as mobility and interference. This has led to learning-based predictors that combine physics priors with data-driven pattern recognition, maintaining accuracy as conditions shift~\cite{rel_11, rel_02, rel_01}. Early learning-based work targeted scenarios with clear reliability drivers, such as persistent-state learning for \gls{BER} in railway systems~\cite{rel_ber_01} and regression models for weather-induced outages in high-frequency links~\cite{rel_ber_02}. More recent studies integrate deep transfer learning with finite-blocklength models for \gls{URLLC}, achieving near-analytical \gls{BLER} prediction at lower complexity~\cite{rel_09}. In commercial 5G, analytics frameworks correlate radio counters and traffic features with reliability KPIs, revealing operational drivers of errors~\cite{rel_05, rel_09}.

As networks became denser and more heterogeneous, scalable tabular learners emerged as the standard for packet-loss and \gls{PER} prediction, balancing accuracy and efficiency. Supervised models using transport-level metrics forecast near-future loss for real-time control, with ensembles and recurrent models outperforming static baselines on bursty traces~\cite{basikolo2023towards, rel_per_07, rel_per_05, rel_per_01}. Measurement studies confirmed the bursty, temporally dependent nature of packet loss, motivating sequence-aware models~\cite{rel_per_02, rel_per_03}. In programmable networks, in-band monitoring with \gls{ML} enables rapid loss detection and localization~\cite{rel_per_04}, while surveys highlight best practices for large-scale reliability modeling~\cite{rel_06}.

Beyond tabular models, multimodal and topology-aware learning, such as \glspl{GNN}, jointly predict latency and packet loss, capturing spatial dependencies that flat models miss~\cite {rel_gnn_01}. In cellular \glspl{RAN}, optimization that includes packet drop rates as objectives improves reliability by co-optimizing scheduling and resource allocation~\cite{rel_per_06, rel_10}. At the service edge, application-centric \gls{CNN} models in \gls{O-RAN} map network indicators to perceived reliability, enabling SLA-aware control for cloud services~\cite{rel_cnn_01}. These advances align with AI-native RAN life-cycle processes that keep reliability estimators calibrated as networks evolve~\cite{rel_03}.

With the advent of network slicing, reliability prediction expanded to end-to-end service levels. Operators now forecast SLA violations for slices and use these predictions to maintain reliability targets. Deep reinforcement learning has been applied to proactively adjust slice resources, as seen in 5G-LEO satellite networks~\cite{rel_rl_01}. Multi-task sequence and graph-based models anticipate slice-level KPI drifts, though real-world multi-slice failure data remains limited~\cite{rel_01}.

Temporal dynamics are central to reliability prediction. Sequence models like \gls{LSTM}, \gls{GRU}, and temporal \glspl{CNN} improve short-term forecasts of \gls{BLER}/\gls{PER} and packet loss by leveraging autocorrelation and regime shifts, outperforming static models in real-time communications~\cite{rel_per_07, rel_per_02, rel_per_05}. At the system level, AI-powered estimators jointly predict latency and reliability, supporting SLA- and \gls{URLLC}-aware control~\cite{rel_08}.
At the physical layer, advanced channel estimation and impairment correction, using cascaded \glspl{CNN}, attention-guided multitask networks, and universal-approximation frameworks, stabilize \gls{BER}/\gls{BLER} under mobility and hardware offsets~\cite{rel_dl_03, rel_dl_05, rel_dl_02, rel_dl_01, rel_07, rel_dl_04}. Learning-based \gls{CSI} feedback and prediction further reduce \gls{BLER}, especially at high mobility~\cite{rel_dl_07, rel_dl_08, rel_dl_10, rel_dl_09, rel_dl_11}. End-to-end learned \gls{PHY} abstractions accelerate multi-cell optimization and reliability-aware planning~\cite{rel_dl_06}.

To address extreme \gls{PER} or packet-loss events, prediction shifts from average loss to the risk of imminent bursts. Sequence-aware models analyze short sequences of transport metrics to forecast the likelihood of entering high-loss regimes~\cite{rel_per_02, rel_per_03, basikolo2023towards}. These models, including \gls{LSTM} and \gls{GRU}, excel at detecting regime shifts and outperform static baselines on bursty data~\cite{rel_per_07, rel_per_05, rel_per_01}. Topology-aware predictors like \glspl{GNN} capture spatial dependencies, ensuring correlated loss events are not treated as independent~\cite{rel_gnn_01}. Rapid monitoring and tail-sensitive objectives in control loops further enhance robustness and guide decision-making by risk, not just averages~\cite{rel_per_04, rel_per_06, rel_10, rel_08}.

Comparative studies show that sequence models consistently improve burst detection and short-term forecasts over static regressors~\cite{rel_per_07, rel_per_02, rel_per_05, rel_per_01, rel_per_03}, while \glspl{GNN} add value at network scale by modeling spatial coupling~\cite{rel_gnn_01}. At the PHY, model-aided deep estimators and multitask networks outperform classical baselines, and learned \gls{CSI} feedback reduces \gls{BLER} under challenging conditions~\cite{rel_dl_03, rel_dl_05, rel_dl_01, rel_dl_02, rel_07, rel_dl_07, rel_dl_08, rel_dl_10, rel_dl_09, rel_dl_11}. End-to-end learned abstractions accelerate optimization and adapt better to non-ideal channels~\cite{rel_dl_06}. RL-based schedulers and fairness-aware learning improve reliability and reduce edge drops~\cite{rel_10, rel_per_06}, while life-cycle-managed models and in-band monitors support robust field operation~\cite{rel_03, rel_09, rel_per_04}. Remaining challenges include cross-domain generalization, calibrated uncertainty, and multi-objective optimization, now being tackled by joint latency-reliability estimators and AI-native RAN processes~\cite{rel_08, rel_03}.

The latest research tightly integrates reliability prediction with network control and resource allocation, often via reinforcement learning or closed-loop optimization~\cite{rel_rl_01, rel_10, rel_per_06}. Here, prediction is coupled with decision-making for slicing, scheduling, and proactive monitoring, as seen in deep RL-based slice mapping for 5G-\gls{LEO} networks~\cite{rel_rl_01}, sequence models for reliability forecasting~\cite{rel_per_07, rel_per_05, rel_per_02}, and SLA-aware control for \gls{URLLC}\cite{rel_08}. These approaches, leveraging operational data and advanced architectures~\cite{rel_05, rel_09, rel_per_07, rel_gnn_01, rel_01}, represent the forefront of the field, pointing toward autonomous, self-optimizing networks enabled by AI-native RAN life-cycle management~\cite{rel_03}.

\begin{table*}[h!]
\centering
\caption{Matrix Taxonomy of Reliability-Related KPI Prediction Approaches.}
\label{tab:rel_matrix}
\scriptsize
\renewcommand{\arraystretch}{0.9}
\setlength{\tabcolsep}{1pt}
\begin{tabular}{@{} l | c c | c c | c c c | c c | c c c @{}}
\toprule
\multirow{2}{*}{\textbf{Reference}} & \multicolumn{2}{c|}{\textbf{Data Source}} & \multicolumn{2}{c|}{\textbf{Network Type}} & \multicolumn{3}{c|}{\textbf{Target KPI}} & \multicolumn{2}{c|}{\textbf{Horizon}} & \multicolumn{3}{c}{\textbf{Model Family}} \\
\cmidrule(lr){2-3} \cmidrule(lr){4-5} \cmidrule(lr){6-8} \cmidrule(lr){9-10} \cmidrule(lr){11-13}
 & \textbf{Real Trace} & \textbf{Sim./Synth.} & \textbf{Cellular/5G} & \textbf{IP/SDN/TCP} & \textbf{PHY/CE} & \textbf{Net/Loss} & \textbf{Slice/QoE} & \textbf{Static/Point} & \textbf{Time-Seq.} & \textbf{Class. ML} & \textbf{Deep NN} & \textbf{RL} \\
\midrule
\cite{rel_dl_01, rel_dl_02, rel_dl_03, rel_dl_04, rel_dl_05, rel_dl_07, rel_dl_08, rel_dl_10, rel_dl_11, rel_07} & & $\checkmark$ & $\checkmark$ & & $\checkmark$ & & & $\checkmark$ & & & $\checkmark$ & \\
\cite{rel_per_01, rel_per_05} & $\checkmark$ & & & $\checkmark$ & & $\checkmark$ & & $\checkmark$ & & $\checkmark$ & & \\
\cite{rel_per_02} & $\checkmark$ & & & $\checkmark$ & & $\checkmark$ & & & $\checkmark$ & $\checkmark$ & $\checkmark$ & \\
\cite{rel_10} & & $\checkmark$ & $\checkmark$ & & & $\checkmark$ & & & $\checkmark$ & & & $\checkmark$ \\
\cite{rel_rl_01, rel_per_06} & & $\checkmark$ & $\checkmark$ & & & & $\checkmark$ & & $\checkmark$ & $\checkmark$ & & $\checkmark$ \\
\cite{rel_11, rel_08} & $\checkmark$ & $\checkmark$ & $\checkmark$ & & & & $\checkmark$ & $\checkmark$ & & $\checkmark$ & $\checkmark$ & \\
\cite{rel_ber_01} & $\checkmark$ & & $\checkmark$ & & $\checkmark$ & & & & $\checkmark$ & & $\checkmark$ & \\
\cite{rel_gnn_01} & & $\checkmark$ & & $\checkmark$ & & $\checkmark$ & & $\checkmark$ & & & $\checkmark$ & \\
\cite{rel_09} & & $\checkmark$ & $\checkmark$ & & $\checkmark$ & & $\checkmark$ & & $\checkmark$ & $\checkmark$ & $\checkmark$ & \\
\cite{rel_cnn_01} & $\checkmark$ & & $\checkmark$ & & & & $\checkmark$ & $\checkmark$ & & & $\checkmark$ & \\
\cite{rel_ber_02} & & $\checkmark$ & $\checkmark$ & & $\checkmark$ & & & $\checkmark$ & & $\checkmark$ & & \\
\cite{rel_per_04, rel_per_07} & & $\checkmark$ & & $\checkmark$ & & $\checkmark$ & & & $\checkmark$ & $\checkmark$ & & \\
\cite{rel_dl_09} & & $\checkmark$ & $\checkmark$ & & $\checkmark$ & & & & $\checkmark$ & & $\checkmark$ & \\
\bottomrule
\multicolumn{13}{p{17.5cm}}{\textit{Note:} 'PHY/CE' includes Channel Estimation, CSI feedback, BER, and attenuation. 'Net/Loss' denotes frame/packet loss rate (PER), transport retransmissions, and MAC-level drops. 'Slice/QoE' includes Slice availability, gaming/video reliability mapping, and NF anomalies. 'Static/Point' covers snapshot predictions and classification, whereas 'Time-Seq.' implies dynamic multi-step forecasting or episodic RL states.}
\end{tabular}
\end{table*}

\section{System Aspects of Enabling Data-driven KPI Predictions in Mobile Networks}
\label{sec:enablers}
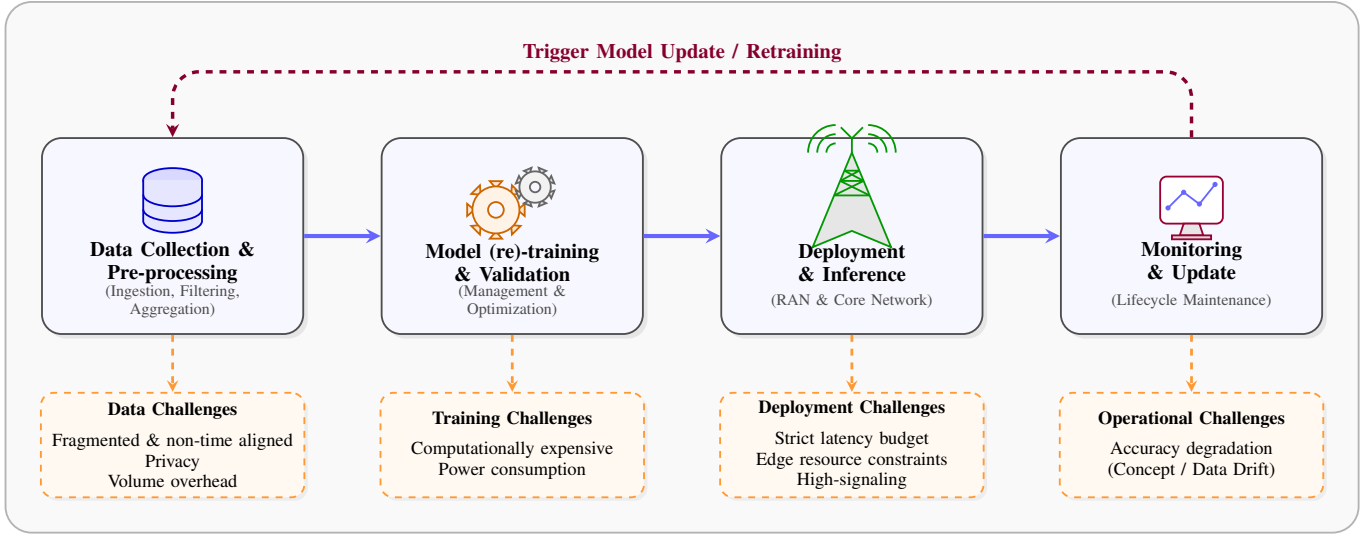
\begin{figure*}[htbp]
\centering
\resizebox{\textwidth}{!}{
\begin{tikzpicture}[
    >=stealth,
    % Balanced minimum height: 3.0cm (original was 3.6, over-shrunk was 2.6)
    stagebox/.style={rectangle, draw=black!70, fill=blue!3, thick, rounded corners=3mm, 
                     minimum width=4cm, minimum height=3.0cm,
                     drop shadow={shadow xshift=2pt, shadow yshift=-2pt, opacity=0.15}},
    % Balanced challenge box: 1.6cm height
    chalbox/.style={rectangle, draw=orange!80, thick, dashed, fill=orange!5, 
                    text width=3.8cm, align=center, font=\footnotesize, 
                    rounded corners=2mm, minimum height=1.6cm,
                    drop shadow={shadow xshift=1.5pt, shadow yshift=-1.5pt, opacity=0.1}},
    % --- Base Station Antenna Styles ---
    antenna/.style={insert path={-- coordinate (ant#1) ++(0,0.25) -- +(135:0.25) + (0,0) -- +(45:0.25)}},
    station/.style={align=center, font=\small, draw, shape=dart, shape border rotate=90, 
                    minimum width=12mm, minimum height=12mm, outer sep=0pt, inner sep=3pt},
    radiation/.style={{decorate, decoration={expanding waves, angle=90, segment length=4pt}}}
]

% ==========================================
% 1. MAIN PIPELINE NODES
% ==========================================

% Base containers for the stages
\node[stagebox] (box0) at (0, 0) {};
\node[stagebox] (box1) at (5.2, 0) {};
\node[stagebox] (box2) at (10.4, 0) {};
\node[stagebox] (box3) at (15.6, 0) {};

% --- STAGE 1: Data Collection (Coin / Database Shape) ---
\begin{scope}[shift={(box0.center)}]
    \begin{scope}[yshift=0.5cm, scale=0.9] % Scaled graphic down slightly
        \draw[thick, draw=blue!80!black, fill=blue!15] (0, 0.4) ellipse (0.5 and 0.2);
        \draw[thick, draw=blue!80!black, fill=blue!10] (-0.5, 0.4) -- (-0.5, -0.3) arc (180:360:0.5 and 0.2) -- (0.5, 0.4);
        \draw[thick, draw=blue!80!black] (-0.5, 0.15) arc (180:360:0.5 and 0.2);
        \draw[thick, draw=blue!80!black] (-0.5, -0.1) arc (180:360:0.5 and 0.2);
    \end{scope}
    % Text is spaced out appropriately to prevent merging
    \node[align=center, font=\small\bfseries] at (0, -0.4) {Data Collection \&\\Pre-processing};
    \node[align=center, font=\scriptsize, text=black!70] at (0, -1.0) {(Ingestion, Filtering,\\Aggregation)};
\end{scope}

% --- STAGE 2: Model Training (Gears / Cogs Shape) ---
\begin{scope}[shift={(box1.center)}]
    \begin{scope}[shift={(-0.25, 0.4)}, scale=0.3] % Scaled gears to fit better
        \foreach \i in {0,45,...,315} {
            \filldraw[thick, draw=orange!80!black, fill=orange!20] (\i:1.1) -- (\i-12:1.5) -- (\i+12:1.5) -- (\i:1.1) -- cycle;
        }
        \draw[thick, draw=orange!80!black, fill=orange!10, even odd rule] (0,0) circle (1.2) (0,0) circle (0.4);
    \end{scope}
    \begin{scope}[shift={(0.35, 0.75)}, scale=0.2] 
        \foreach \i in {22.5,67.5,...,337.5} {
            \filldraw[thick, draw=gray!80!black, fill=gray!20] (\i:1.1) -- (\i-12:1.5) -- (\i+12:1.5) -- (\i:1.1) -- cycle;
        }
        \draw[thick, draw=gray!80!black, fill=gray!10, even odd rule] (0,0) circle (1.2) (0,0) circle (0.4);
    \end{scope}
    \node[align=center, font=\small\bfseries] at (0, -0.4) {Model (re)-training\\\& Validation};
    \node[align=center, font=\scriptsize, text=black!70] at (0, -1.0) {(Management \&\\Optimization)};
\end{scope}

% --- STAGE 3: Deployment (New Base Station Antenna Shape) ---
\begin{scope}[shift={(box2.center)}]
    \begin{scope}[shift={(0, 0.35)}, scale=0.9] 
        % The Base Station Dart
        \node[station, thick, draw=green!60!black, fill=gray!20] (base) {};
        
        % The structural cross-beams
        \draw[thick, draw=green!60!black, line join=bevel] (base.100) -- (base.80) -- (base.110) -- (base.70) -- (base.north west) -- (base.north east);
        \draw[thick, draw=green!60!black, line join=bevel] (base.100) -- (base.70) (base.110) -- (base.north east);
        
        % The top antenna
        \draw[thick, draw=green!60!black, line cap=rect] ([yshift=-0.2pt]base.north) [antenna=1];
        
        % The expanding radiation waves
        \draw[thick, draw=green!60!black, radiation, decoration={angle=45}] ([xshift=.2768cm]base.north) -- +(45:0.5);
        \draw[thick, draw=green!60!black, radiation, decoration={angle=45}] ([xshift=-.2768cm]base.north) -- +(135:0.5);
    \end{scope}
    \node[align=center, font=\small\bfseries] at (0, -0.4) {Deployment\\\& Inference};
    \node[align=center, font=\scriptsize, text=black!70] at (0, -1.0) {(RAN \& Core Network)};
\end{scope}

% --- STAGE 4: Monitoring (Screen & Chart Shape) ---
\begin{scope}[shift={(box3.center)}]
    \begin{scope}[shift={(0, 0.55)}, scale=0.6] % Shrunk chart to give text room
        \draw[thick, draw=purple!80!black, fill=white, rounded corners=1mm] (-0.8, -0.6) rectangle (0.8, 0.6);
        \draw[thick, draw=purple!80!black, fill=gray!20] (-0.2, -0.6) -- (-0.3, -0.9) -- (0.3, -0.9) -- (0.2, -0.6) -- cycle;
        \draw[thick, draw=purple!80!black, fill=gray!30] (-0.5, -0.9) rectangle (0.5, -1.0);
        \draw[thick, draw=blue!60] (-0.6, -0.2) -- (-0.2, 0.2) -- (0.2, -0.1) -- (0.6, 0.4);
        \filldraw[blue!60] (-0.6, -0.2) circle (1.5pt) (-0.2, 0.2) circle (1.5pt) (0.2, -0.1) circle (1.5pt) (0.6, 0.4) circle (1.5pt);
    \end{scope}
    \node[align=center, font=\small\bfseries] at (0, -0.4) {Monitoring\\\& Update};
    \node[align=center, font=\scriptsize, text=black!70] at (0, -1.0) {(Lifecycle Maintenance)};
\end{scope}

% ==========================================
% 2. WORKFLOW CONNECTIONS
% ==========================================

% Forward flow arrows
\draw[->, thick, draw=blue!60, line width=1.8pt] (box0.east) -- (box1.west);
\draw[->, thick, draw=blue!60, line width=1.8pt] (box1.east) -- (box2.west);
\draw[->, thick, draw=blue!60, line width=1.8pt] (box2.east) -- (box3.west);

% Cross-layer Feedback Loop 
\draw[->, thick, draw=purple!70!black, dashed, line width=1.5pt, rounded corners=4mm] 
     (box3.north) -- +(0, 1.0) -| (box0.north)
     node[pos=0.25, above, font=\small\bfseries, text=purple!70!black] {Trigger Model Update / Retraining};

% ==========================================
% 3. ASSOCIATED CHALLENGES (Dashed Boxes)
% ==========================================

% Perfectly balanced y-shift to prevent overlapping with connecting arrows
\node[chalbox] (chal0) at ([yshift=-3.2cm]box0.center) {\textbf{Data Challenges}\\[1.5mm] Fragmented \& non-time aligned \\ Privacy \\ Volume overhead};
\node[chalbox] (chal1) at ([yshift=-3.2cm]box1.center) {\textbf{Training Challenges}\\[1.5mm] Computationally expensive \\ Power consumption};
\node[chalbox] (chal2) at ([yshift=-3.2cm]box2.center) {\textbf{Deployment Challenges}\\[1.5mm] Strict latency budget \\ Edge resource constraints \\ High-signaling};
\node[chalbox] (chal3) at ([yshift=-3.2cm]box3.center) {\textbf{Operational Challenges}\\[1.5mm] Accuracy degradation \\ (Concept / Data Drift)};

% Arrows pointing to challenges
\draw[->, thick, draw=orange!80, dashed, line width=1.2pt] (box0.south) -- (chal0.north);
\draw[->, thick, draw=orange!80, dashed, line width=1.2pt] (box1.south) -- (chal1.north);
\draw[->, thick, draw=orange!80, dashed, line width=1.2pt] (box2.south) -- (chal2.north);
\draw[->, thick, draw=orange!80, dashed, line width=1.2pt] (box3.south) -- (chal3.north);

% ==========================================
% 4. BACKGROUND GROUPING
% ==========================================
\begin{scope}[on background layer]
    % Surrounding boundary with tightened ysep
    \node[rectangle, draw=black!30, thick, fill=gray!5, rounded corners=4mm, 
          inner xsep=15pt, inner ysep=15pt, fit=(box0) (box3) (chal0) (chal3) (current bounding box.north)] (bg) {};
\end{scope}

\end{tikzpicture}
}
\vspace{-2mm}
\caption{\textcolor{black}{Lifecycle Management of ML models, outlining the key stages from data collection to updates, alongside the challenges encountered at each phase within the RAN and Core network.}}
\label{fig:lcm}
\end{figure*}

Data-driven \gls{KPI} prediction requires tight integration with the underlying management and control architecture of 5G and future 6G systems. In this context, \emph{lifecycle management of machine learning} models, shown in Fig.~\ref{fig:lcm}, refers to the coordinated processes through which ML models are created, trained, validated, deployed, monitored, and updated throughout their operational lifetime as shown in Fig.~\ref{fig:lcm}. Unlike traditional software components, \gls{ML} models are dynamic entities whose performance depends on continuously evolving network conditions, user behavior, traffic patterns, and environmental factors. Consequently, they require systematic workflows for data ingestion, retraining, drift detection, versioning, and deployment to ensure that predictions remain accurate and reliable across domains. To support these requirements, modern mobile networks must provide an end-to-end \gls{ML} lifecycle management framework that supports \gls{ML} tasks across the \gls{RAN}, Core, and management layers. 
To this end, different standardization bodies (e.g., \gls{3GPP} and the \gls{O-RAN} Alliance) have introduced complementary frameworks that define architectural and operational principles for \gls{ML} lifecycle management in mobile systems. In \gls{3GPP}, these capabilities are implemented through the \gls{MnS} paradigm, in which standardized management services orchestrate the end-to-end \gls{AI}/\gls{ML} lifecycle across the \gls{RAN}, core, and management domains. Concretely, the \gls{AI}/\gls{ML} management framework in TS~28.105 specifies domain-agnostic procedures and interfaces for model onboarding, training, deployment, monitoring (including quality and drift indicators), and lifecycle control~\cite{3gpp-ts-28105}. In parallel, the \gls{O-RAN} Alliance specifies a \gls{RAN}-centric realization of \gls{ML} lifecycle management through the \gls{RIC} architecture, with a functional split between the \gls{Non-RT} \gls{RIC}, responsible for model training, policy management, and long-term optimization, and the \gls{Near-RT} \gls{RIC}, which hosts \glspl{xApp} for \gls{Near-RT} inference and control~\cite{oran-architecture}.

In the following, we outline the key lifecycle processes required to enable \gls{ML}-based \gls{KPI} prediction in operational mobile networks, highlighting how these processes are realized within \gls{3GPP} and \gls{O-RAN} frameworks.

\subsection{Data Collection and Preprocessing}

Data-driven KPI prediction relies on rich measurement streams collected across the \gls{RAN}, core, and transport domains. In \gls{3GPP}, this capability is supported by the standardized \gls{PM} framework in the management plane~\cite{3gpp-ts-32401}. TS~32.401 defines the \gls{PM} concept and requirements for administering measurements and collecting measurement results from network elements via management systems (e.g., \gls{OSS}). Moreover, TS~28.552 specifies standardized 5G performance measurements~\cite{3gpp-ts-28552} that can serve as consistent inputs for \gls{KPI} model training. From an interface perspective, \gls{3GPP} SA5 specifies \gls{MnS} for data reporting, enabling controlled configuration of measurement collection and bulk or streaming export of measurement files/streams to analytics applications for downstream preprocessing and \gls{ML} training~\cite{3gpp-ts-28532}.

In parallel, \gls{O-RAN} specifies complementary \gls{RAN}-centric telemetry interfaces. \gls{Near-RT} cell- and \gls{UE}-related measurements are exposed to the \gls{Near-RT} \gls{RIC} via the E2 interface for low-latency ingestion by \glspl{xApp}, while the \gls{Non-RT} \gls{RIC} (within the \gls{SMO}) aggregates longer-term data via O1 and supports offline training and policy/model refinement, with policies and model-related guidance distributed towards the \gls{Near-RT} \gls{RIC} over A1 \cite{oran-architecture}. In operational deployments, these standardized mechanisms are often complemented by vendor- or operator-specific telemetry pipelines (e.g., streaming telemetry, Prometheus exporters, in-house probes) to increase feature richness or reduce integration friction.

\subsection{Model Training, Deployment, and Inference}
Structured workflows for training, validation, deployment, and execution of \gls{ML} models in operational networks are discussed next. In \gls{3GPP}, the management of the \gls{AI}/\gls{ML} lifecycle is standardized through the \gls{MnS} and the \gls{AI}/\gls{ML} management architecture in TS~28.105. Within this framework, model training jobs can be instantiated, monitored, and repeated, with support for dataset selection, objective configuration, and lifecycle control across domains. Validation capabilities expose model-quality indicators (e.g., accuracy) and drift-related metrics via dedicated management \glspl{API}, enabling controlled onboarding and safe operation of \gls{ML} models in production environments. Deployed models and associated analytics capabilities can be realized in \gls{NWDAF}, which may be logically decomposed into an \gls{AnLF} and a \gls{MTLF}~\cite{3gpp-ts-23288}, or in the \gls{MDAF}, as two examples.
In \gls{O-RAN}, the corresponding intelligence pipeline is implemented through the \gls{RIC}/\gls{SMO} ecosystem: model training and higher-level policy generation are performed by \glspl{rApp} in the \gls{Non-RT} \gls{RIC} (within the \gls{SMO}), typically using offline or aggregated datasets, followed by validation before deployment. Validated policies and \gls{ML} models are then distributed to the \gls{Near-RT} \gls{RIC} via the A1 interface, where \glspl{xApp} perform \gls{Near-RT} inference and control based on telemetry streams and measurements obtained through the E2 interface~\cite{oran-architecture}.

\subsection{Monitoring and Model Update}
Continuous monitoring and model updates are important for providing reliable \gls{KPI} prediction in dynamic mobile environments. In the \gls{3GPP} framework, inference outputs can be evaluated in real time, and performance degradation can automatically trigger retraining actions according to the drift mitigation and retraining policies defined in TS~28.105~\cite{3gpp-ts-28105}. These monitoring capabilities also include support for explainability reporting, version management, and accuracy assessment through dedicated \glspl{API}, ensuring that model evolution remains traceable and safe as conditions change. % \textcolor{red}{\cite{3gpp-aiml}}. 
In \gls{O-RAN}, the \gls{Non-RT} \gls{RIC} supervises the performance of deployed \glspl{xApp} and determines when retraining or model refinement is required, based on accuracy degradation or environmental shifts. Updated models and policies can be periodically provisioned to the \gls{Near-RT} \gls{RIC} via the A1 interface, enabling the \gls{RAN} to adapt to changes in traffic patterns, interference conditions, user mobility, and slice-level \gls{QoS} requirements~\cite{oran-architecture}.

\subsection{Challenges}
As mentioned above, accurate KPI prediction relies on the ability to collect and deliver high-quality measurement data across different network domains. Data-driven \gls{ML} predictors require time-aligned, fine-grained, and cross-domain visibility to capture the dynamics that impact end-to-end \glspl{KPI} such as throughput or latency. However, observability in mobile networks is inherently fragmented: relevant measurements originate from heterogeneous elements with different reporting granularities and may not be time-aligned across domains. A key challenge is the need for a sufficiently accurate time synchronization across these data sources located in different domains. Throughput, delay, interference, \gls{HARQ} statistics, or scheduling information may be timestamped differently across the network, and even small misalignments can degrade model performance, particularly for latency prediction.
In this context, \gls{3GPP} has standardized architectural support for time-sensitive communication and time synchronization through the \gls{TSCTSF}, which enables exposure and configuration of time synchronization services~\cite{3gpp-ts-29565}; improving end-to-end time alignment across domains is therefore a prerequisite not only for high-fidelity \gls{ML} training, but also for interoperable operation of standardized time synchronization services.

Telemetry granularity and volume introduce additional constraints. Fine-grained measurements at millisecond or sub-millisecond resolution, required for short-term \gls{KPI} prediction, can generate substantial overhead both at the measurement point and within the network. To address this overhead, raw data can be aggregated into performance-management counters at variable intervals (e.g., 1–15 minutes). Such aggregation, however, may be insufficient for prediction tasks that depend on capturing transient phenomena such as fast fading, queue build-up, etc~\cite{lindstaahl2024rome}. These challenges are further amplified when incorporating UE-level data: \gls{UE}-reported measurements may carry privacy-sensitive information, requiring privacy-preserving processing. Furthermore, frequent data reporting can increase battery consumption and increase signaling overhead~\cite{taghia2022policy}. Existing studies highlight related issues such as detecting network changes under incomplete observability, integrating multi-segment data streams, and designing adaptive pipelines that selectively collect measurements needed for specific \gls{ML} tasks~\cite{lindstaahl2023change}.

Another set of challenges stems from system-level constraints that shape how \gls{ML} models are (re-)trained, deployed, and executed in telecom environments. When time-to-insight is critical, the model should be deployed in an execution environment that minimizes aggregate latency across the data and prediction pipeline, including data collection, model inference, and delivery of model output to the desired location. Often, this means model inference may need to be brought closer to the network edge. Further, the strictness of latency budgets is directly impacted by the prediction horizon. Shorter horizons (e.g., sub-millisecond) leave a diminishing window for the entire pipeline, requiring deployment at \gls{Near-RT} or edge locations~\cite{polese2023understanding,koufos2021trends}, while longer-horizon forecasting can remain centralized, e.g., in the cloud or the core-network~\cite{gkonis2024leveraging}. Having model inference close to the edge network puts more demand on scarce resources.  For example, compute, memory, power, and footprint limitations at or near base stations constrain the size, complexity, and update frequency of models that can be hosted and operated there. Moreover, telemetry handling and inference introduce non-negligible bandwidth and energy costs; for example, KPI subscription traffic has been shown to scale linearly with \gls{RIC} power consumption, and eliminating redundant subscriptions can reduce energy usage by nearly 87\%~\cite{lima2025power}. Similar considerations apply to core-side analytics, where the data-collection entity must filter and downsample measurements to avoid overwhelming the system with large volumes of data. Finally, updating the deployed \gls{ML} models is essential, as network conditions can evolve with load, mobility, interference, and scheduling configurations, leading to a degradation in accuracy. Retraining from scratch is typically computationally expensive, especially at the edge, so operators must balance accuracy and operational cost. Current architectures support both continual learning (e.g., through \gls{NWDAF}'s \gls{MTLF}) and on-demand retraining triggered by accuracy degradation or drift detection~\cite{gudepu2023adaptive}.

\subsection{Practical Guidelines for \gls{KPI} Prediction Design}
\begin{figure*}[htbp]
\centering
\begin{tikzpicture}[
    >=Stealth,
    font=\small,
    % Tightened node distance to compress vertical and horizontal layout
    node distance=0.4cm and 0.4cm,
    % Flattened the diamonds (aspect=2.2) to save significant vertical space
    decision/.style={diamond, aspect=2.2, draw=gray!70, fill=blue!5, thick, text width=2.6cm, align=center, inner sep=0pt, drop shadow={opacity=0.15}},
    % Reduced minimum heights and inner padding
    block/.style={rectangle, draw=gray!70, fill=Green!5, thick, text width=2.5cm, align=center, rounded corners=3pt, minimum height=0.8cm, inner sep=3pt, drop shadow={opacity=0.15}},
    action/.style={rectangle, draw=orange!80, fill=orange!5, thick, dashed, text width=2.8cm, align=center, rounded corners=3pt, minimum height=0.8cm, inner sep=3pt, drop shadow={opacity=0.15}},
    startblock/.style={rectangle, draw=black!80, fill=white, thick, text width=5cm, align=center, rounded corners=5pt, minimum height=0.9cm, drop shadow={opacity=0.15}},
    edge label/.style={fill=gray!10, inner sep=1.5pt, font=\scriptsize\bfseries, text=black!70},
    flowarrow/.style={->, thick, draw=Gray!90, rounded corners=2pt},
    mlarrow/.style={->, thick, draw=orange!80, dashed, rounded corners=2pt}
]

% 1. Start Node
\node [startblock] (start) {\textbf{Define KPI \& Control Loop Task}};

% Distance heavily reduced
\node [decision, below=0.35cm of start] (timescale) {{Timescale \& Deployment?}};

% 2. First Branch (Timescale)
\node [decision, left=0.4cm of timescale] (fast) {{Sub-second} \\[1pt] \scriptsize (Near-RT / UE)};
\node [decision, right=0.4cm of timescale] (slow) {{> 1 Second} \\[1pt] \scriptsize (Non-RT / Cloud)};

\node [decision, below=0.35cm of fast] (sla_fast) {{Strict SLA / URLLC?}};
\node [decision, below=0.35cm of slow] (data_type) {{Primary Data Structure?}};

% 3. Fast / Near-RT branch (Leaf nodes)
% Shrunk the X-shift slightly and moved them vertically closer
\node [action, below=0.5cm of sla_fast, xshift=-1.6cm] (fast_point) {\textbf{Deterministic} \\[1pt] \textit{\scriptsize LightGBM, GRU, RF}};
\node [action, below=0.5cm of sla_fast, xshift=1.6cm] (fast_prob) {\textbf{Probabilistic} \\[1pt] \textit{\scriptsize Conformal Pred., BDL}};

% 4. Slow / NWDAF branch (Intermediate Data nodes)
\node [block, below=0.5cm of data_type] (spatial) {\textbf{Gridded Spatial} \\[1pt] \scriptsize (Coverage Maps)};
\node [block, left=0.45cm of spatial] (temporal) {\textbf{Sequential} \\[1pt] \scriptsize (Load, Traffic)};
\node [block, right=0.45cm of spatial] (topology) {\textbf{Topology-aware} \\[1pt] \scriptsize (IP Paths, Cells)};

% 5. ML actions for Slow branch (Leaf nodes)
\node [action, below=0.35cm of temporal] (transformer) {\textbf{Deep Sequence} \\[1pt] \textit{\scriptsize Transformers, LSTMs}};
\node [action, below=0.35cm of spatial] (cnn) {\textbf{Deep Spatial} \\[1pt] \textit{\scriptsize CNNs, U-Net}};
\node [action, below=0.35cm of topology] (gnn) {\textbf{Graph Models} \\[1pt] \textit{\scriptsize GNNs, Graph-Trans.}};

% --- EDGES AND ROUTING ---
\draw [flowarrow] (start) -- (timescale);

\draw [flowarrow] (timescale) -- node[edge label, above] {$< 1$s} (fast);
\draw [flowarrow] (timescale) -- node[edge label, above] {$\geq 1$s} (slow);

\draw [flowarrow] (fast) -- (sla_fast);
\draw [flowarrow] (slow) -- (data_type);

% Orthogonal drop reduced from -0.4 to -0.2 to save space
\draw [mlarrow] (sla_fast.south) -- ++(0,-0.2) -| node[edge label, pos=0.25, above] {No} (fast_point.north);
\draw [mlarrow] (sla_fast.south) -- ++(0,-0.2) -| node[edge label, pos=0.25, above] {Yes} (fast_prob.north);

% Orthogonal drop reduced from -0.5 to -0.25
\draw [flowarrow] (data_type.south) -- ++(0,-0.25) -| (temporal.north);
\draw [flowarrow] (data_type.south) -- ++(0,-0.25) -| (topology.north);
\draw [flowarrow] (data_type.south) -- (spatial.north);

\draw [mlarrow] (temporal) -- (transformer);
\draw [mlarrow] (spatial) -- (cnn);
\draw [mlarrow] (topology) -- (gnn);

% --- BACKGROUND SHADOW BOX ---
\begin{scope}[on background layer]
    \node[
        fill=gray!10, 
        draw=gray!30, 
        thick, 
        rounded corners=8pt, 
        drop shadow={opacity=0.15, shadow xshift=3pt, shadow yshift=-3pt}, 
        fit=(current bounding box), 
        inner sep=8pt % Tightened padding
    ] {};
\end{scope}

\end{tikzpicture}
\caption{Decision flowchart for selecting the KPI prediction model based on control timescale, deployment architecture, data structure, and SLA needs. The deployment nodes and method choices are illustrated from an O-RAN perspective, though similar hierarchical data analytics functions and features are also present in standard 3GPP architectures.}
\label{fig:model_selection_flowchart}
\end{figure*}
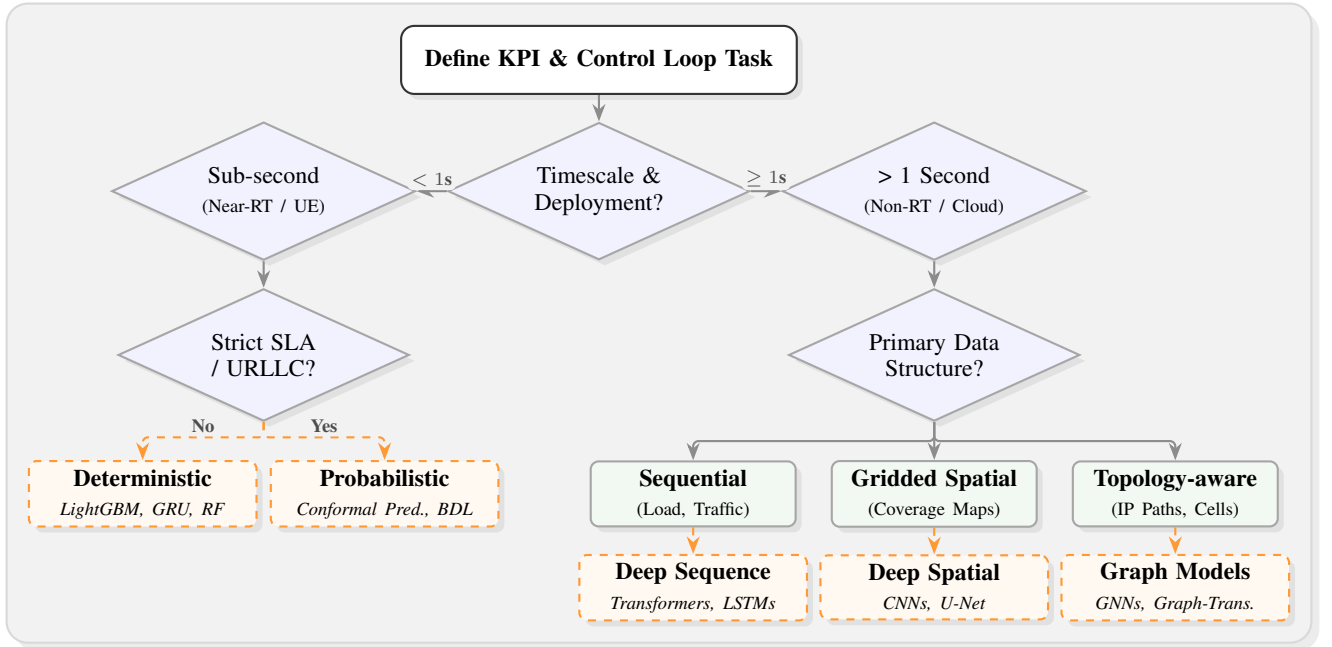

Selecting the optimal predictive model for 6G networks requires navigating a systematic decision framework, as illustrated in Fig. \ref{fig:model_selection_flowchart}, that is primarily driven by deployment timescale, \gls{SLA} criticality, and data morphology. For edge deployments operating in fast, sub-second control loops (e.g., \gls{UE} or Near-RT \gls{RIC}), inference latency is a strict bottleneck, meaning the architectural choice hinges entirely on \gls{SLA} requirements. In these constrained environments, non-critical tasks can rely on computationally efficient, deterministic point forecasting using lightweight tree ensembles (e.g., LightGBM, Random Forest) or shallow \glspl{GRU}, whereas safety-critical \gls{URLLC} applications mandate uncertainty-aware, probabilistic approaches to explicitly manage tail risks. Conversely, in centralized management domains operating on slower timescales exceeding one second (e.g., Non-RT \gls{RIC}), computational resources are abundant, allowing decisions to be guided by the intrinsic structure of the telemetry data. Within this domain, sequential traffic traces with strong temporal dependencies are best addressed by deep sequence models such as Transformers or LSTMs; gridded spatial phenomena, such as coverage maps, are ideally suited to deep spatial models such as \glspl{CNN}; and tasks involving multi-cell coordination or routing paths require \glspl{GNN} to explicitly capture topological dependencies. By following this structured routing, network engineers can ensure that the selected machine learning family perfectly aligns with the operational constraints, data realities, and reliability demands of the target use case.

\section{Limitations, Challenges, and Future Works}
\label{sec:challenges}
In this section, we outline the key themes and open challenges that emerge across the literature: (i) active development across all \gls{KPI} families, (ii) limited reproducibility and weak comparability motivating shared open benchmarks and systematic comparative studies, (iii) the need to validate generalization and robustness claims on open datasets under realistic domain shifts, (iv) the underdeveloped integration of predictions into live systems, (v) unresolved trade‑offs among inference latency, accuracy, and interpretability under real-time constraints, (vi) lifecycle issues such as drift, retraining, and sustainable data collection, (vii) the need for explainability that supports accountability and action, (viii) sustainability concerns related to measurement overhead, computational cost, and long‑term maintainability of \gls{ML} pipelines, and (ix) privacy and security risks when learning from \gls{UE}‑level and user‑related measurements.

Table~\ref{tab:kpi_maturity} shows the maturity of predictive models and open problems for key \gls{KPI} families in future 6G networks. For coverage and capacity, predictive modeling is well-established for short‑horizon forecasts using scalable tabular approaches and increasingly advanced spatio‑temporal deep learning models, though challenges remain in out‑of‑domain generalization (e.g., cross‑cell, cross‑operator, and cross‑frequency transfer) and in deployment issues related to metadata availability, calibration, and drift. In latency and reliability, the field is shifting toward tail‑ and risk‑aware prediction, with difficulties centered on modeling rare events, producing calibrated uncertainty estimates, and maintaining stability when predictions influence schedulers, congestion control, or \gls{SLA}/slice‑assurance mechanisms.

\begin{table*}[ht]
\centering
\caption{Maturity of Predictive Models and Open Problems for Key KPI Families in Future 6G Networks.}
\label{tab:kpi_maturity}
\footnotesize
\setlength{\tabcolsep}{6pt}
\renewcommand{\arraystretch}{1.1}
\begin{tabular}{p{1.2cm} p{7.5cm} p{7cm}}
\hline
\textbf{KPI} & \textbf{Maturity of Models} & \textbf{Open Problems} \\
\hline

\textbf{Coverage} &
\begin{itemize}\setlength\itemsep{0pt}
\vspace{-0.2cm}
\item Pointwise coverage prediction is strong: tree/boosting models with engineered radio + environment features perform well at scale.
\item Spatial deep models are maturing: map- and imagery-driven CNNs enable radio-map-style inference and denser coverage reconstruction.
\item Risk-aware prediction is emerging: uncertainty/quantile objectives increasingly support conservative planning decisions.
\end{itemize}
&
\begin{itemize}\setlength\itemsep{0pt}
\vspace{-0.2cm}
\item Cross-area generalization: robust performance under geo-disjoint, cross-city, and cross-operator tests remains limited.
\item Deployability with limited metadata: many methods assume rich BS/layout inputs that are unavailable in practice.
\item Reliable uncertainty at scale: calibrated UQ and standardized REM benchmarks (grid/trajectory/layout shift) are still lacking.
\end{itemize}
\\
\hline

\textbf{Capacity} &
\begin{itemize}\setlength\itemsep{0pt}
\vspace{-0.2cm}
\item Short-term prediction is mature: tree ensembles and compact RNNs are strong, scalable baselines for near-future throughput.
\item Temporal modeling is widely used: LSTM/GRU-type predictors are common for seconds-ahead forecasting.
\item Spatio-temporal modeling is growing: transformers/graph-based models are increasingly used to capture coupling and longer horizons.
\end{itemize}
&
\begin{itemize}\setlength\itemsep{0pt}
\vspace{-0.2cm}
\item Domain shift robustness: sensitivity to band/vendor/config/mobility changes remains a key gap.
\item Fine-time unpredictability: sub-second capacity can be inherently hard to predict due to conflicting histories and hidden scheduler effects.
\item From forecasts to actions: uncertainty calibration and safe integration into rate control/scheduling/slicing loops remain underdeveloped.
\end{itemize}
\\
\hline

\textbf{Latency} &
\begin{itemize}\setlength\itemsep{0pt}
\vspace{-0.2cm}
\item Near-term prediction is strong: tabular models/ensembles and sequence models handle average delay and regime shifts effectively.
\item Tail/distribution modeling is advancing: mixture and EVT-style methods better represent heavy tails for risk-aware latency.
\item Network-scale inference is emerging: topology-aware models (GNN/graph transformers) enable digital-twin style delay prediction.
\end{itemize}
&
\begin{itemize}\setlength\itemsep{0pt}
\vspace{-0.2cm}
\item URLLC-grade tails: calibrated prediction of rare extremes under sparse tail data is still difficult.
\item Label quality: scalable, accurate one-way-delay ground truth and robustness to noise/nonstationarity remain challenging.
\item Closed-loop stability: joint latency--throughput--reliability objectives and deployment-time adaptation without feedback instability.
\end{itemize}
\\
\hline

\textbf{Reliability} &
\begin{itemize}\setlength\itemsep{0pt}
\vspace{-0.2cm}
\item Loss/PER prediction is established: tabular learners are common; sequence models help with bursty loss dynamics.
\item PHY-oriented learning is active: learned channel/CSI/impairment blocks and BLER surrogates improve reliability enablers.
\item Control integration is increasing: RL/optimization approaches appear in scheduling and slice/resource orchestration.
\end{itemize}
&
\begin{itemize}\setlength\itemsep{0pt}
\vspace{-0.2cm}
\item Rare failures and imbalance: extreme-event prediction and well-calibrated confidence remain hard.
\item Generalization and drift: robustness across cells/bands/vendors and lifecycle monitoring are under-addressed.
\item Realistic benchmarks: scarcity of public end-to-end failure logs and consistent evaluation setups limits reproducibility.
\end{itemize}
\\
\hline

\end{tabular}
\end{table*}

\subsection{Maturity of prediction across \glspl{KPI}}
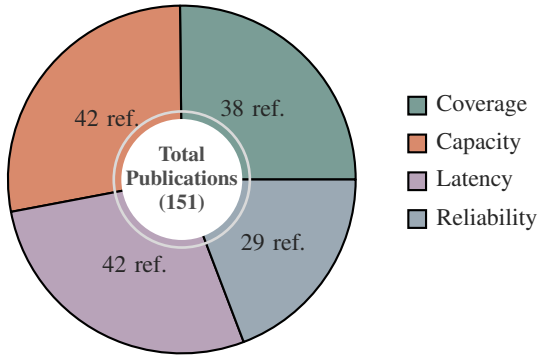
\begin{figure}[ht]
\centering

% Define the natural, earthy, and muted colors to match previous figures
\definecolor{covcol}{HTML}{7A9D96} % Mist
\definecolor{capcol}{HTML}{D98A6C} % Muted Salmon
\definecolor{latcol}{HTML}{B8A3B9} % Lavender Gray
\definecolor{relcol}{HTML}{9BA9B4} % Light Blue Grey

\begin{tikzpicture}[font=\small, text=black!85]
    
    % Draw the main pie chart
    \pie[
        text=legend,
        radius=2.3, % Slightly larger for a premium feel
        color={covcol, capcol, latcol, relcol},
        % Thick white borders give a clean, separated "exploded" look without geometry distortion
        pie/.style={draw=white, line width=2.5pt}, 
        sum=auto,
        after number = {~ref.}
    ]
    {38/Coverage, 42/Capacity, 42/Latency, 29/Reliability}

    % Overlay a white circle in the center to create a modern Donut Chart
    \fill[white] (0,0) circle (0.8cm);
    
    % Add a subtle, soft inner border to the donut hole
    \draw[black!15, line width=1pt] (0,0) circle (0.90cm);

    % Add descriptive text perfectly centered inside the donut hole
    \node[align=center, text=black!70, font=\footnotesize\bfseries] at (0,0) {Total\\Publications\\(151)};

\end{tikzpicture}

\caption{Distribution of references per KPI.}
\label{fig:pie}
\end{figure}

\begin{figure}[htbp]
    \centering

    % Define the natural, earthy, and muted colors for the first plot
    \definecolor{covcol}{HTML}{7A9D96} % Mist
    \definecolor{capcol}{HTML}{D98A6C} % Muted Salmon
    \definecolor{latcol}{HTML}{B8A3B9} % Lavender Gray
    \definecolor{relcol}{HTML}{9BA9B4} % Light Blue Grey

    % ==========================================
    % Subfloat (a): Publications per KPI
    % ==========================================
    \subfloat[]{
        \begin{tikzpicture}[font=\small]
        \begin{axis}[
            width=0.98\linewidth,  
            height=5.0cm, 
            ybar stacked,
            bar width=14pt, 
            ymin=0,
            ylabel={Number of publications},
            xlabel={Year of publication},
            ylabel style={text=black!90, font=\small\bfseries}, 
            xlabel style={text=black!90, font=\small\bfseries}, 
            axis line style={black!80, thick}, 
            grid=major,
            major grid style={dashed, black!15}, 
            symbolic x coords={{<2018},2018,2019,2020,2021,2022,2023,2024,2025},
            xtick=data,
            x tick label style={rotate=45, anchor=east, font=\small, text=black!85}, 
            y tick label style={font=\small, text=black!85},
            enlarge x limits=0.08,
            % Legend INSIDE the plot (Top-Left)
            legend style={
                at={(0.02,0.96)}, 
                anchor=north west, 
                legend columns=2, 
                font=\footnotesize, 
                text=black!90, 
                draw=black!60, 
                fill=white, 
                fill opacity=0.9, 
                text opacity=1
            }
        ]

        % Coverage
        \addplot+[fill=covcol, draw=black!60, line width=0.6pt] coordinates {
          (<2018,2) (2018,1) (2019,7) (2020,6) (2021,5) (2022,7) (2023,2) (2024,8) (2025,2)
        };
        
        % Capacity
        \addplot+[fill=capcol, draw=black!60, line width=0.6pt] coordinates {
          (<2018,4) (2018,3) (2019,0) (2020,5) (2021,4) (2022,6) (2023,7) (2024,11) (2025,3)
        };
        
        % Latency
        \addplot+[fill=latcol, draw=black!60, line width=0.6pt] coordinates {
          (<2018,6) (2018,1) (2019,1) (2020,4) (2021,3) (2022,3) (2023,6) (2024,13) (2025,5)
        };
        
        % Reliability
        \addplot+[fill=relcol, draw=black!60, line width=0.6pt] coordinates {
          (<2018,1) (2018,0) (2019,2) (2020,5) (2021,2) (2022,4) (2023,4) (2024,7) (2025,6)
        };

        \legend{Coverage, Capacity, Latency, Reliability}

        \end{axis}
        \end{tikzpicture}
        \label{fig:timeline_kpi}
    }\\ % End of first subfloat

    %\vspace{1em} % Spacing between the subfloats % Forces the second subfloat to appear on a new line

    % ==========================================
    % Subfloat (b): Publications per Model
    % ==========================================
    \subfloat[]{
        \begin{tikzpicture}[font=\small]
        \begin{axis}[
            width=0.98\linewidth,
            height=5.0cm, 
            ybar stacked,
            bar width=14pt,
            ymin=0,
            ylabel={Number of publications},
            xlabel={Year of publication},
            ylabel style={text=black!90, font=\small\bfseries}, 
            xlabel style={text=black!90, font=\small\bfseries}, 
            axis line style={black!80, thick}, 
            grid=major,
            major grid style={dashed, black!15}, 
            symbolic x coords={{<2018},2018,2019,2020,2021,2022,2023,2024,2025},
            xtick=data,
            x tick label style={rotate=45, anchor=east, font=\small, text=black!85}, 
            y tick label style={font=\small, text=black!85},
            enlarge x limits=0.08,
            % Legend INSIDE the plot (Top-Left)
            legend style={
                at={(0.02,0.96)}, 
                anchor=north west, 
                legend columns=2, 
                font=\footnotesize, 
                text=black!90, 
                draw=black!60, 
                fill=white, 
                fill opacity=0.9, 
                text opacity=1
            }
        ]
        
        % Unsupervised learning
        \addplot+[fill=model1col, draw=black!50, line width=0.6pt] coordinates {
          (<2018,0) (2018,0) (2019,0) (2020,1) (2021,1) (2022,0) (2023,0) (2024,0) (2025,0)
        };
        \addlegendentry{Unsupervised}
        
        % Classical supervised
        \addplot+[fill=model2col, draw=black!50, line width=0.6pt] coordinates {
          (<2018,10) (2018,4) (2019,4) (2020,10) (2021,4) (2022,5) (2023,8) (2024,15) (2025,6)
        };
        \addlegendentry{Classical supervised}
        
        % Ensemble
        \addplot+[fill=model3col, draw=black!50, line width=0.6pt] coordinates {
          (<2018,1) (2018,1) (2019,1) (2020,0) (2021,3) (2022,2) (2023,1) (2024,5) (2025,1)
        };
        \addlegendentry{Ensemble}
        
        % Deep Spatial (CNN)
        \addplot+[fill=model4col, draw=black!50, line width=0.6pt] coordinates {
          (<2018,0) (2018,0) (2019,1) (2020,2) (2021,1) (2022,4) (2023,2) (2024,9) (2025,0)
        };
        \addlegendentry{Deep Spatial (CNN)}
        
        % Deep Sequence
        \addplot+[fill=model5col, draw=black!50, line width=0.6pt] coordinates {
          (<2018,1) (2018,0) (2019,1) (2020,9) (2021,2) (2022,8) (2023,3) (2024,11) (2025,6)
        };
        \addlegendentry{Deep Sequence}
        
        % Federated learning / Other
        \addplot+[fill=model6col, draw=black!50, line width=0.6pt] coordinates {
          (<2018,0) (2018,0) (2019,0) (2020,0) (2021,1) (2022,1) (2023,3) (2024,3) (2025,2)
        };
        \addlegendentry{Other}
        
        \end{axis}
        \end{tikzpicture}
        \label{fig:timeline_model}
    } % End of second subfloat

    \caption{Chronological analysis of publications broken down by (a) KPI category and (b) ML/AI model type. "Other" in (b) refers to both Federated learning and GNN.}
    \label{fig:timeline_combined}
\end{figure}
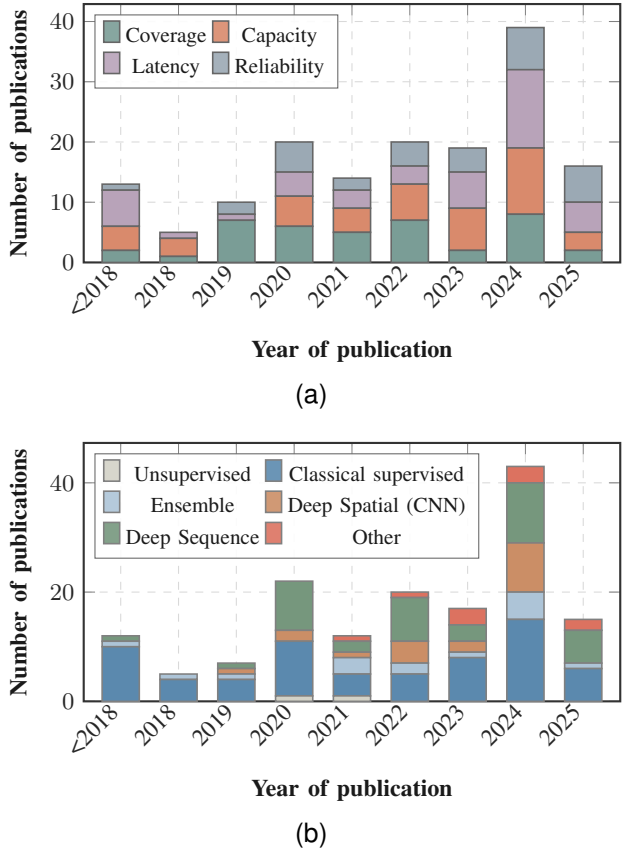

\gls{KPI} prediction has matured into a diverse and multifaceted research area, rather than one centered around a single dominant \gls{KPI}. The existing literature is distributed relatively evenly across \glspl{KPI} as shown in Fig.~\ref{fig:pie} for a total of 151 studies: \emph{coverage} (38 references), \emph{capacity} (42), and \emph{latency} (42), while \emph{reliability} is comparatively less represented (29). This balanced focus is also reflected over time, as shown in Fig.~\ref{fig:timeline_kpi}. According to this figure, all four \gls{KPI} categories have demonstrated sustained research activity spanning multiple years, with a marked increase in publications between 2023 and 2025. This trend underscores ongoing momentum and continuous innovation, rather than the maturation or saturation of any individual \gls{KPI} domain. We also observe a relative slowdown during 2020-2022. While bibliographic counts alone cannot prove causality, this dip is consistent with the COVID-19 period, during which limited access to campuses and facilities and longer review/production cycles may have reduced experimental throughput and delayed publication.  

The model timelines, in Fig.~\ref{fig:timeline_model}, further illustrate the field's dynamic nature. Classical supervised learning approaches appeared early and continue to serve as robust baselines. As datasets with richer temporal structures and shorter prediction horizons were introduced, deep sequence models gained prominence. Deep spatial (\gls{CNN}-style) models have become particularly relevant for tasks that require capturing spatial dependencies, such as coverage and radio-map prediction. More recently, paradigms such as \gls{FL} and \glspl{GNN} have begun to emerge, albeit sporadically and only in the later years, suggesting that while methodological innovation is occurring across all \gls{KPI} families, these newer approaches have yet to become mainstream.

\subsection{Reproducibility and Comparability}
Table~\ref{tab:integrated_public_data_code_clean} highlights the studies that make publicly accessible artifacts available across different \gls{KPI} families, while also illustrating a broader methodological issue: reproducibility and comparability remain limited. Although some papers release data, many do not provide the associated code, preprocessing steps, or precise train/validation/test splits, which prevents full end‑to‑end replication.

More critically, from a survey standpoint, direct comparisons across studies are often not possible because many studies introduce new datasets alongside new methods or rely on partially released derivatives of public datasets. This makes it difficult to determine whether reported improvements arise from the model itself, the data curation process, the labeling approach, or the evaluation design. Such fragmentation slows methodological progress. Without a small set of shared, openly available benchmark datasets, with standardized splits and metrics, together with well‑documented baseline models, advancements tend to appear as isolated case studies rather than as cumulative scientific evidence. Therefore, beyond encouraging broader artifact release, an essential open challenge is the creation of common datasets and comparative benchmark studies. These would enable iterative model development on stable reference tasks and facilitate validation, replication, and fair comparison across KPI families and model classes.

\begin{figure}[htbp] 
\centering

% Define the natural, earthy, and muted colors for KPIs (Subfigure a)
\definecolor{covcol}{HTML}{7A9D96} % Mist
\definecolor{capcol}{HTML}{D98A6C} % Muted Salmon
\definecolor{latcol}{HTML}{B8A3B9} % Lavender Gray
\definecolor{relcol}{HTML}{9BA9B4} % Light Blue Grey

% Variables
\def\nmeas{94}
\def\nsim{58}
\def\nsyn{3}
\def\nhyb{5}
\def\ntotal{160} % sum of above

% ---------------- Aggregated counts per KPI ----------------
\def\codeCov{1}    
\def\codeCap{4}    
\def\codeLat{4}    
\def\codeRel{4}    
\def\codeTotal{13}  

\def\dataCov{3}    
\def\dataCap{14}    
\def\dataLat{10}    
\def\dataRel{6}    
\def\dataTotal{33} 

% Master Layout Variables (Optimized for 1-column)
\def\barwidth{8.2}  % Standard 1-column width
\def\barheight{0.7} % Slightly sleeker height

\pgfmathsetmacro{\lcodeCov}{\barwidth*\codeCov/\codeTotal}
\pgfmathsetmacro{\lcodeCap}{\barwidth*\codeCap/\codeTotal}
\pgfmathsetmacro{\lcodeLat}{\barwidth*\codeLat/\codeTotal}
\pgfmathsetmacro{\lcodeRel}{\barwidth*\codeRel/\codeTotal}

\pgfmathsetmacro{\ldataCov}{\barwidth*\dataCov/\dataTotal}
\pgfmathsetmacro{\ldataCap}{\barwidth*\dataCap/\dataTotal}
\pgfmathsetmacro{\ldataLat}{\barwidth*\dataLat/\dataTotal}
\pgfmathsetmacro{\ldataRel}{\barwidth*\dataRel/\dataTotal}

% ---------------- Aggregated counts per model ----------------
\def\codeUns{2}
\def\codeCML{8}
\def\codeEns{3}
\def\codeDSP{4}
\def\codeDSeq{6}
\def\codeFed{3}
\def\codeTotalAll{26}

\def\dataUns{4}
\def\dataCML{27}
\def\dataEns{15}
\def\dataDSP{6}
\def\dataDSeq{16}
\def\dataFed{2}
\def\dataTotalAll{70}

\pgfmathsetmacro{\lcodeUns}{\barwidth*\codeUns/\codeTotalAll}
\pgfmathsetmacro{\lcodeCML}{\barwidth*\codeCML/\codeTotalAll}
\pgfmathsetmacro{\lcodeEns}{\barwidth*\codeEns/\codeTotalAll}
\pgfmathsetmacro{\lcodeDSP}{\barwidth*\codeDSP/\codeTotalAll}
\pgfmathsetmacro{\lcodeDSeq}{\barwidth*\codeDSeq/\codeTotalAll}
\pgfmathsetmacro{\lcodeFed}{\barwidth*\codeFed/\codeTotalAll}

\pgfmathsetmacro{\ldataUns}{\barwidth*\dataUns/\dataTotalAll}
\pgfmathsetmacro{\ldataCML}{\barwidth*\dataCML/\dataTotalAll}
\pgfmathsetmacro{\ldataEns}{\barwidth*\dataEns/\dataTotalAll}
\pgfmathsetmacro{\ldataDSP}{\barwidth*\dataDSP/\dataTotalAll}
\pgfmathsetmacro{\ldataDSeq}{\barwidth*\dataDSeq/\dataTotalAll}
\pgfmathsetmacro{\ldataFed}{\barwidth*\dataFed/\dataTotalAll}

% ====================================================================
% Subfigure A: KPIs
% ====================================================================
\subfloat[]{
\begin{tikzpicture}[font=\small]

% --------- First bar: Data availability ---------
\node[anchor=south west, text=black!90, font=\footnotesize\bfseries, inner sep=0pt] at (0, 0.8) {Data availability:};

\fill[covcol] (0,0) rectangle ++(\ldataCov,\barheight);
\node[align=center, text=black!85, font=\footnotesize] at ({0.5*\ldataCov},0.35) {1.87\%};

\fill[capcol] (\ldataCov,0) rectangle ++(\ldataCap,\barheight);
\node[align=center, text=black!85, font=\footnotesize] at ({\ldataCov+0.5*\ldataCap},0.35) {8.75\%};

\fill[latcol] ({\ldataCov+\ldataCap},0) rectangle ++(\ldataLat,\barheight);
\node[align=center, text=black!85, font=\footnotesize] at ({\ldataCov+\ldataCap+0.5*\ldataLat},0.35) {6.25\%};

\fill[relcol] ({\ldataCov+\ldataCap+\ldataLat},0) rectangle ++(\ldataRel,\barheight);
\node[align=center, text=black!85, font=\footnotesize] at ({\ldataCov+\ldataCap+\ldataLat+0.5*\ldataRel},0.35) {3.75\%};

\draw[thick, draw=black!60] (0,0) rectangle (\barwidth,\barheight);

% --------- Second bar: Code availability ---------
\node[anchor=south west, text=black!90, font=\footnotesize\bfseries, inner sep=0pt] at (0, -0.7) {Code availability:};

\fill[covcol] (0,-1.5) rectangle ++(\lcodeCov,\barheight);
\node[align=center, text=black!85, font=\footnotesize] at ({0.5*\lcodeCov},-1.15) {0.6\%};

\fill[capcol] (\lcodeCov,-1.5) rectangle ++(\lcodeCap,\barheight);
\node[align=center, text=black!85, font=\footnotesize] at ({\lcodeCov+0.5*\lcodeCap},-1.15) {2.5\%};

\fill[latcol] ({\lcodeCov+\lcodeCap},-1.5) rectangle ++(\lcodeLat,\barheight);
\node[align=center, text=black!85, font=\footnotesize] at ({\lcodeCov+\lcodeCap+0.5*\lcodeLat},-1.15) {2.5\%};

\fill[relcol] ({\lcodeCov+\lcodeCap+\lcodeLat},-1.5) rectangle ++(\lcodeRel,\barheight);
\node[align=center, text=black!85, font=\footnotesize] at ({\lcodeCov+\lcodeCap+\lcodeLat+0.5*\lcodeRel},-1.15) {2.5\%};

\draw[thick, draw=black!60] (0,-1.5) rectangle (\barwidth,-1.5+\barheight);

% --------- Legend (Harmonized tight 1-row layout) ---------
\fill[covcol, draw=black!60, thick] (0.0,-2.5) rectangle +(0.3, 0.3);
\node[right, text=black!90, font=\small] at (0.3,-2.35) {Coverage};

\fill[capcol, draw=black!60, thick] (2.1,-2.5) rectangle +(0.3, 0.3);
\node[right, text=black!90, font=\small] at (2.4,-2.35) {Capacity};

\fill[latcol, draw=black!60, thick] (4.1,-2.5) rectangle +(0.3, 0.3);
\node[right, text=black!90, font=\small] at (4.4,-2.35) {Latency};

\fill[relcol, draw=black!60, thick] (6.1,-2.5) rectangle +(0.3, 0.3);
\node[right, text=black!90, font=\small] at (6.4,-2.35) {Reliability};

\end{tikzpicture}
}
\\[1.5em] % Forces a vertical break between subfigures for 1-column layout
% ====================================================================
% Subfigure B: ML/AI Models 
% ====================================================================
\subfloat[]{
\begin{tikzpicture}[font=\small]

% --------- First bar: Data availability ---------
\node[anchor=south west, text=black!90, font=\footnotesize\bfseries, inner sep=0pt] at (0, 0.8) {Data availability:};

\fill[model1col] (0,0) rectangle ++(\ldataUns,\barheight);
\node[align=center, text=black, font=\footnotesize] at ({0.5*\ldataUns},0.35) {2.5\%};

\fill[model2col] (\ldataUns,0) rectangle ++(\ldataCML,\barheight);
\node[align=center, text=black, font=\footnotesize] at ({\ldataUns+0.5*\ldataCML},0.35) {16.8\%};

\fill[model3col] ({\ldataUns+\ldataCML},0) rectangle ++(\ldataEns,\barheight);
\node[align=center, text=black, font=\footnotesize] at ({\ldataUns+\ldataCML+0.5*\ldataEns},0.35) {9.3\%};

\fill[model4col] ({\ldataUns+\ldataCML+\ldataEns},0) rectangle ++(\ldataDSP,\barheight);
\node[align=center, text=black, font=\footnotesize] at ({\ldataUns+\ldataCML+\ldataEns+0.5*\ldataDSP},0.35) {4.6\%};

\fill[model5col] ({\ldataUns+\ldataCML+\ldataEns+\ldataDSP},0) rectangle ++(\ldataDSeq,\barheight);
\node[align=center, text=black, font=\footnotesize] at ({\ldataUns+\ldataCML+\ldataEns+\ldataDSP+0.5*\ldataDSeq},0.35) {10\%};

\fill[model6col] ({\ldataUns+\ldataCML+\ldataEns+\ldataDSP+\ldataDSeq},0) rectangle ++(\ldataFed,\barheight);
\node[align=center, text=black, font=\footnotesize] at ({\ldataUns+\ldataCML+\ldataEns+\ldataDSP+\ldataDSeq+0.5*\ldataFed},0.35) {1\%};

\draw[thick, draw=black!60] (0,0) rectangle (\barwidth,\barheight);

% --------- Second bar: Code availability ---------
\node[anchor=south west, text=black!90, font=\footnotesize\bfseries, inner sep=0pt] at (0, -0.7) {Code availability:};

\fill[model1col] (0,-1.5) rectangle ++(\lcodeUns,\barheight);
\node[align=center, text=black, font=\footnotesize] at ({0.5*\lcodeUns},-1.15) {0.8\%};

\fill[model2col] (\lcodeUns,-1.5) rectangle ++(\lcodeCML,\barheight);
\node[align=center, text=black, font=\footnotesize] at ({\lcodeUns+0.5*\lcodeCML},-1.15) {3.1\%};

\fill[model3col] ({\lcodeUns+\lcodeCML},-1.5) rectangle ++(\lcodeEns,\barheight);
\node[align=center, text=black, font=\footnotesize] at ({\lcodeUns+\lcodeCML+0.5*\lcodeEns},-1.15) {1.2\%};

\fill[model4col] ({\lcodeUns+\lcodeCML+\lcodeEns},-1.5) rectangle ++(\lcodeDSP,\barheight);
\node[align=center, text=black, font=\footnotesize] at ({\lcodeUns+\lcodeCML+\lcodeEns+0.5*\lcodeDSP},-1.15) {1.5\%};

\fill[model5col] ({\lcodeUns+\lcodeCML+\lcodeEns+\lcodeDSP},-1.5) rectangle ++(\lcodeDSeq,\barheight);
\node[align=center, text=black, font=\footnotesize] at ({\lcodeUns+\lcodeCML+\lcodeEns+\lcodeDSP+0.5*\lcodeDSeq},-1.15) {2.3\%};

\fill[model6col] ({\lcodeUns+\lcodeCML+\lcodeEns+\lcodeDSP+\lcodeDSeq},-1.5) rectangle ++(\lcodeFed,\barheight);
\node[align=center, text=black, font=\footnotesize, inner sep=0pt] at ({\lcodeUns+\lcodeCML+\lcodeEns+\lcodeDSP+\lcodeDSeq+0.5*\lcodeFed},-1.15) {1\%};

\draw[thick, draw=black!60] (0,-1.5) rectangle (\barwidth,-1.5+\barheight);

% --------- Legend for ML/AI Models (Stacked 2-row layout) ---------
% Row 1
\fill[model1col, draw=black!60, thick] (0.0,-2.5) rectangle +(0.3, 0.3);
\node[right, text=black!90, font=\small] at (0.3,-2.35) {Unsupervised};

\fill[model2col, draw=black!60, thick] (2.9,-2.5) rectangle +(0.3, 0.3);
\node[right, text=black!90, font=\small] at (3.2,-2.35) {Classical ML};

\fill[model3col, draw=black!60, thick] (5.8,-2.5) rectangle +(0.3, 0.3);
\node[right, text=black!90, font=\small] at (6.1,-2.35) {Ensemble};

% Row 2
\fill[model4col, draw=black!60, thick] (0.0,-3.1) rectangle +(0.3, 0.3);
\node[right, text=black!90, font=\small] at (0.3,-2.95) {Deep Spatial};

\fill[model5col, draw=black!60, thick] (2.9,-3.1) rectangle +(0.3, 0.3);
\node[right, text=black!90, font=\small] at (3.2,-2.95) {Deep Sequence};

\fill[model6col, draw=black!60, thick] (5.8,-3.1) rectangle +(0.3, 0.3);
\node[right, text=black!90, font=\small] at (6.1,-2.95) {Other};

\end{tikzpicture}
}

\caption{Percentage distribution of references providing code or data per KPI and ML model relative to the total dataset: (a) Number of references with code/data available per KPI, and (b) Number of references with code/data available per model.}
\label{fig:availability}
\end{figure}
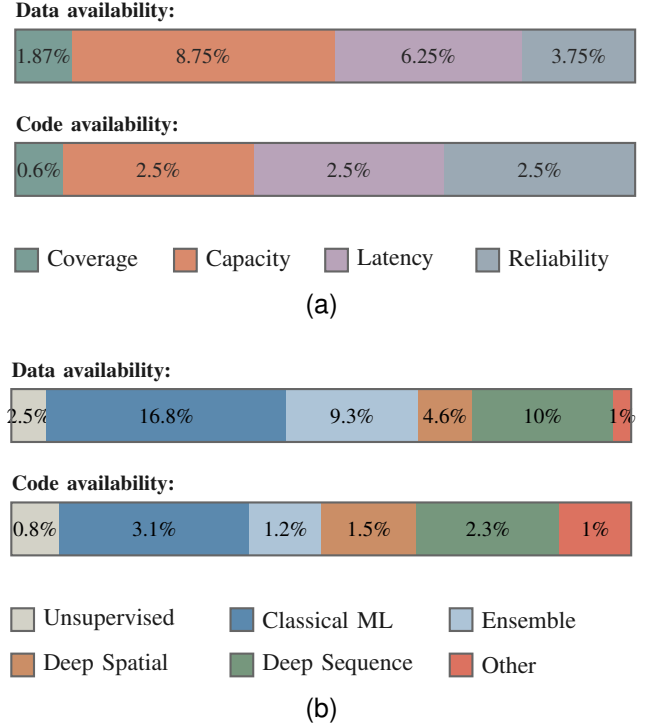

Fig.~\ref{fig:availability} shows the limited publicly available artifacts, where only 13 references provide code (about 8.1\% as shown in the figure), while 33 provide data (about 20\%). The \gls{KPI}-wise breakdown indicates that code and data availability are skewed toward capacity and latency, whereas coverage is the least supported. The model-wise breakdown further suggests that released artifacts are concentrated in widely used model families, notably classical supervised learning and deep sequence models, while other paradigms are represented only marginally.

\begin{figure}[htbp]
\centering

% Data (edit as needed)
\def\nmeas{94}
\def\nsim{58}
\def\nsyn{3}
\def\nhyb{5}
\def\ntotal{160} % sum of above

% Measured breakdown
\def\measCov{20}
\def\measCap{35}
\def\measLat{33}
\def\measRel{6}

% Simulated breakdown
\def\simCov{20}
\def\simCap{10}
\def\simLat{12}
\def\simRel{16}

% Synthetic breakdown
\def\synCov{1}
\def\synCap{0}
\def\synLat{1}
\def\synRel{1}

% Hybrid breakdown
\def\hybCov{2}
\def\hybCap{0}
\def\hybLat{0}
\def\hybRel{3}

% Adjusted layout variables for 1-column fit
\def\barwidth{8.2} % Scaled down to fit standard single-column width (~8.5cm)
\def\subgap{0.15}  % Reduced gap to save horizontal space

% Compute segment lengths (main bars)
\pgfmathsetmacro{\lmeas}{\barwidth*\nmeas/\ntotal}
\pgfmathsetmacro{\lsim}{\barwidth*\nsim/\ntotal}
\pgfmathsetmacro{\lsyn}{\barwidth*\nsyn/\ntotal}
\pgfmathsetmacro{\lhyb}{\barwidth*\nhyb/\ntotal}

% Compute sub-bar lengths (for each data source)
\pgfmathsetmacro{\lmeasCov}{\lmeas*\measCov/\nmeas}
\pgfmathsetmacro{\lmeasCap}{\lmeas*\measCap/\nmeas}
\pgfmathsetmacro{\lmeasLat}{\lmeas*\measLat/\nmeas}
\pgfmathsetmacro{\lmeasRel}{\lmeas*\measRel/\nmeas}

\pgfmathsetmacro{\lsimCov}{\lsim*\simCov/\nsim}
\pgfmathsetmacro{\lsimCap}{\lsim*\simCap/\nsim}
\pgfmathsetmacro{\lsimLat}{\lsim*\simLat/\nsim}
\pgfmathsetmacro{\lsimRel}{\lsim*\simRel/\nsim}

\pgfmathsetmacro{\lsynCov}{\lsyn*\synCov/\nsyn}
\pgfmathsetmacro{\lsynCap}{\lsyn*\synCap/\nsyn}
\pgfmathsetmacro{\lsynLat}{\lsyn*\synLat/\nsyn}
\pgfmathsetmacro{\lsynRel}{\lsyn*\synRel/\nsyn}

\pgfmathsetmacro{\lhybCov}{\lhyb*\hybCov/\nhyb}
\pgfmathsetmacro{\lhybCap}{\lhyb*\hybCap/\nhyb}
\pgfmathsetmacro{\lhybLat}{\lhyb*\hybLat/\nhyb}
\pgfmathsetmacro{\lhybRel}{\lhyb*\hybRel/\nhyb}

% Calculate sub-bar starting positions with gaps
\pgfmathsetmacro{\submeasstart}{0}
\pgfmathsetmacro{\subsimstart}{\lmeas + \subgap}
\pgfmathsetmacro{\subsynstart}{\lmeas + \lsim + 2*\subgap}
\pgfmathsetmacro{\subhybstart}{\lmeas + \lsim + \lsyn + 3*\subgap}

% Define natural, earthy, and muted colors using HTML hex codes
% Main bar colors
\definecolor{meascol}{HTML}{2A4B7C} % Midnight Blue
\definecolor{simcol}{HTML}{5D7736} % Olive Green
\definecolor{syncol}{HTML}{4A4A4A} % Charcoal Grey
\definecolor{hybcol}{HTML}{A88771} % Pastel Brown / Beige

% Sub-bar colors
\definecolor{covcol}{HTML}{7A9D96} % Mist
\definecolor{capcol}{HTML}{D98A6C} % Muted Salmon
\definecolor{latcol}{HTML}{B8A3B9} % Lavender Gray
\definecolor{relcol}{HTML}{9BA9B4} % Light Blue Grey

\begin{tikzpicture}[font=\small, >=Latex]

% Main bar: Data source types
\node[anchor=east] at (-0.2,0) {};

% Measured
\fill[meascol!70!white] (0,0) rectangle ++(\lmeas,0.7);
% Simulated
\fill[simcol!70!white] (\lmeas,0) rectangle ++(\lsim,0.7);
% Synthetic
\fill[syncol] ({\lmeas+\lsim},0) rectangle ++(\lsyn,0.7);
% Hybrid
\fill[hybcol] ({\lmeas+\lsim+\lsyn},0) rectangle ++(\lhyb,0.7);

% Outline
\draw[thick, draw=black!80] (0,0) rectangle (\barwidth,0.7);

% Labels above segments (White text for contrast against darker natural tones)
% Measured and Simulated fit inside
\node[align=center, text=white, font=\footnotesize\bfseries] at ({0.5*\lmeas},0.35) {Meas\\\nmeas};
\node[align=center, text=white, font=\footnotesize\bfseries] at ({\lmeas+0.5*\lsim},0.35) {Sim\\\nsim};

% Synthetic and Hybrid are too small for inner labels at this scale; pin them from above
\draw[draw=black!80, thin] ({\lmeas+\lsim+0.5*\lsyn}, 0.7) -- ({\lmeas+\lsim+0.5*\lsyn}, 0.95);
\node[align=center, text=black!90, font=\scriptsize\bfseries, anchor=south east, inner sep=1pt] at ({\lmeas+\lsim+0.5*\lsyn + 0.1}, 0.95) {Sy\\\nsyn};

\draw[draw=black!80, thin] ({\lmeas+\lsim+\lsyn+0.5*\lhyb}, 0.7) -- ({\lmeas+\lsim+\lsyn+0.5*\lhyb}, 0.95);
\node[align=center, text=black!90, font=\scriptsize\bfseries, anchor=south west, inner sep=1pt] at ({\lmeas+\lsim+\lsyn+0.5*\lhyb - 0.1}, 0.95) {Hy\\\nhyb};

% Sub-bars: breakdown by KPI (separated with gaps)
\def\subbary{-1.2}
\def\subbarheight{0.4}

% Measured sub-bar
\fill[covcol] (\submeasstart,\subbary) rectangle ++(\lmeasCov,\subbarheight);
\fill[capcol] ({\submeasstart+\lmeasCov},\subbary) rectangle ++(\lmeasCap,\subbarheight);
\fill[latcol] ({\submeasstart+\lmeasCov+\lmeasCap},\subbary) rectangle ++(\lmeasLat,\subbarheight);
\fill[relcol] ({\submeasstart+\lmeasCov+\lmeasCap+\lmeasLat},\subbary) rectangle ++(\lmeasRel,\subbarheight);
\draw[thick, draw=black!60] (\submeasstart,\subbary) rectangle ({\submeasstart+\lmeas},\subbary+\subbarheight);

% Simulated sub-bar
\fill[covcol] (\subsimstart,\subbary) rectangle ++(\lsimCov,\subbarheight);
\fill[capcol] ({\subsimstart+\lsimCov},\subbary) rectangle ++(\lsimCap,\subbarheight);
\fill[latcol] ({\subsimstart+\lsimCov+\lsimCap},\subbary) rectangle ++(\lsimLat,\subbarheight);
\fill[relcol] ({\subsimstart+\lsimCov+\lsimCap+\lsimLat},\subbary) rectangle ++(\lsimRel,\subbarheight);
\draw[thick, draw=black!60] (\subsimstart,\subbary) rectangle ({\subsimstart+\lsim},\subbary+\subbarheight);

% Connecting arrows/dashed lines from main bars to sub-bars
\draw[black!60, dashed, thick, ->] ({0.5*\lmeas}, 0) -- ({0.5*\lmeas}, -0.15) -- ({\submeasstart+0.5*\lmeas}, -0.65) -- ({\submeasstart+0.5*\lmeas}, \subbary+\subbarheight);
\draw[black!60, dashed, thick, ->] ({\lmeas+0.5*\lsim}, 0) -- ({\lmeas+0.5*\lsim}, -0.15) -- ({\subsimstart+0.5*\lsim}, -0.65) -- ({\subsimstart+0.5*\lsim}, \subbary+\subbarheight);

% Sub-bar labels (KPI for all sources) - Shrunk to \scriptsize to fit tighter widths
% Measured
\node[align=center, text=black!85, font=\scriptsize] at ({\submeasstart+0.5*\lmeasCov},\subbary+0.2) {\measCov};
\node[align=center, text=black!85, font=\scriptsize] at ({\submeasstart+\lmeasCov+0.5*\lmeasCap},\subbary+0.2) {\measCap};
\node[align=center, text=black!85, font=\scriptsize] at ({\submeasstart+\lmeasCov+\lmeasCap+0.5*\lmeasLat},\subbary+0.2) {\measLat};
\node[align=center, text=black!85, font=\scriptsize] at ({\submeasstart+\lmeasCov+\lmeasCap+\lmeasLat+0.5*\lmeasRel},\subbary+0.2) {\measRel};

% Simulated
\node[align=center, text=black!85, font=\scriptsize] at ({\subsimstart+0.5*\lsimCov},\subbary+0.2) {\simCov};
\node[align=center, text=black!85, font=\scriptsize] at ({\subsimstart+\lsimCov+0.5*\lsimCap},\subbary+0.2) {\simCap};
\node[align=center, text=black!85, font=\scriptsize] at ({\subsimstart+\lsimCov+\lsimCap+0.5*\lsimLat},\subbary+0.2) {\simLat};
\node[align=center, text=black!85, font=\scriptsize] at ({\subsimstart+\lsimCov+\lsimCap+\lsimLat+0.5*\lsimRel},\subbary+0.2) {\simRel};

% Horizontal Legend (Placed under the plot for 1-column fit)
\def\legendy{-2.0}
\def\boxsize{0.25}

\fill[covcol, draw=black!60, thick] (0.0, \legendy) rectangle +(\boxsize, \boxsize);
\node[right, text=black!90, font=\footnotesize] at (0.0+\boxsize, \legendy+0.125) {\textbf{Coverage}};

\fill[capcol, draw=black!60, thick] (2.1, \legendy) rectangle +(\boxsize, \boxsize);
\node[right, text=black!90, font=\footnotesize] at (2.1+\boxsize, \legendy+0.125) {\textbf{Capacity}};

\fill[latcol, draw=black!60, thick] (4.2, \legendy) rectangle +(\boxsize, \boxsize);
\node[right, text=black!90, font=\footnotesize] at (4.2+\boxsize, \legendy+0.125) {\textbf{Latency}};

\fill[relcol, draw=black!60, thick] (6.1, \legendy) rectangle +(\boxsize, \boxsize);
\node[right, text=black!90, font=\footnotesize] at (6.1+\boxsize, \legendy+0.125) {\textbf{Reliability}};

\end{tikzpicture}

\caption{Data source composition and KPI breakdown.}
\label{fig:barplot_datasource}
\end{figure}
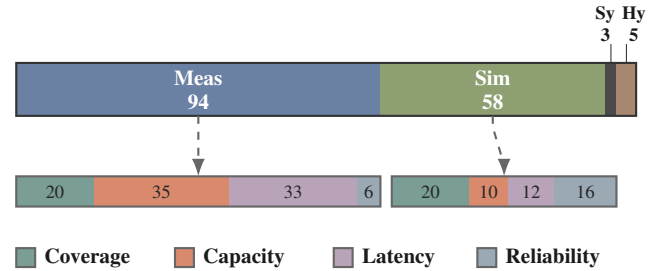

Fig.~\ref{fig:barplot_datasource} summarizes data-source usage and shows that measurement-based studies dominate overall (94 references), followed by simulation-based studies (58), while synthetic (3) and hybrid (5) datasets are rare. The \gls{KPI} breakdown in the sub-bars reveals a more nuanced picture. For latency, work is overwhelmingly grounded in measurements with only a modest contribution from simulation (5), suggesting that researchers tend to evaluate delay behavior directly in real testbeds and operational networks. Capacity prediction shows a similar, though slightly less extreme, bias, with no capacity-oriented hybrids or synthetic-only datasets in the current corpus.
By contrast, coverage and reliability show a more balanced or even simulation-leaning profile. Coverage studies are split almost evenly between measurement and simulation, reflecting the dual role of drive tests and ray-tracing/link-level simulation in radio-coverage modeling; only a couple of studies use hybrid data, and a single study is synthetic-only. Reliability is the most ``virtual'' \gls{KPI} in terms of data: only 6 measurement-based studies versus 16 simulation-based, complemented by 3 hybrid and 1 synthetic dataset. This aligns with the difficulty and cost of collecting large, labeled reliability datasets (e.g., very low \gls{BLER}/packet-loss ratio events, extreme tail behavior), which are often easier to explore through controlled simulators and analytical models. Overall, the figure highlights a strong measurement-centric practice for capacity and especially latency, a mixed measurement/simulation ecosystem for coverage, and a simulation- and hybrid-heavy approach for reliability. It also shows that hybrid and synthetic datasets remain rare, pointing to opportunities for more systematic use of combined measurement–simulation pipelines and synthetic data generation, particularly for reliability and cross-\gls{KPI} studies where real-world data are scarce or difficult to label.

\subsection{Evaluation Metrics}
\begin{figure}[htbp]
    \centering
    \vspace{-0.5em} % Reduces white space above the figure
    
    % Defined the natural, earthy, and muted color palette 
    \definecolor{covcol}{HTML}{7A9D96} % Mist
    \definecolor{capcol}{HTML}{D98A6C} % Muted Salmon
    \definecolor{latcol}{HTML}{B8A3B9} % Lavender Gray
    \definecolor{relcol}{HTML}{9BA9B4} % Light Blue Grey
    \definecolor{centercol}{HTML}{4A4A4A} % Charcoal Grey

    \begin{tikzpicture}[
        >=Stealth, 
        font=\footnotesize, % Scaled down from \small for a tighter fit
        % Core line style for organic routing
        arrow/.style={->, line width=1.4pt, rounded corners=4pt}, 
        % Central node styling
        center/.style={circle, draw=centercol!90!black, line width=1pt, 
                       fill=centercol!10, text=centercol!90!black,
                       drop shadow={opacity=0.15, shadow xshift=1.5pt, shadow yshift=-1.5pt},
                       minimum size=0.4cm, align=center, inner sep=4pt, font=\footnotesize\bfseries},
        % Card styling for the KPIs (narrowed for 1-column fit)
        kpicard/.style={rectangle, rounded corners=6pt, draw, line width=1pt,
                        drop shadow={opacity=0.15, shadow xshift=1.5pt, shadow yshift=-1.5pt},
                        text width=3.5cm, align=center, inner sep=5pt}
    ]

    %------------------------------------------------
    % 1. Central Node
    %------------------------------------------------
    \node[center] (center) at (0,0) {Eval.\\[1pt]Metric};

    %------------------------------------------------
    % 2. KPI Cards (Compact X-Layout)
    %------------------------------------------------
    
    % Coverage Card (Top Right)
    \node[kpicard, draw=covcol!90!black, fill=covcol!15] (cov) at (2.2, 1.8) {
        \textbf{\textcolor{covcol!80!black}{Coverage}} \\[0.5mm]
        \textcolor{covcol!80!black}{\rule{2.4cm}{0.8pt}} \\[1mm]
        MSE $\cdot$ RMSE $\cdot$ MAE \\
        $R^2$ $\cdot$ NRMSE $\cdot$ NMSE \\
        MAPE $\cdot$ SMAPE $\cdot$ STD \\
        Corr.
    };

    % Capacity Card (Bottom Right)
    \node[kpicard, draw=capcol!90!black, fill=capcol!15] (cap) at (2.2, -1.8) {
        \textbf{\textcolor{capcol!80!black}{Capacity}} \\[0.5mm]
        \textcolor{capcol!80!black}{\rule{2.4cm}{0.8pt}} \\[1mm]
        MSE $\cdot$ RMSE $\cdot$ MAE \\
        MAPE $\cdot$ NRMSE $\cdot$ ARE \\
        Rel.\ err.\ \% $\cdot$ $R^2$
    };

    % Reliability Card (Top Left)
    \node[kpicard, draw=relcol!90!black, fill=relcol!15] (rel) at (-2.2, 1.8) {
        \textbf{\textcolor{relcol!80!black}{Reliability}} \\[0.5mm]
        \textcolor{relcol!80!black}{\rule{2.4cm}{0.8pt}} \\[1mm]
        BER $\cdot$ BLER $\cdot$ PLR \\
        Loss\ \% $\cdot$ Outage $P$ \\
        Reliab.\ $R$ $\cdot$ F1 \\
        Accuracy $\cdot$ ROC-AUC
    };

    % Latency Card (Bottom Left)
    \node[kpicard, draw=latcol!90!black, fill=latcol!15] (lat) at (-2.2, -1.8) {
        \textbf{\textcolor{latcol!80!black}{Latency}} \\[0.5mm]
        \textcolor{latcol!80!black}{\rule{2.4cm}{0.8pt}} \\[1mm]
        MSE $\cdot$ RMSE $\cdot$ MAE \\
        NMSE $\cdot$ NMAE $\cdot$ MAPE \\
        NRMSE $\cdot$ p95 $\cdot$ p99 \\
        CDF err.\ $\cdot$ NLL
    };

    %------------------------------------------------
    % 3. Elegant Bezier S-Curve Routing
    %------------------------------------------------
    
    % Right side curves
    \draw[arrow, draw=covcol!90!black] (center.35) to[out=45, in=270] (cov.south);
    \draw[arrow, draw=capcol!90!black] (center.325) to[out=315, in=90] (cap.north);
    
    % Left side curves
    \draw[arrow, draw=relcol!90!black] (center.145) to[out=135, in=270] (rel.south);
    \draw[arrow, draw=latcol!90!black] (center.215) to[out=225, in=90] (lat.north);

    %------------------------------------------------
    % 4. Background Shadow Box 
    %------------------------------------------------
    \begin{scope}[on background layer]
        \node[
            fill=gray!8, 
            draw=gray!30, 
            thick, 
            rounded corners=10pt, 
            drop shadow={opacity=0.1, shadow xshift=2pt, shadow yshift=-2pt}, 
            fit=(current bounding box), 
            inner sep=6pt % Reduced margin around the elements
        ] {};
    \end{scope}

    \end{tikzpicture}
    
    %\vspace{-0.5em} % Pulls the caption slightly closer to the image
    \caption{Evaluation metrics broken down by KPI category.}
    \label{fig:predoutput}
    \vspace{-1.5em} % Reduces the white space below the entire figure block
\end{figure}
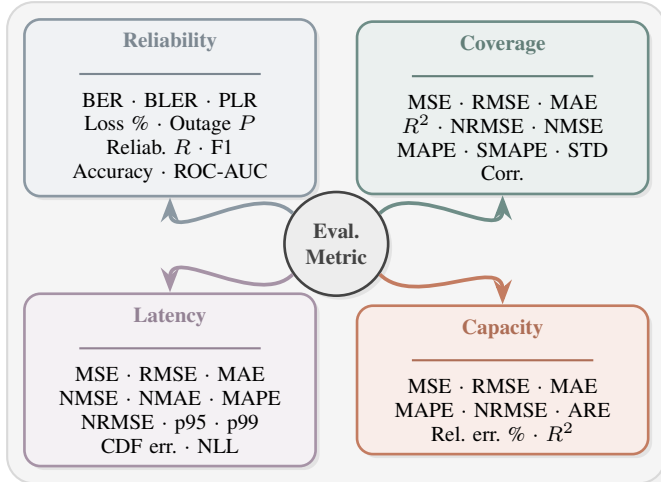

A second, equally practical obstacle to benchmarking \gls{KPI}‑prediction methods concerns the inconsistency of evaluation metrics. Even when studies focus on the same \gls{KPI} family, they often report different, and sometimes incompatible, metrics, making cross‑paper comparison ambiguous. As illustrated in Fig.~\ref{fig:predoutput}, coverage and capacity studies predominantly rely on point‑error measures such as \gls{MAE}, \gls{RMSE}, or \gls{MSE} (including normalized variants). In contrast, latency- and reliability-related studies frequently combine point‑error metrics with tail‑sensitive measures (such as p95/p99, \gls{CDF} error, or \gls{NLL} for probabilistic models) or with classification metrics such as F1 and \gls{ROC}‑\gls{AUC} when predicting violations or outages. Consequently, two methods cannot be compared fairly when one is optimized and evaluated using \gls{RMSE}, capturing average‑case accuracy, while the other is tuned to minimize tail‑exceedance risk or improve distributional fit. In some cases, the choice of metric can even reverse conclusions: for example, a model might achieve lower \gls{RMSE} yet consistently underpredict extreme values, rendering it unsuitable for \gls{URLLC}‑type scenarios.
Therefore, beyond releasing data and code, the field requires comparative studies conducted on common open datasets, where competing models are evaluated under a unified metric suite, for instance, \gls{MAE}/\gls{RMSE} combined with calibrated uncertainty, tail‑quantile errors for latency and reliability, and explicit out-of-domain stress tests. Such an approach would enable direct, reproducible, and decision‑relevant comparisons across models.

\subsection{Generalizability and Robustness}
Beyond metrics, generalization claims are often underspecified. In operational networks, the dominant failure mode is not \gls{iid}\ test error, but \emph{domain shift}. Three forms are particularly important and should be treated explicitly:
\subsubsection{Cross-cell (or cross-route) shift}
A model trained on a subset of cells may fail on unseen cells due to differences in propagation conditions, antenna configurations, neighbor topology, traffic mix, and local optimization settings. Accordingly, cross-cell prediction should be assessed using protocols that hold out entire cells or spatial regions, not random record-level splits, and it often benefits from architectures that condition on cell metadata and topology (e.g., neighbor graphs) in addition to time series.
\subsubsection{Cross-operator shift}
Models that appear robust within one operator can degrade when transferred to another operator with different vendor stacks, parameter settings, measurement pipelines, subscriber populations, or operational policies. Credible cross-operator claims, therefore, require multi-operator datasets or clearly defined transfer studies (e.g., pretraining on one domain and adapting using limited target data), with transparent reporting of adaptation cost, performance recovery, and calibration.
\subsubsection{Cross-band shift}
For coverage and capacity tasks, changing frequency bands alters path loss, penetration, interference patterns, and scheduler/link-adaptation behavior; even at the same site, the feature-to-\gls{KPI} mapping can change materially. Models should therefore incorporate frequency-aware inputs and normalization, learn shared representations with explicit domain labels, or employ adaptation/calibration layers. Evaluation should include leave-one-band-out testing (or low-shot adaptation), rather than only within-band splits.

\subsection{Offline Prediction}
Another notable trend across the surveyed studies is the predominant focus on offline prediction accuracy, with limited attention paid to how these predictions are actually leveraged by real-time decision-makers, such as schedulers, congestion controllers, bitrate selectors, slice orchestrators, admission controllers, or SON functions. Critical operational details, including actuation constraints and stability guarantees, are often left unspecified. In cases where integration is explored, it is typically through trace-driven emulation or narrowly scoped prototypes, rather than through persistent in-network deployments. As a result, several important practical questions remain inadequately addressed. For example, how frequently are predictive models queried in live systems? What are the specific costs and consequences of incorrect predictions, both overprediction and underprediction, for various controllers? How do these controllers respond to challenges such as concept drift, network outages, or partial observability? Furthermore, it is often unclear how safety constraints are maintained when the model's predictions are erroneous.

These concerns are especially pressing for latency and reliability \glspl{KPI}, where the impact of rare tail events is significant, and the costs of failure are inherently asymmetric. To advance beyond the current paradigm of ``predictions as a dashboard'', the field must prioritize closed-loop evaluations with well-defined actuation policies. It is equally important to conduct ablation studies that disentangle the contributions of the predictor from those of the controller and to systematically report failure scenarios, such as system oscillations, overly conservative resource allocation, or masked congestion, in which predictive models may inadvertently degrade overall system performance. This shift will be essential for translating advances in prediction accuracy into tangible, reliable benefits in real-world network operations.

\subsection{Trade-offs among Inference Latency and Accuracy in Real-time \gls{RAN}}
\textcolor{black}{A review of the literature indicates that model selection is frequently driven by training convenience rather than deployment suitability. In practice, evaluating a model requires considering the full end-to-end ``sense-predict-act'' pipeline, which encompasses data collection, preprocessing, hardware-constrained inference, and actuation delays. This broader pipeline can easily dominate overall latency. While some studies evaluate lightweight models, many propose complex architectures (e.g., Transformers, large \glspl{CNN}, or ensembles) without addressing deployment budgets, hardware limits, or worst-case end-to-end delays. Ultimately, high offline accuracy is insufficient if the system cannot support timely decision-making. Since control decisions often operate on millisecond timescales, average inference time is inadequate; operators require bounded tail latency (e.g., p99) and jitter to prevent stale predictions and maintain closed-loop stability.}

\textcolor{black}{This trade-off forms an accuracy-latency frontier that must be explicitly reported. For short prediction horizons, minor inference delays can severely degrade practical value, yet overly compact models risk underfitting critical, localized anomalies (e.g., interference spikes). To bridge this gap, researchers should systematically leverage and report practical design patterns such as (i) cascaded predictors, (ii) early-exit inference, (iii) model distillation, pruning, and quantization, (iv) feature gating, and (v) caching for slowly varying embeddings.}

\begin{figure}[ht]
\centering

% Define the natural, earthy, and muted colors
\definecolor{covcol}{HTML}{7A9D96} % Mist
\definecolor{capcol}{HTML}{D98A6C} % Muted Salmon
\definecolor{latcol}{HTML}{B8A3B9} % Lavender Gray
\definecolor{relcol}{HTML}{9BA9B4} % Light Blue Grey

\begin{tikzpicture}[font=\small]
\begin{axis}[
    view={135}{25},
    width=0.48\textwidth, % Slightly widened for better text breathing room
    height=6cm,
    ylabel={Prediction horizon (s)},
    zlabel={Density},
    ylabel style={sloped, rotate=-1, text=black!90, font=\small\bfseries},
    zlabel style={text=black!90, font=\small\bfseries},
    % KPI axis (x):
    xmin=0.5, xmax=4.5,
    xtick={1,2,3,4},
    xticklabels={Capacity,Latency,Coverage,Reliability},
    xticklabel style={text=black!90, font=\small\bfseries},
    % Horizon axis (y): LOG SCALE with actual seconds values
    ymode=log,
    ymin=0.0001, ymax=10000,
    ytick={0.001,0.01,0.1,1,10,100,1000},
    yticklabels={0.001,0.01,0.1,1,10,100,1000},
    yticklabel style={font=\scriptsize, text=black!85},
    % Density axis (z): NORMALIZED so sum per KPI = 1
    zmin=0, zmax=0.35,
    zticklabel style={text=black!85, font=\footnotesize},
    zmajorgrids,
    grid=both,
    grid style={dashed, black!15}, % Softer grid lines for a cleaner look
    axis line style={black!80, thick}, % Harmonized axis lines
    ticklabel style={font=\small},
]

% ------------------------------------------------------------
% NORMALIZED DATA: Each count divided by total studies per KPI
% Capacity total = 40, Latency total = 20, Coverage total = 12, Reliability total = 3
% Sum of all bars per KPI = 1.0 (probability density)
% Using actual time values in seconds for log scale
% ------------------------------------------------------------

% Capacity (x=1) - normalized by total=40
% Added darker draw outlines for better 3D depth perception
\addplot3[ybar, bar width=0.2, bar shift=0, fill=capcol!95, draw=capcol!60!black, line width=0.4pt]
coordinates {
  (1, 0.0001, 0.050)  % 0 s: 2/40 = 0.050
  (1, 0.1, 0.075)     % 0.1 s: 3/40 = 0.075
  (1, 1, 0.175)       % 1 s: 7/40 = 0.175
  (1, 3, 0.100)       % 3 s: 4/40 = 0.100
  (1, 4, 0.125)       % 4 s: 5/40 = 0.125
  (1, 5, 0.125)       % 5 s: 5/40 = 0.125
  (1, 7, 0.100)       % 7 s: 4/40 = 0.100
  (1, 10, 0.075)      % 10 s: 3/40 = 0.075
  (1, 12, 0.075)      % 12 s: 3/40 = 0.075
  (1, 20, 0.050)      % 20 s: 2/40 = 0.050
  (1, 48, 0.025)      % 48 s: 1/40 = 0.025
  (1, 300, 0.025)     % 300 s: 1/40 = 0.025
};

% Latency (x=2) - normalized by total=20
\addplot3[ybar, bar width=0.2, bar shift=0, fill=latcol!95, draw=latcol!60!black, line width=0.4pt]
coordinates {
  (2, 0.01, 0.050)    % 0.01 s: 1/20 = 0.050
  (2, 0.032, 0.050)   % 0.032 s: 1/20 = 0.050
  (2, 0.08, 0.050)    % 0.08 s: 1/20 = 0.050
  (2, 0.1, 0.050)     % 0.1 s: 1/20 = 0.050
  (2, 0.5, 0.100)     % 0.5 s: 2/20 = 0.100
  (2, 1, 0.050)       % 1 s: 1/20 = 0.050
  (2, 3, 0.100)       % 3 s: 2/20 = 0.100
  (2, 4, 0.050)       % 4 s: 1/20 = 0.050
  (2, 5, 0.050)       % 5 s: 1/20 = 0.050
  (2, 7, 0.100)       % 7 s: 2/20 = 0.100
  (2, 10, 0.100)      % 10 s: 2/20 = 0.100
  (2, 12, 0.050)      % 12 s: 1/20 = 0.050
  (2, 20, 0.050)      % 20 s: 1/20 = 0.050
  (2, 48, 0.050)      % 48 s: 1/20 = 0.050
  (2, 900, 0.050)     % 900 s: 1/20 = 0.050
  (2, 7200, 0.050)    % 7200 s: 1/20 = 0.050
};

% Coverage (x=3) - normalized by total=12
\addplot3[ybar, bar width=0.2, bar shift=0, fill=covcol!95, draw=covcol!60!black, line width=0.4pt]
coordinates {
  (3, 0.0001, 0.167)  % 0 s: 2/12 ≈ 0.167
  (3, 0.001, 0.250)   % 0.001 s: 3/12 = 0.250
  (3, 0.005, 0.167)   % 0.005 s: 2/12 ≈ 0.167
  (3, 0.01, 0.167)    % 0.01 s: 2/12 ≈ 0.167
  (3, 0.032, 0.083)   % 0.032 s: 1/12 ≈ 0.083
  (3, 0.08, 0.083)    % 0.08 s: 1/12 ≈ 0.083
  (3, 0.1, 0.083)     % 0.1 s: 1/12 ≈ 0.083
};

% Reliability (x=4) - normalized by total=3
\addplot3[ybar, bar width=0.2, bar shift=0, fill=relcol!95, draw=relcol!60!black, line width=0.4pt]
coordinates {
  (4, 0.5, 0.333)     % 0.5 s: 1/3 ≈ 0.333
  (4, 900, 0.333)     % 900 s: 1/3 ≈ 0.333
  (4, 9000, 0.333)    % 9000 s: 1/3 ≈ 0.333
};

\end{axis}
\end{tikzpicture}
\caption{Normalized prediction horizon distribution for deep sequence-modeling works per KPI. The density is normalized to the number of included studies per KPI.}
\label{fig:pred_horizon}
\end{figure}
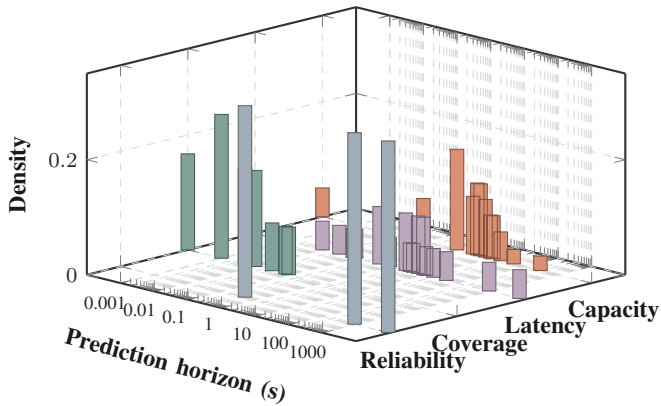

Fig.~\ref{fig:pred_horizon} illustrates how maximum prediction horizons for deep sequence models vary across \glspl{KPI}, tightly reflecting their operational timescales. Coverage forecasts generally operate at 1–10~ms horizons, capturing fine-grained, subframe-level \gls{PHY} dynamics (e.g., \gls{SINR}, \gls{CQI}). Conversely, capacity predictions typically span 1–10~s to support throughput adaptation. Latency and reliability prediction horizons are more varied: latency spans from tens of milliseconds for real-time queueing effects to multi-minute horizons for cell-level planning. Similarly, reliability exhibits a bimodal distribution, ranging from sub-second forecasts for \gls{URLLC} behaviors to day-scale predictions for long-term \gls{QoE} stability.

\subsection{Data Lifecycle}
Across all \gls{KPI} families, the data-generating environment is inherently dynamic, shaped by factors such as new user equipment chipsets, evolving scheduler configurations, the introduction of new frequency bands, changing traffic patterns, seasonal effects, and modifications to network topology. Despite this, most studies operate under the assumption of a static training set and conduct only one-off evaluations. Only a limited subset of research explicitly addresses lifecycle challenges such as continual learning, model drift, and retraining frequency, and even then, these issues are typically explored in controlled environments, using a single testbed or a narrow range of device types and tasks, rather than in long-term, real-world deployments.

This disconnect creates a practical gap for operators and system builders, who require more than just an accurate model; they need a comprehensive maintenance strategy. This includes clear triggers for data collection, robust labeling strategies (particularly for latency and reliability tail events), monitoring metrics that detect silent performance degradation, and safe mechanisms for updating models, such as A/B testing, canary deployments, and reliable rollback procedures.

Furthermore, several existing approaches rely on data sources that are costly or fragile to obtain, such as rooted devices, proprietary logging tools, or active probing, which complicates the process of continuously refreshing datasets. A key open direction for the field is to design predictive models with sustainable observability in mind: leveraging features that are realistically available in production environments and adopting update strategies, such as federated or edge training, weak supervision, or self-supervision, that minimize dependence on curated labels. This shift is essential to ensure that KPI prediction systems remain robust, maintainable, and effective as real-world conditions evolve.

\subsection{Explainability}
\textcolor{black}{The studies reviewed in this survey indicate that explainability is frequently treated as an auxiliary add-on (e.g., reporting global feature importance) rather than as a core requirement tied to operational decision-making. In networking, explainability is not solely about trust; it is fundamentally about enabling action. When a predicted \gls{KPI} deteriorates, operators or automated controllers must understand which levers to adjust, whether that involves power settings, scheduling weights, handover parameters, slice resources, congestion control strategies, or routing decisions.}

\textcolor{black}{For coverage and capacity, global feature rankings can be useful, but they are often insufficient for diagnosing localized issues (e.g., why a specific cell, route segment, or time period underperforms). For latency and reliability, particularly when models are designed to capture tail behavior or \gls{SLA} violations, explanations must go beyond point predictions to clarify risk drivers, uncertainty, and the expected cost asymmetry of errors (over- vs.\ under-prediction). The challenge becomes even more pronounced in cross-layer or multi-\gls{KPI} scenarios, where relationships are intertwined across the \gls{RAN}, transport, and application layers and where spurious correlations can lead to incorrect interventions.}

\textcolor{black}{A critical next step is to align explanation outputs with network operations by providing: (i) per-prediction uncertainty estimates (not only point predictions), (ii) localized causal hypotheses or operator-interpretable contributing factors at the right granularity (cell/\gls{UE}/flow/time window), (iii) sensitivity to controllable parameters (what-if or counterfactual analyses), and (iv) clear distinctions between directly measured and inferred variables. Studies should also document who the explanation is for (engineer, operator, controller), what decision it supports, and how explanation fidelity is validated (e.g., through intervention tests, ablations, or consistency checks), rather than relying solely on generic saliency maps.}

\subsection{Sustainability}
Finally, the studies reviewed in this survey highlight that sustainability often represents an overlooked cost in \gls{KPI} prediction research. Here, sustainability encompasses more than just environmental impact; it includes the ongoing operational burden of (i) collecting measurements, whether through active probing or specialized logging, (ii) storing and processing high-frequency data traces, which in some cases reach tens of milliseconds granularity or involve very large data volumes, (iii) training and retraining models, particularly those relying on deep learning or ensemble methods, and (iv) ensuring reliable operation through monitoring, drift detection, and incident response.

Notably, the most accurate approaches in several \gls{KPI} families are also the most resource-intensive to maintain. For example, multi-modal coverage models may require image data and semantic segmentation; fine-grained capacity predictors often depend on low-level \gls{PHY}/\gls{MAC} counters; latency models might need synchronized one-way delay ground truth; and reliability predictors can rely on long-term labeled failure logs. These requirements can make continuous operation costly and complex. To ensure a sustainable trajectory for both research and deployment, greater emphasis is needed on (a) parsimonious feature sets that reflect what is realistically observable in production environments, (b) compact models with predictable inference costs, (c) standardized benchmarks that enable cumulative progress rather than repeated reinvention with each new dataset, and (d) transparent reporting of measurement overhead and lifecycle costs. Without these considerations, the field risks developing increasingly sophisticated predictors that are not economically or operationally viable for continuous use.

\subsection{Privacy and Security Considerations}
\textcolor{black}{A final limitation that cuts across \gls{KPI} families is that many high-performing predictors implicitly depend on \gls{UE}-level or otherwise user-related measurements (e.g., fine-grained radio measurements, per-device mobility traces, application/\gls{QoE} indicators, or detailed session logs). These data can be sensitive because they may enable re-identification through spatio-temporal patterns, reveal behavioral attributes, or be linkable to external information even when explicit identifiers are removed. In addition, \gls{ML} systems introduce privacy risks beyond raw data access, including membership inference (whether a user/device participated in training), model inversion (recovering features from model outputs), and leakage through overly detailed explanations or logs.}

\textcolor{black}{These concerns create a practical tension: richer \gls{UE}-level features often improve accuracy and tail performance, particularly for latency and reliability, but also increase privacy exposure and compliance burden. As a result, privacy must be treated as a first-class design constraint rather than an afterthought. Promising directions include: (i) data minimization (collect only what is necessary; reduce granularity where possible), (ii) aggregation and anonymization with explicit threat models (recognizing that naive anonymization can fail), (iii) on-device inference or edge-local processing to avoid centralizing raw UE traces, (iv) federated learning combined with secure aggregation to limit server visibility into individual updates, (v) differential privacy mechanisms (at the data, gradient, or output level) when training on user-related measurements, and (vi) access control and auditing for both training pipelines and deployed model endpoints.}

\textcolor{black}{From a research perspective, an open challenge is to report privacy-relevant details with the same rigor as accuracy: Answer what \gls{UE}-level signals are used, what level of aggregation is assumed, what adversary is considered, and what accuracy degradation is incurred when privacy protections are enabled. Without this, it is difficult to assess whether a KPI prediction method is deployable in realistic operational and regulatory environments.}

\begin{table*}[ht]
\centering
\caption{References with Publicly Available Data and/or Code Across KPI Families.}
\label{tab:integrated_public_data_code_clean}
\footnotesize
\setlength{\tabcolsep}{8pt}
\renewcommand{\arraystretch}{1.1}
\begin{tabular}{@{}p{1.5cm}|p{1.2cm}p{0.8cm}p{1cm}p{1cm}p{8.5cm}@{}}

\hline
\textbf{KPI family} & \textbf{Reference} & \textbf{Year} & \textbf{Public data} & \textbf{Public code} & \textbf{Repository} \\
\hline

% =========================================================
% Capacity (sorted by year)
% =========================================================
\multirow{13}{*}{\textbf{Capacity}}
& \cite{elsherbiny20204g} & 2020 & \cmark & \xmark & \url{https://doi.org/10.5683/SP2/EQWKO1} \\

& \cite{cap_04} & 2020 & \cmark & \xmark &
\url{https://datasets.simula.no/hsdpa/}; \url{https://github.com/NYU-METS/Main} \\

& \cite{azmin2022bandwidth} & 2022 & \cmark & \xmark &
\url{https://github.com/uccmisl/5Gdataset} \\

& \cite{cap_03} & 2022 & \cmark & \xmark &
\url{https://github.com/NYU-METS/Main}; \url{https://github.com/uccmisl/5Gdataset} \\

& \cite{cap_10} & 2022 & \cmark & \xmark &
\url{https://vehicle2x.net/v2x-measurements/} \\

& \cite{kim5G23} & 2023 & \cmark & \xmark &
\url{https://networking.umn.edu/lumos5g} \\

& \cite{biernacki2024throughput} & 2024 & \cmark & \xmark &
\textbf{\url{https://github.com/uccmisl/5Gdataset}}\newline
\url{https://github.com/arczello/5g_traces} \\

& \cite{cap_02} & 2024 & \cmark & \cmark &
\textbf{\url{https://github.com/fraunhoferhhi/BerlinV2X}}; \url{https://github.com/cedric-cnam/5G3E-dataset}; \textbf{\url{https://github.com/NYU-METS/Main}}; code \url{https://github.com/ds-kiel/bandseer} \\

& \cite{cap_07} & 2024 & \cmark & \xmark &
\url{https://github.com/uccmisl/5Gdataset} \\

& \cite{cap_11} & 2024 & \cmark & \xmark &
\url{https://github.com/NUWiNS/MASS_2024_Throughput_Prediction} \\

& \cite{cap_27} & 2024 & \cmark & \xmark &
\url{https://github.com/uccmisl/5Gdataset} \\

& \cite{cap_32} & 2024 & \cmark & \xmark &
\url{https://github.com/fraunhoferhhi/BerlinV2X} \\

& \cite{cap_01} & 2025 & \cmark & \cmark &
\url{https://github.com/laiguokun/multivariate-time-series-data}; \url{https://github.com/QData/spacetimeformer}; \url{https://github.com/zhouhaoyi/Informer2020}; code: \url{https://github.com/ds-kiel/FORESEE} \\

\hline
% =========================================================
% Latency (sorted by year)
% =========================================================
\multirow{10}{*}{\textbf{Latency}}
& \cite{khatouni2019machine} & 2019 & \cmark & \xmark &
\url{https://github.com/MONROE-PROJECT/data-exporter} \\

& \cite{lat_05} & 2021 & \cmark & \xmark &
\url{https://www.cs.cornell.edu/people/egs/meridian/data.php}; \url{https://ci.nii.ac.jp/naid/80016337478/en/}; \url{http://www.pdos.lcs.mit.edu/~strib/pl_app}; \url{https://labs.ripe.net/datarepository/data-sets/nlanr-amp-data}; \url{https://ci.nii.ac.jp/naid/10024634329/en/} \\

& \cite{lat_06} & 2021 & \cmark & \xmark &
\url{https://github.com/foroughsh/KTH-traces} \\

& \cite{mostafavi2023data} & 2023 & \cmark & \cmark &
\url{https://github.com/samiemostafavi/wireless-pr3d} \\

& \cite{fadhil2023estimation} & 2023 & \cmark & \xmark &
\url{https://ieee-dataport.org/open-access/5g-campus-networks-measurement-traces} \\

& \cite{larsson2023domain} & 2023 & \cmark & \xmark &
\url{https://github.com/foroughsh/KTH-traces} \\

& \cite{taghia2024congruent} & 2024 & \cmark & \cmark &
\url{https://github.com/EricssonResearch/congruent-federated-learning} \\

& \cite{cai2024adaptive} & 2024 & \xmark & \cmark &
\url{https://github.com/MerelyBreeze/CCSND} \\

& \cite{mostafavi2025probabilistic} & 2025 & \cmark & \cmark &
\url{https://github.com/samiemostafavi/wireless-tpp} \\

& \cite{lat_23} & 2025 & \cmark & \xmark &
\url{https://github.com/filipkrasniqi/QoSML}; \url{https://github.com/BNN-UPC/NetworkModelingDatasets/tree/master/datasets_v0} \\
\hline

% =========================================================
% Coverage (sorted by year)
% =========================================================
\multirow{3}{*}{\textbf{Coverage}}
& \cite{rsrp_rsrq_ml_01_16} & 2021 & \cmark & \xmark &
\url{https://dx.doi.org/10.21227/vmw5-c226} \\

& \cite{snr_dl_radiounet_2021} & 2021 & \cmark & \cmark &
\url{https://radiomapseer.github.io/}; code \url{https://github.com/RonLevie/RadioUNet} \\

& \cite{rsrp_dnn_01_13} & 2025 & \cmark & \xmark &
\textbf{\url{https://radiomapseer.github.io/}} \\
\hline

% =========================================================
% Reliability (sorted by year)
% =========================================================
\multirow{4}{*}{\textbf{Reliability}}
& \cite{rel_gnn_01} & 2021 & \cmark & \cmark &
\url{https://knowledgedefinednetworking.org/} \\

& \cite{rel_07} & 2024 & \cmark & \xmark &
\url{https://github.com/lisixu-JLAU/DeepMIMO-V1} \\

& \cite{rel_per_07} & 2024 & \cmark & \cmark &
\url{https://github.com/safiqul/packet-loss-prediction} \\

& \cite{rel_per_02} & 2025 & \cmark & \cmark &
\url{https://smartdata.polito.it/rtc-classification/}; code \url{https://github.com/gianlucaperna/Retina} \\
\hline
\end{tabular}\\
\vspace{0.2cm}
{Note: Only references with a working repository link and accessible data are included in this table. References that provide a link but lack available data or code in the repository are excluded (by our last check on 13 Feb. 2026). \textbf{Bolded} repositories highlight datasets cited by multiple authors}
\end{table*}

\subsection{Future Works}
\label{sec:futureworks}
Despite advances in \gls{KPI} predictors such as classical \gls{ML}, deep learning, and structure-aware models, challenges remain in cross-domain generalization amid configuration and traffic drift, data efficiency with limited labels, uncertainty-aware decision support, and deployment constraints such as latency and energy use. Addressing these requires AI-native architectures to improve transferability, representation learning, and probabilistic modeling. Future work will systematically evaluate KPI prediction based on foundation-model paradigms.

As AI-native networking evolves, the evaluation, deployment, and comparison of advanced KPI prediction approaches are fragmented. Clarifying the roles of generative and foundational architectures is crucial. For instance, while LLMs aren't designed for time-series forecasting, they can augment KPI pipelines by enhancing model selection, feature generation, telemetry standardization, and analytics workflows. Integrating LLMs into closed-loop systems presents challenges like ensuring reliability and controllability.
Pretrained sequence models trained on diverse telemetry data can improve transferability across networks, reducing retraining needs. Open questions include pretraining on heterogeneous data, adapting with limited labels, and maintaining stability in changing environments.
Extending pretraining to incorporate spatial relationships via topology-aware models can address multi-entity KPI challenges such as cell interactions and interference. Practical issues include managing dynamic topologies, addressing vendor differences, and enabling scalable inference.
For uncertainty-aware decision-making, diffusion-based methods enable probabilistic forecasting, scenario analysis, and uncertainty quantification, especially in the tails of the distribution. Their deployment in real-time systems requires balancing computational cost, calibration, and operational fit.

\section{Conclusions}
\label{sec:conclusions}
This survey demonstrates that KPI prediction in cellular networks has evolved significantly, transitioning from isolated, single-layer estimators to sophisticated, cross-layer, data-driven forecasting frameworks that closely reflect the realities of modern network engineering and operations. For coverage, capacity, latency, and reliability, a clear trend emerges: while scalable tabular models remain effective in stable environments with well-instrumented features, the most advanced solutions increasingly leverage sequence modeling to address non-stationary dynamics and burstiness, spatial and topology-aware learning to capture inter-cell and path dependencies, and probabilistic or tail-aware outputs to meet the stringent requirements of SLA and URLLC scenarios where rare events are critical. The primary challenges now lie less in incremental accuracy improvements and more in ensuring deployment robustness, specifically, achieving reliable generalization across frequency bands, vendors, geographies, and traffic patterns; providing calibrated uncertainty estimates under sparse failure or tail data; and integrating predictive models safely into closed-loop control systems such as scheduling, congestion management, and slice assurance, all while maintaining effective monitoring and lifecycle management. Ultimately, KPI prediction is poised to become a foundational component of AI-native network operations, but its practical value depends on developing robust, uncertainty-aware, and maintainable models that can withstand distributional shifts and operational constraints.

%\section*{Acknowledgments}
%This should be a simple paragraph before the References to thank those individuals and institutions who have supported your work on this article.

\bibliographystyle{IEEEtran}
\bibliography{KPI_Prediction_ML.bib}

{\appendix[]

{\appendices
\section*{Literature collection, screening, and coding protocol \label{app:screening}}
\subsection{Search strategy and sources}
\label{app:search}
We conducted a structured search in major digital libraries and indexing services for wireless networking and ML research, using combinations of keywords from three groups: (i) mobile networks and architecture (e.g., 4G/5G/5G-Advanced/6G, RAN, O-RAN, NWDAF), (ii) KPI terminology (e.g., throughput/capacity, latency/delay, coverage/RSRP/SINR, reliability/BLER/PLR), and (iii) prediction/ML terminology (e.g., forecasting, machine learning, DL, LSTM/Transformer, GNN, FL).

We queried both publisher-hosted digital libraries and indexing/search services. The main publication/issuing-body groups represented in our search include IEEE, ACM, Elsevier, Springer Nature (e.g., SpringerLink), MDPI, Wiley, Nature Publishing Group, AAAS (Science), and arXiv. In addition, we used cross-venue indexing/search services commonly employed for comprehensive coverage (e.g., Google Scholar, Scopus, and DBLP). We also searched standards and technical bodies relevant to KPI definition and network architecture, including 3GPP, ETSI, ITU-T, and IETF (RFCs). Finally, we included selected industry/grey-literature sources when they contained primary empirical or technical material relevant to KPI prediction, including (non-exhaustively) Ericsson, Zenodo, 5G PPP, 5GAA, SBC, IEICE, and KICS.

\subsection{Inclusion and exclusion criteria}
\label{app:criteria}
A study is included if it
\begin{itemize}
    \item targets a mobile/cellular-network context and predicts at least one KPI-relevant metric;
    \item uses a data-driven method (ML/DL/statistical learning) with a defined prediction setup (inputs/targets);
    \item reports sufficient information to identify, at a minimum, the KPI family and the learning/model family.
\end{itemize}

A study is excluded if it
\begin{itemize}
    \item is purely reactive optimization/control without an explicit KPI prediction task; or
    \item is a survey/tutorial/editorial (used only as background rather than counted as a KPI-prediction study).
\end{itemize}

The final corpus comprises 151 studies.

\subsection{Data extraction and coding template}
\label{app:coding}
Each included study is annotated using a structured coding template aligned with the taxonomy in Fig.~\ref{fig:taxonomy}. For every paper, we extract the following attributes.

\subsubsection{Target Predicted-KPI}
\label{subsubsec:kpi_type}
Each study is assigned to a single primary KPI family, including capacity, latency, coverage, or reliability, based on its main prediction target, i.e., the specific KPI $y$ being estimated or forecasted.

In cases where a study predicts multiple KPIs spanning different families (for example, both latency and loss), we determine the primary KPI family based on the study's explicitly stated objective or optimization target. 

\subsubsection{Prediction ML/AI Model}
\label{subsubsec:model_family}
We map each study to an ML/AI model family (e.g., classical supervised learning, ensemble methods, deep spatial models, deep sequence models, unsupervised learning, and other/emerging paradigms). 

\subsubsection{Data source type}
\label{subsubsec:data_source}
We categorize the training/evaluation data as \emph{measurement-based}, \emph{simulation-based}, \emph{synthetic}, or \emph{hybrid}. A study is classified as hybrid only when both measurements and simulations play an active and explicit role in the learning or evaluation process.

\subsubsection{Prediction attributes}
\label{subsubsec:pred_attributes}

We extract three prediction attributes:
\begin{itemize}
    \item Protocol stack: We record the network layer at which prediction is formulated (PHY, RAN, network, application, or cross-layer).
    \item Output and evaluation metrics: We record whether a study reports point regression, probabilistic/density estimation, or classification formulations (e.g., threshold violation), as well as the evaluation metrics used. 
    \item Horizon: We record the prediction horizon when explicitly stated. 
\end{itemize}

\subsubsection{Reproducibility (code/data availability)}
\label{subsubsec:reproducibility}

To assess reproducibility, we record whether authors provide public code and/or public data. 
}

\end{document}